\documentclass[a4paper]{JHEP3}

\usepackage{amsmath,amsfonts,amssymb,graphicx,rotfloat}
\usepackage[tight]{subfigure}
\usepackage[numbers,sort&compress]{natbib}

\def\be#1#2\ee{\begin{equation}\label{#1}#2\end{equation}}
\def\ben#1\een{\begin{equation}\nonumber #1\end{equation}}
\def\bea#1#2\eea{\begin{eqnarray}\label{#1}#2\end{eqnarray}}
\def\bean#1\eean{\begin{eqnarray}\nonumber#1\end{eqnarray}}
\def\muc{\multicolumn}
\def\note#1{\marginpar{\raggedright\if@twoside\ifodd\c@page\raggedleft\fi\fi\sf\scriptsize Note: #1}}

\def\bN{\mathbb{N}}
\def\bZ{\mathbb{Z}}
\def\cA{\mathcal{A}}
\def\cN{\mathcal{N}}
\def\cO{\mathcal{O}}

\def\C{\text{C}}
\def\V{\text{V}}
\def\Sym{\mathbf{Sym}}
\def\Anti{\mathbf{Anti}}
\def\Adj{\mathbf{Adj}}
\def\cc{c.c.}
\def\Im{\mathrm{Im}\,}
\def\Re{\mathrm{Re}\,}
\def\rep#1{\mathbf{#1}}

\def\half{\frac{1}{2}}

\def\incfig#1#2{\def\a{#1}\def\b{}\ifx\a\b{\bf\LARGE EMPTY IMAGE}\else\includegraphics[width=#2\linewidth,trim=0mm 5mm 0mm 5mm,clip]{#1}\fi}

\def\widefig#1#2#3{\begin{figure}[ht]\centering\incfig{#1}{1}\caption{#3}\label{#2}\end{figure}}
\def\twofig#1#2#3#4{\begin{figure}[ht]\centering\subfigure[\label{#3_a}]{\incfig{#1}{.5}}\hfill\subfigure[\label{#3_b}]{\incfig{#2}{.5}}\caption{#4}\label{#3}\end{figure}}

\def\tabfig#1#2#3#4#5#6{\begin{figure}[ht]\begin{minipage}[b]{#5\linewidth}\incfig{#1}{1}\par\vspace{0pt}\end{minipage}\hfill\begin{minipage}[b]{#6\linewidth}\centering #4\par\vspace{3.5mm}\end{minipage}\caption{#3}\label{#2}\end{figure}}

\renewcommand{\arraystretch}{1.3}
\title{Mapping an Island in the Landscape}

\author{%
Florian Gmeiner${}^1$ and Gabriele Honecker${}^2$\vspace{2ex}\\%
\vspace{1ex}{}$^1$NIKHEF, Kruislaan 409, 1098 SJ Amsterdam, The Netherlands\\%
\vspace{3ex}{}$^2$PH-TH Division, CERN, 1211 Geneva 23, Switzerland%
}

\abstract{We provide a complete classification and statistical analysis of all type IIA orientifold compactifications with
intersecting D6--branes on the orbifold $T^6/\bZ'_6$. The total number of four dimensional
N=1 supersymmetric models is found to be $\cO(10^{23})$.
After a statistical analysis of the gauge sector properties of all possible solutions, we study three subsets
of configurations which contain the chiral matter sector of the standard model, a Pati--Salam or $SU(5)$ GUT model,
respectively. We find $\cO(10^{15})$ compactifications with an MSSM and $\cO(10^{11})$ models with a Pati--Salam sector.
Along the way we derive an explicit algebraic formulation for the computation of the non--chiral matter spectrum
for all $\bZ_N$ orbifolds.}

\preprint{CERN-PH-TH/2007-139\\NIKHEF/2007-017}

\begin{document}

%
%
\section{Introduction}\label{intro}
One of the most important issues for string theory today is to make contact
with reality. Concretely one would like to make testable predictions
either for cosmology or particle physics.
We are facing two major problems in this endeavour, however. Firstly, there
is no known explicit construction of a string theory model that resembles our
universe. Secondly, there exists a huge amount of possible solutions,
too big to be ever computed completely in an explicit manner.

The latter issue, known as the ''string theory landscape``~\cite{su03,sc06},
could however not only be regarded as a problem, but on the contrary as a tool to
actually make predictions\footnote{See also~\cite{lu07} for a discussion of the
relation between the landscape and standard model physics.}.
To do so, one has to employ methods different from standard string theory
model building. Using a statistical approach~\cite{do03} to analyse distributions
of properties in large subsets of the landscape, one might hope to
find patterns in the huge space of solutions.
If present, these patterns might give important insights into the overall shape
of the landscape. On the one hand, they could be a valuable guide for model
building, hinting at interesting regions that should be investigated more closely.
On the other hand, the issue of correlations of properties within the ensemble of models
is of great importance. Finding correlations between low energy observables
in several distinct corners of the landscape could not only be
interpreted as a sign for a more fundamental principle of string theory yet to be
discovered, but might also be used to make concrete predictions for
experiments, thereby assuming that these correlations exist everywhere in the
landscape.

Up to now our understanding of the landscape is very limited and only a few
broad studies have been carried out. One can distinguish two possible approaches
to the problem. In a true statistical approach one can try to find general features of large
classes of string compactifications without explicitly constructing
them~\cite{dedo04,bghlw04,dedo05,adv05,ku06} by trying to find a good measure on the
space of solutions. Alternatively, in a more direct approach, one might try
to construct as many solutions as possible explicitly and apply statistical
methods to analyse this ensemble. This is the method that we
use for our present analysis in the context of type II orientifold
compactifications. For this class of models there exist already several
studies on different backgrounds~\cite{kuwe05,gbhlw05,gm05,gmst06,dota06,gls07,ftz07},
but similar methods have also been applied to Gepner models~\cite{dhs04a,dhs04b,adks06}
and heterotic compactifications~\cite{di06,le06,dlsw07}.
With this work we would like to make a contribution to this exploration, adding
a new class of vacua obtained from type II orientifold compactifications
with intersecting D--branes\footnote{For a general review on these constructions
see~\cite{bkls06}, for a summary of statistical work in particular backgrounds
see~\cite{gm06}.} on the specific toroidal background $T^6/\bZ'_6$.
Compared to earlier studies, we will provide new tools to compute the
non--chiral matter sector which has not been studied statistically so far. 

Dealing with statistics, there are several caveats not to be overlooked.
One of them concerns the finiteness of solutions~\cite{acdo06}. This turns out to be not
a problem in our case since it can be shown explicitly that the number of
solutions is finite.
Moreover, one has to be extremely careful if one decides to make statistical
predictions for a larger class of models based on smaller, explicitly analysed
examples. In this context it is not a priori clear if the subset, usually
chosen by a random method, can be used to make valid statements about all
possible solutions due to unwanted correlations~\cite{dile06}.
Fortunately this will also not be of our concern since we have been able
to explicitly construct and classify all possible supersymmetric
compactifications on this background. 

In~\cite{gls07} the statistics of models on $T^6/\bZ_6$ has
been considered. This is a closely related variant of the orbifold background studied here,
which differs only in the way the $\bZ_6$ orbifold group acts on the torus lattice.
One drawback of this geometry is the enforced simultaneous 
absence of symmetric and antisymmetric representations from the spectrum, which makes it
impossible to obtain phenomenologically interesting $SU(5)$ GUT models.
Due to the different embedding of the orbifold action, this is not the case
for the $\bZ'_6$ variety.
For earlier work on type II orientifold
models on the $\bZ'_6$ orbifold, see~\cite{balo06,balo07}, where a specific model
with the gauge group of the standard model has been studied.
However, due to the correction of a sign in the orientifold projection of exceptional cycles,
we do not recover their exact model.

This article is organised as follows. In Section~\ref{setup} we explain
the geometric setup of the $T^6/\bZ'_6$ orientifold and the constraints from
supersymmetry, tadpole cancellation and K--theory. We also discuss the computation of the 
complete (non--chiral) spectrum for $T^6/\bZ_M$ orbifolds.
In Section~\ref{methods} we give an analytic proof of the finiteness of possible
solutions to the constraining equations and explain our methods of statistical
analysis. The results of a systematic study of the
distribution of gauge sector properties are presented in Section~\ref{results}.
In particular we look for the frequency distribution of models with a
standard model, Pati--Salam or $SU(5)$ gauge group and the appropriate
chiral matter content.
Finally we sum up our results and give an outlook to further directions of research in
Section~\ref{conclusions}.
Some technical details are collected in the appendix.

%
%
\section{Setup}\label{setup}
In this section we review the geometric setup of the $T^6/\bZ'_6$
orientifold and the possible D6--brane configurations.
Furthermore we summarise the consistency conditions for supersymmetric
models and give algebraic formulae for the complete matter spectrum in
terms of the intersection numbers of three--cycles.
Our notation and conventions are similar to those of~\cite{hoot04,gls07}
to simplify the comparison between both geometries, but differ
from~\cite{balo06}.

\subsection{Geometry}\label{geometry}
We assume a factorisation of $T^6$ into three two--tori, which can be described
by complex variables $z^j, j=1,2,3$. The $\bZ'_6$ orbifold group action is
generated by
\ben
  \theta: z^j \rightarrow e^{2\pi i v_j}z^j,
\een
with shift vector $\vec{v}=\frac{1}{6}(1,2,-3)$.\footnote{Note the difference to the
shift vector for the $\bZ_6$ action of~\cite{gls07}, which reads
$\vec{v}=\frac{1}{6}(1,1,-2)$. In particular, there is no 
permutation symmetry among the two--tori in the present case.}
The orbifold has twelve $\bZ_6$ fixed points in the origin of $T_1^2$
multiplied by different fixed points on $T^2_2 \times T^2_3$, 
nine $\bZ_3$ fixed points on $T^2_1 \times T^2_2$ and 16 $\bZ_2$ fixed points on $T^2_1 \times T^2_3$.

\begin{figure}[ht]  
\begin{center}
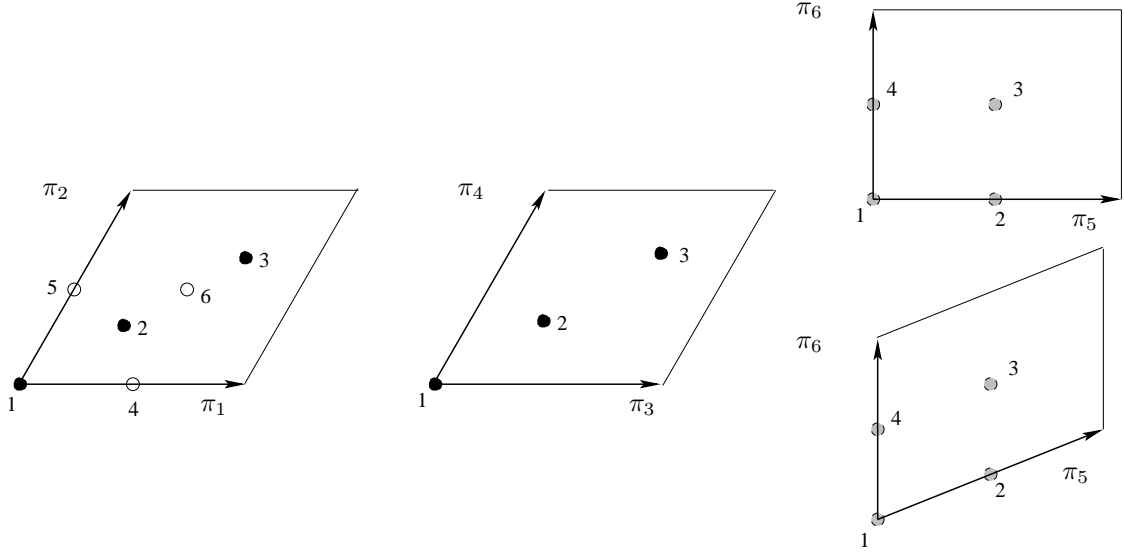
\end{center}
\caption{Fixed points of the $T^6/\bZ'_6$ orbifold. Circles on $T^2_2 \times T^2_3$
denote fixed points of $\theta$. $T^2_3$ is fixed under $\theta^2$, $T^2_2$ is
fixed under $\theta^3$.
On $T^2_1$, point 1 is fixed under $\theta$, points 4,5,6 are fixed under $\theta^3$ 
and points 2,3 are fixed under $\theta^2$.
The horizontal radius along $\pi_5 - b \pi_6$ on $T^2_3$ is called $R_1$, the vertical extension along $\pi_6$ is 
denoted by $R_2$. Both options for an untilted ({\bf a}) and tilted ({\bf b}) shape of $T^2_3$, parametrised
by $b=0,1/2$, are shown.}
\label{Fig:Z6p_geo}
\end{figure}

The complex structures on  $T^2_1 \times T^2_2$ 
are fixed by the orbifold action, whereas on $T^2_3$, the complex structure is 
parameterised by the continuous ratio of radii $R_2/R_1$ along the torus one--cycles
$\pi_6$ and $\pi_5 - b\pi_6$. The discrete variable
$b=0, 1/2$ corresponds to the two different possible choices of shapes for $T^2_3$.
This geometric setup is summarised in Figure~\ref{Fig:Z6p_geo}.

\subsection{Three--cycles}\label{cycles}
In order to implement an orientifold action and add D6--branes to this orbifold,
we need to know the number of homologically different three--cycles.
The Hodge numbers for the untwisted (U) and twisted sectors are given by (cf.~\cite{klra00}),
\be{eqhodge}\begin{array}{rclrclrclrcl}
  h^{U}_{1,1} &=& 3,&\quad h^{\theta+\theta^5}_{1,1}&=&12,&\quad
  h^{\theta^2+\theta^4}_{1,1} &=& 12,&\quad h^{\theta^3}_{1,1} &=& 8,\\
  h^{U}_{2,1} &=& 1,&\quad h^{\theta+\theta^5}_{2,1}&=&0,&\quad
  h^{\theta^2+\theta^4}_{2,1} &=& 6,&\quad h^{\theta^3}_{2,1} &=& 4.
\end{array}\ee
This gives us a total dimension of the space of three--cycles on $T^6/\bZ'_6$ of
$2(1+h_{2,1})=24$. They can be split into three groups.
Firstly there are four cycles inherited from the underlying six--torus ($h^{U}_{2,1}$).
Secondly two groups of exceptional cycles arise from the $\theta^2+\theta^4$ and $\theta^3$
sectors. We will have a close look at all of them in the following.

\subsubsection{Bulk cycles}
We begin with the three--cycles inherited from the torus, which will be called
``bulk cycles'' in the rest of this paper. They can be expanded in terms of the basis
\ben\begin{array}{rcl}
\rho_1 = 2(1+\theta+\theta^2)\pi_{1,3,5} &=&2\left(\pi_{1,3,5}+\pi_{2,4-3,-5}+\pi_{2-1,-4,5}\right)\\
&=& 2\left(\pi_{1,3,5}-2 \pi_{2,4,5}+\pi_{2,3,5}+\pi_{1,4,5}\right),\\
\rho_2 = 2(1+\theta+\theta^2)\pi_{2,3,5} &=&2\left(\pi_{1,4,5}+\pi_{2,-3,-5}+\pi_{2-1,3-4,5}\right)\\
&=& 2\left(2\pi_{1,4,5}+2 \pi_{2,3,5}-\pi_{2,4,5}-\pi_{1,3,5}\right),\\
\rho_3 = 2(1+\theta+\theta^2)\pi_{1,3,6} &=&2\left(\pi_{1,3,6}+\pi_{2,4-3,-6}+\pi_{2-1,-4,6}\right)\\
&=& 2\left(\pi_{1,3,6}-2 \pi_{2,4,6}+\pi_{2,3,6}+\pi_{1,4,6}\right),\\
\rho_4 = 2(1+\theta+\theta^2)\pi_{2,3,6} &=&2\left(\pi_{1,4,6}+\pi_{2,-3,-6}+\pi_{2-1,3-4,6}\right)\\
&=& 2\left(2\pi_{1,4,6}+2 \pi_{2,3,6}-\pi_{2,4,6}-\pi_{1,3,6}\right).
\end{array}\een
An arbitrary bulk cycle is characterised by the wrapping numbers $(n_i,m_i)$ along $\pi_{2i-1}$ and 
$\pi_{2i}$ on $\otimes_{i=1}^3 T^2_i$
which transform under the $\bZ'_6$ generator $\theta$ according to
\be{eqnmtransform}
\left(\begin{array}{cc}
n_1 & m_1 \\ n_2 & m_2 \\n_3 & m_3
 \end{array}\right)
\stackrel{\theta}{\longrightarrow}
\left(\begin{array}{cc}
-m_1 & n_1+m_1 \\ -(n_2+m_2) & n_2 \\-n_3 & -m_3
\end{array}\right)
\stackrel{\theta}{\longrightarrow}
\left(\begin{array}{cc}
-(n_1+m_1) & n_1 \\ m_2 & -(n_2+m_2) \\ n_3 & m_3
\end{array}\right).
\ee
The intersection numbers among the bulk three cycles are given by\footnote{The bulk intersection numbers 
given here differ from those in~\cite{balo06} by a sign whereas the intersection numbers of exceptional cycles 
agree. The choice of absolute signs is just a convention, but the relative sign between bulk and exceptional
cycles is important.
An extensive computer analysis reveals that the relative sign in~\cite{balo06} leads to half--integer 
multiplicities of (anti)symmetric representations on some D6--branes. Moreover, the generic formulae for
the non--chiral spectra displayed in the present article  for the first time
in section~\protect\ref{Sec:OpenSpectrum} only reproduce the non--chiral 
spectra of~\cite{bgk00,bcs04} for the present choice of relative signs of intersection numbers.
}
\ben
\begin{aligned}
\rho_1 \circ \rho_2 &= \rho_3 \circ \rho_4 = 0 ,\\
\rho_2 \circ \rho_3 &= \rho_1 \circ \rho_4 = 2 ,\\
\rho_1 \circ \rho_3 &= \rho_2 \circ \rho_4 = 4 .
\end{aligned}
\een
In order to shorten the notation, we define wrapping numbers along bulk three--cycles,
\be{eqpquvdef}
\begin{aligned}
P & \equiv  \left(n_1 n_2- m_1 m_2 \right)n_3 , \\
Q & \equiv \left(n_1 m_2 + m_1  n_2 + m_1 m_2 \right)n_3 ,\\ 
U & \equiv \left(n_1 n_2- m_1 m_2 \right)m_3 ,\\ 
V & \equiv \left(n_1 m_2 + m_1  n_2 + m_1 m_2 \right)m_3 , 
\end{aligned}
\ee
such that any bulk cycle can be expanded as
\be{eqbulkexp}
  \Pi^{bulk} = P \rho_1 + Q \rho_2 + U \rho_3 + V \rho_4.
\ee
The intersection number of two bulk cycles reads
\ben
\Pi_a^{bulk} \circ \Pi_b^{bulk} 
= 2(P_a V_b -P_b V_a +Q_a U_b -Q_b U_a ) + 4 (P_a U_b - P_b U_a + Q_a V_b -Q_b V_a).
\een

\subsubsection{Exceptional cycles}
We obtain two classes of exceptional three--cycles. One from the $\bZ_3$ fixed points on $T^2_1 \times T^2_2$
times a one--cycle on $T^2_3$ and the other one from $\bZ_2$ fixed points on $T^2_1 \times T^2_3$ times
a one--cycle on $T^2_2$.

\paragraph{Exceptional cycles from $\mathbf{\bZ_2}$ fixed points:}

The fixed points on $T^2_1$ transform as
\ben
4 \stackrel{\theta}{\longrightarrow} 5 \stackrel{\theta}{\longrightarrow} 6  \stackrel{\theta}{\longrightarrow} 4,
\een
whereas $T^2_3$ is only subject to a $\bZ_2$ rotation and the fixed points are invariant under 
the $\bZ_6'$ operation, $\theta (j) =j$; $j=1 \ldots 4$.
Combined with the transformation of the one--cycles on $T^2_2$,
\ben
\theta(\pi_{3})=\pi_4-\pi_3, \qquad \theta(\pi_{4})=-\pi_3.
\een
we obtain eight independent exceptional three--cycles, given by
\ben\begin{array}{rcl}
\delta_j &=&\left(d_{4j}-d_{5j}\right) \otimes \pi_3 +\left(d_{5j}-d_{6j}\right) \otimes \pi_4, \\
\Tilde{\delta}_j &=&\left(d_{6j}-d_{5j}\right) \otimes \pi_3 +\left(d_{4j}-d_{6j}\right) \otimes \pi_4,
\end{array}\een
where $j=1 \ldots 4$ and $d_{ij} \circ d_{kl} = -2 \, \delta_{ik} \delta_{jl}$. The number of cycles is
in accordance with $h^{\theta^3}_{2,1}=4$. Starting from a particular fixed point times a one--cycle, the
$\bZ_6'$ invariant orbits are listed in Table~\ref{Z_2_Orbits}.
The intersection numbers among the $\bZ_2$ exceptional cycles are then given by
\ben
\delta_i \circ \tilde{\delta}_j = -2 \delta_{ij},
\quad\quad\quad
\delta_i \circ \delta_j = \tilde{\delta}_i\circ \tilde{\delta}_j =0.
\een

\renewcommand{\arraystretch}{1.3}
\begin{table}[ht]
  \begin{center}
    \begin{equation*}
      \begin{array}{|c|c||c|c|} \hline
        \multicolumn{4}{|c|}{\rule[-3mm]{0mm}{8mm}
\text{\bf Orbits of exceptional cycles for } T^6/\bZ'_6} \\ \hline\hline
 & \text{ Orbit}  & & \text{ Orbit} \\ \hline\hline
d_{4j} \otimes \pi_3 & \delta_j
& d_{4j} \otimes \pi_4 & \tilde{\delta}_j 
\\ \hline
d_{5j} \otimes \pi_3 & -\tilde{\delta}_j
& d_{5j} \otimes \pi_4 & \delta_j - \tilde{\delta}_j 
\\ \hline
d_{6j} \otimes \pi_3 & \tilde{\delta}_j - \delta_j
& d_{6j} \otimes \pi_4 & - \delta_j
\\ \hline
     \end{array}
    \end{equation*}
  \end{center}
\caption{Orbits of $\bZ_2$ fixed points times one--cycles.}
\label{Z_2_Orbits}
\end{table}
\renewcommand{\arraystretch}{1.3}

\paragraph{Exceptional cycles from $\mathbf{\bZ_3}$ fixed points:}
The third torus is fixed under $\theta^2$, and we have
$\theta (\pi_k) =-\pi_k, k=5,6$. 
The fixed points on $T^2_1 \times T^2_2$ have the form $c_{ij}$ with
$i,j=1,2,3$. On $T^2_1$, the points transform in the following way,
\ben
1 \stackrel{\theta}{\rightarrow} 1,\qquad  2 \stackrel{\theta}{\leftrightarrow} 3,
\een
whereas on $T^2_2$, all fixed
points of $\theta^2$ are also fixed under $\theta$.

For the exceptional cycles in the $\theta^2+\theta^4$ sector, we use the ansatz
\ben\begin{array}{rcl}
\gamma_j^{(\alpha)} &=& \left( c_{2j}^{(\alpha)}-c_{3j}^{(\alpha)}\right) \otimes \pi_5,\\
\Tilde{\gamma}_j^{(\alpha)} &=& \left( c_{2j}^{(\alpha)}-c_{3j}^{(\alpha)}\right) \otimes \pi_6,
\end{array}\een
with $j=1,2,3$ and $\alpha=1,2$. The parameter $\alpha$ corresponds to the fact that each fixed point supports two
cycles, associated to the $\theta^2$ and $\theta^4$ twisted sectors, respectively.
The intersection matrix of $c^{(\alpha)}_{ij}$ 
for fixed $ij$ is given by minus the Cartan matrix of $A_2$, which leads to the
non--trivial intersections of $\bZ_3$ exceptional cycles,
\ben
\begin{aligned}
\gamma^{(\alpha)}_j \circ \tilde{\gamma}^{(\alpha)}_j &=  -2,
\gamma^{(\alpha)}_j \circ \tilde{\gamma}^{(\beta)}_j &= 1, \qquad \mbox{with}\quad\{\alpha,\beta\}\in\{1,2\}. 
\end{aligned}
\een

\subsubsection{Fractional cycles}\label{fraccycsec}

The intersection form for pure bulk and exceptional cycles is not uni--modular, 
and therefore fractional cycles exist. The uni--modular lattice of three--cycles
consists of combinations of bulk cycles and all kinds of exceptional cycles.
A possible choice of basis is displayed in Appendix~\ref{App:Basis}. 
However, only bulk cycles and  exceptional cycles at $\bZ_2$ fixed points 
have a known interpretation in terms of partition functions~\cite{bbkl02}, and therefore
we only work with a sublattice consisting of these cycles,
\ben
  \Pi^{frac} = \frac{1}{2} \Pi^{buk} + \frac{1}{2} \Pi^{\bZ_2},
\een
where as usual the exceptional cycles consists of a sum of all $\bZ_2$
fixed point orbits traversed by the bulk cycle. The correspondence among 
even and odd wrapping numbers on $T^2_1 \times T^2_3$ and fixed points $d_{ij}$
 is given in Table~\ref{T:Cycles_FixedPoints} 
\renewcommand{\arraystretch}{1.3}
\begin{table}[ht]
  \begin{center}
    \begin{equation*}
      \begin{array}{|c||c|c|c|} \hline
        \multicolumn{4}{|c|}{\rule[-3mm]{0mm}{8mm}
\text{\bf Wrapping numbers and fixed points}} \\ \hline\hline
(n_3,m_3)& ({\rm odd, odd}) & ({\rm odd, even}) & ({\rm even, odd})\\ \hline\hline
(n_1,m_1)& \multicolumn{3}{|c|}{\rule[-3mm]{0mm}{8mm}{(\sigma_1,\sigma_2)=(0,0),(\sigma_5,\sigma_6)=(0,0)}}
\\\hline 
({\rm odd, odd}) & (d_{11}), (d_{13}), d_{61}, d_{63} 
& (d_{11}), (d_{12}), d_{61}, d_{62} 
&  (d_{11}), (d_{14}), d_{61}, d_{64}
\\\hline\hline
& \multicolumn{3}{|c|}{\rule[-3mm]{0mm}{8mm}{(\sigma_1,\sigma_2)=(0,0),(\sigma_5,\sigma_6)=(0,1/2)}}
\\\hline
({\rm odd, odd}) & (d_{12}), (d_{14}), d_{62}, d_{64} 
& (d_{13}), (d_{14}), d_{63}, d_{64} 
& (d_{11}), (d_{14}), d_{61}, d_{64} 
\\\hline \hline 
& \multicolumn{3}{|c|}{\rule[-3mm]{0mm}{8mm}{(\sigma_1,\sigma_2)=(\underline{1/2,0}),(\sigma_5,\sigma_6)=(0,0)}}
\\\hline 
({\rm odd, odd}) &  d_{41}, d_{43}, d_{51}, d_{53} 
& d_{41}, d_{42}, d_{51}, d_{52} 
& d_{41}, d_{44}, d_{51}, d_{54}
\\\hline\hline
& \multicolumn{3}{|c|}{\rule[-3mm]{0mm}{8mm}{(\sigma_1,\sigma_2)=(\underline{1/2,0}),(\sigma_5,\sigma_6)=(0,1/2)}}
\\\hline
({\rm odd, odd}) &  d_{42}, d_{44}, d_{52}, d_{54} 
&  d_{43}, d_{44}, d_{53}, d_{54} 
&  d_{41}, d_{44}, d_{51}, d_{54} 
\\\hline \hline
& \multicolumn{3}{|c|}{\rule[-3mm]{0mm}{8mm}{(\sigma_1,\sigma_2)=(0,0),(\sigma_5,\sigma_6)=(1/2,0)}}
\\\hline 
({\rm odd, odd}) & (d_{12}), (d_{14}), d_{62}, d_{64} 
& (d_{11}), (d_{12}), d_{61}, d_{62} 
& (d_{12}), (d_{13}), d_{62}, d_{63}
\\\hline\hline
& \multicolumn{3}{|c|}{\rule[-3mm]{0mm}{8mm}{(\sigma_1,\sigma_2)=(0,0),(\sigma_5,\sigma_6)=(1/2,1/2)}}
\\\hline
({\rm odd, odd}) & (d_{11}), (d_{13}), d_{61}, d_{63} 
& (d_{13}), (d_{14}), d_{63}, d_{64} 
& (d_{12}), (d_{13}), d_{62}, d_{63} 
\\\hline \hline 
& \multicolumn{3}{|c|}{\rule[-3mm]{0mm}{8mm}{(\sigma_1,\sigma_2)=(\underline{1/2,0}),(\sigma_5,\sigma_6)=(1/2,0)}}
\\\hline 
({\rm odd, odd}) &  d_{42}, d_{44}, d_{52}, d_{54} 
&  d_{41}, d_{42}, d_{51}, d_{52} 
&  d_{42}, d_{43}, d_{52}, d_{53} 
\\\hline \hline 
& \multicolumn{3}{|c|}{\rule[-3mm]{0mm}{8mm}{(\sigma_1,\sigma_2)=(\underline{1/2,0}),(\sigma_5,\sigma_6)=(1/2,1/2)}}
\\\hline 
({\rm odd, odd}) &  d_{41}, d_{43}, d_{51}, d_{53} 
&  d_{43}, d_{44}, d_{53}, d_{54} 
&  d_{42}, d_{43}, d_{52}, d_{53} 
\\\hline
     \end{array}
    \end{equation*}
  \end{center}
\caption{Fixed points on $T^2_1 \times T^2_3$ which are traversed by bulk cycles. The case $(\sigma_1,\sigma_2)=(1/2,1/2)$ gives the same result as $(\sigma_1,\sigma_2)=(0,0)$ for wrapping numbers $(n_1,m_1) = ({\rm odd, odd})$.}
\label{T:Cycles_FixedPoints}
\end{table}
\renewcommand{\arraystretch}{1.3}
for all possible combinations of 
 spatial  displacements $\sum_{k=1,2,5,6} \sigma_k \pi_k$  of a given bulk cycle from the origin on $T^2_1 \times T^2_3$
with $\sigma_k \in \{ 0,1/2 \}$.
The resulting exceptional three--cycles obtained from tensoring exceptional two--cycles $d_{ij}$
with the basic one--cycles on $T^2_2$ and taking the $\bZ_6'$ orbit are given in Table~\ref{Z_2_Orbits}. A general
$\bZ_2$ exceptional cycle is given by 
\bea{eq:cycleZ2expansion}\nonumber
  \Pi^{\bZ_2} &=& (-1)^{\tau_0} \sum_{k=0}^2  \theta^k \big[
  \left(d_{i_1 j_1} + (-1)^{\tau_1}d_{i_2 j_1} + (-1)^{\tau_3}d_{i_1 j_2} + (-1)^{\tau_1+\tau_3}d_{i_2 j_2}\right)\\
&& \otimes  \left(n_2 \pi_3 + m_2 \pi_4 \right) \big].
\eea
In the following, it will be convenient to expand the $\bZ_2$ exceptional cycles as
\be{eq:delta_expansion}
  \Pi^{\bZ_2} = \sum_{i=1}^4 \left(d_i \, \delta_i + e_i \, \tilde{\delta}_i\right),
\ee
with coefficients $d_i, e_i \in \bZ$.

As an example and consistency check, the fractional cycles with bulk parts along the O6--planes and neither displacements 
nor Wilson lines which reproduce the non--chiral models of~\cite{bgk00,bcs04} are listed in~\eqref{Cycles_nonchiral}.

\subsection{RR tadpoles}\label{secrrtadpoles}

The worldsheet parity $\Omega$ is accompanied by a complex conjugation ${\cal R}$ 
\ben
z^i \stackrel{\cal R}{\longrightarrow} \overline{z}^i,
\een
which leads to O6--planes. The lattices on $T^2_1 \times T^2_2$ can have two different orientations
{\bf A} and {\bf B} with respect to complex conjugation. In the first case $\pi_{2i-1}$ lies along the 
invariant axis, in the latter $\pi_{2i-1} + \pi_{2i}$ stays invariant. The notation for $T^2_1 \times T^2_2$
is the same as in~\cite{hoot04,gls07}. Similarly for $T^2_3$ according to the standard notation 
first introduced in~\cite{blkFB02}, $(\pi_5 - b\pi_6)/(1-b)$ is positioned on the 
${\cal R}$ invariant axis with $b=0,1/2$ for the {\bf a} and {\bf b} shape, respectively. 

\renewcommand{\arraystretch}{1.3}
\begin{table}[ht]
  \begin{center}
    \begin{equation*}
      \begin{array}{|c||c|c|c|c|} \hline
        \multicolumn{5}{|c|}{\rule[-3mm]{0mm}{8mm}
\text{\bf $\Omega{\cal R}$ images of cycles for } T^6/\bZ'_6} \\ \hline\hline
\text{lattice}  &  \Omega {\cal R}:\rho_1 &  \Omega{\cal R}:\rho_2  &  \Omega{\cal R}:\rho_3 &  \Omega{\cal R}:\rho_4\\ \hline\hline
{\bf AAa/b} &\rho_1-(2b)\rho_3 & \rho_1 - \rho_2+(2b)\left(\rho_4-\rho_3 \right) 
& -\rho_3  & \rho_4 - \rho_3 \\\hline  
{\bf ABa/b}, {\bf BAa/b}   & \rho_2-(2b)\rho_4 & \rho_1-(2b)\rho_3 & - \rho_4 & -\rho_3 \\\hline  
{\bf BBa/b}  & \rho_2 - \rho_1-(2b)\left(\rho_4-\rho_3 \right) &  \rho_2-(2b)\rho_4
& \rho_3 - \rho_4  & - \rho_4 \\\hline  
      \end{array}
    \end{equation*}
  \end{center}
\caption{$\Omega{\cal R}$ images of the cycles inherited from the torus for $T^6/\bZ'_6$. 
The third torus can be untilted ({\bf a}) or tilted ({\bf b}) as parameterised by $b=0,1/2$.}
\label{RimagesordinarycyclesZ6P}
\end{table}
\renewcommand{\arraystretch}{1.3}

The action on the bulk cycles is summarised in Table~\ref{RimagesordinarycyclesZ6P}.
There exist two orbits of cycles invariant under $\Omega {\cal R}\theta^{2k}$ and  $\Omega {\cal R}\theta^{2k+1}$,
which are wrapped by the O6--planes as displayed in Table~\ref{O-planes}.

\renewcommand{\arraystretch}{1.3}
\begin{table}[ht]
  \begin{center}
    \begin{equation*}
      \begin{array}{|c||c||c|c|c|c||c|} \hline
        \multicolumn{7}{|c|}{\rule[-3mm]{0mm}{8mm}
\text{\bf O6--planes for $T^6/\bZ_6'$}}\\ \hline\hline
\text{lattice}  &  (n_1,m_1;n_2,m_2;n_3,m_3) & P & Q & U & V & \text{cycle}
\\ \hline\hline
{\bf AAa/b} & (1,0;1,0;\frac{1}{1-b},\frac{-b}{1-b}) & \frac{1}{1-b} & 0 & \frac{-b}{1-b} & 0 & \frac{\rho_1-b \rho_3}{1-b}
\\
& (1,1;0,1;0,-1) & 0 & 0 & 1 & -2 & \rho_3 - 2 \rho_4 
\\ \hline
{\bf ABa/b} & (1,0;1,1;\frac{1}{1-b},\frac{-b}{1-b}) & \frac{1}{1-b} & \frac{1}{1-b} & \frac{-b}{1-b} & \frac{-b}{1-b} 
& \frac{\rho_1+\rho_2 -b (\rho_3+\rho_4)}{1-b}
\\
& (1,1;-1,2;0,-1) & 0 & 0 & 3 & -3 & 3(\rho_3 - \rho_4)
\\\hline
{\bf BAa/b} & (1,1;1,0;\frac{1}{1-b},\frac{-b}{1-b}) & \frac{1}{1-b} & \frac{1}{1-b} & \frac{-b}{1-b} & \frac{-b}{1-b} 
& \frac{\rho_1+\rho_2 -b (\rho_3+\rho_4)}{1-b}
\\
& (0,1;0,1;0,-1) & 0 & 0 & 1 & -1 & \rho_3 - \rho_4
\\\hline
{\bf BBa/b} & (1,1;1,1;\frac{1}{1-b},\frac{-b}{1-b}) & 0 & \frac{3}{1-b} & 0 & \frac{-3b}{1-b} & \frac{3(\rho_2-b \rho_4)}{1-b}
\\
& (0,1;-1,2;0,-1) & 0 & 0 & 2 & -1 & 2\rho_3 -\rho_4
\\\hline
     \end{array}
    \end{equation*}
  \end{center}
\caption{O6--planes for $T^6/\bZ'_6$. The first row for each lattice corresponds to the orbit of $\Omega {\cal R} \theta^{-2k}$ 
invariant planes, the second to the $\Omega{\cal R} \theta^{-2k -1} $ invariant ones with the wrapping numbers listed for the $k=0$ 
representatives.
The number of parallel O6--planes depends on the geometry of $T^2_3$, namely $N_{O6} = 2 (1-b)$.
Since the O6--planes are non--dynamical objects which are stuck at the $\bZ_2$ fixed points, an additional factor 1/2 arises in the 
tadpole cancellation condition.}
\label{O-planes}
\end{table}
\renewcommand{\arraystretch}{1.3}

The exceptional cycles at $\bZ_2$ fixed points receive apart from the purely geometric 
${\cal R}$ transformation a global minus sign\footnote{Note that this sign is not present in~\cite{balo06,balo07}.
This has important consequences for the space of solutions, in particular the three generation model presented
in~\cite{balo07} cannot be realised.} from the $\Omega$ action~\cite{bbkl02}.
The resulting orientifold images are given in Table~\ref{RimagesexccyclesZ6P2}.
\renewcommand{\arraystretch}{1.3}
\begin{table}[ht]
  \begin{center}
    \begin{equation*}
      \begin{array}{|c||c|c|c|c|} \hline
\multicolumn{5}{|c|}{\rule[-3mm]{0mm}{8mm}
  \text{\bf $\Omega{\cal R}$ images of $\bZ_2$ fixed--cycles for } T^6/\bZ'_6}
\\\hline\hline
\text{lattice} & \Omega{\cal R}:\delta_1 &  \Omega{\cal R}:\delta_2 & \Omega{\cal R}:\delta_3 & \Omega{\cal R}:\delta_4 \\\hline\hline
{\bf AAa/b} & -\delta_1 & -\delta_{2+2b} & -\delta_{3-2b} & -\delta_4 \\\hline
{\bf ABa/b} & -\Tilde{\delta}_1 & -\Tilde{\delta}_{2+2b} & -\Tilde{\delta}_{3-2b} & -\Tilde{\delta}_4 \\\hline
{\bf BAa/b} & \Tilde{\delta}_1 & \Tilde{\delta}_{2+2b} & \Tilde{\delta}_{3-2b} & \Tilde{\delta}_4 \\\hline
{\bf BBa/b} & \Tilde{\delta}_1 - \delta_1 & \Tilde{\delta}_{2+2b} - \delta_{2+2b}
& \Tilde{\delta}_{3-2b} - \delta_{3-2b}  & \Tilde{\delta}_4 - \delta_4 \\\hline\hline
& \Omega{\cal R}:\Tilde{\delta}_1 & \Omega{\cal R}:\Tilde{\delta}_2
& \Omega{\cal R}:\Tilde{\delta}_3 & \Omega{\cal R}:\Tilde{\delta}_4 \\\hline\hline
{\bf AAa/b} & \Tilde{\delta}_1-\delta_1 & \Tilde{\delta}_{2+2b} -\delta_{2+2b}
& \Tilde{\delta}_{3-2b} - \delta_{3-2b} & \Tilde{\delta}_4 - \delta_4 \\\hline
{\bf ABa/b} & -\delta_1 & -\delta_{2+2b} & -\delta_{3-2b} & -\delta_4\\\hline 
{\bf BAa/b} & \delta_1 & \delta_{2+2b} & \delta_{3-2b} & \delta_4\\\hline
{\bf BBa/b} 
& \Tilde{\delta}_1 & \Tilde{\delta}_{2+2b} & \Tilde{\delta}_{3-2b} & \Tilde{\delta}_4\\\hline 
     \end{array}
    \end{equation*}
  \end{center}
\caption{$\Omega{\cal R}$ images of the $\bZ_2$ exceptional three--cycles for $T^6/\bZ'_6$.}
\label{RimagesexccyclesZ6P2}
\end{table}
\renewcommand{\arraystretch}{1.3}

The tadpole cancellation condition 
\ben
\sum_a N_a \left(\Pi_a + \Pi_a' \right) = 4 \, \Pi_{O6} 
\een
splits into two parts.
One is the toroidal part of the cycles  given by
\be{eqrrtadpoles}
\begin{aligned}
{\bf AAa/b}: &
\left\{\begin{array}{ll}
\frac{R_1}{R_2}: &\;  \sum_a N_a \left(2 P_a + Q_a \right) = 8\\
\frac{R_2}{R_1}: &\;   \sum_a N_a \left(- V_a - b Q_a  \right) =  8 (1-b)
\end{array}\right. , \\
\begin{array}{l}{\bf ABa/b}\\{\bf BAa/b}\end{array}: & 
\left\{\begin{array}{ll}
\frac{R_1}{R_2}: &\;   \sum_a N_a  (P_a + Q_a)  = 8\\
\frac{R_2}{R_1}: &\;   \sum_a N_a \left(U_a - V_a +b(P_a - Q_a) \right)  = k \, 8 (1-b)
\end{array}\right. , \\
{\bf BBa/b:} &
\left\{\begin{array}{ll}
\frac{R_1}{R_2}: &\;   \sum_a N_a \left(P_a + 2 Q_a \right) = 24\\
\frac{R_2}{R_1}: &\;   \sum_a N_a \left(U_a+b P_a \right)  = 8 (1-b)
\end{array}\right.  ,
\end{aligned}
\ee
where $k = 3$ for $\mathbf{ABa/b}$ geometries and $k=1$ in the $\mathbf{BAa/b}$ case.
$R_i/R_j$ labels the ratio of radii to which the divergences in the tree channel amplitude calculation
providing the tadpole cancellation conditions are proportional.  

The second part of the tadpole cancellation condition deals with the exceptional cycles.
Since the O6--planes do not wrap any exceptional cycle, the associated tadpole cancellation condition  
is fulfilled if all exceptional D6--brane contributions cancel among each other. In terms of the 
expansion~\eqref{eq:delta_expansion}, the tadpole conditions for the different geometries can be
written explicitly as
\be{eq:tcc_ex}
\begin{array}{rrl}
  \mathbf{AAa/b}: & \sum_a N_a \big[ &\!\!\!-e^a_1\delta_1 + \left(-e^a_2+2b(d^a_2+e^a_2-d^a_3-e^a_3)\right)\delta_2\\
     & &+ \left(-e^a_3+2b(d^a_3+e^a_3-d^a_2-e^a_2)\right)\delta_3 -e^a_4\delta_4\\
     & &+ 2e^a_1\tilde{\delta}_1 + \left(2e^a_2+2b(e^a_3-e^a_2)\right)\tilde{\delta}_2\\
     & &+ \left(2e^a_3+2b(e^a_2-e^a_3)\right)\tilde{\delta}_3 + 2e^a_4\tilde{\delta}_4 \big] = 0,\\[2ex]
  \mathbf{ABa/b}: & \sum_a N_a \big[ &\!\!\!\left(d^a_1-e^a_1\right)\delta_1 + \left(d^a_2-e^a_2+2b(e^a_2-e^a_3)\right)\delta_2\\
     & &+ \left(d^a_3-e^a_3+2b(e^a_3-e^a_2)\right)\delta_3 + \left(d^a_4-e^a_4\right)\delta_4\\
     & &+ \left(e^a_1-d^a_1\right)\tilde{\delta}_1 + \left(e^a_2-d^a_2+2b(d^a_2-d^a_3)\right)\tilde{\delta}_2\\
     & &+ \left(e^a_3-d^a_3+2b(d^a_3-d^a_2)\right)\tilde{\delta}_3 + \left(e^a_4-d^a_4\right)\tilde{\delta}_4\big] = 0,\\[2ex]
  \mathbf{BAa/b}: & \sum_a N_a \big[ &\!\!\!\left(d^a_1+e^a_1\right)\delta_1 + \left(d^a_2+e^a_2+2b(e^a_3-e^a_2)\right)\delta_2\\
     & &+ \left(d^a_3+e^a_3+2b(e^a_2-e^a_3)\right)\delta_3 + \left(d^a_4+e^a_4\right)\delta_4\\
     & &+ \left(e^a_1+d^a_1\right)\tilde{\delta}_1 + \left(e^a_2+d^a_2+2b(d^a_3-d^a_2)\right)\tilde{\delta}_2\\
     & &+ \left(e^a_3+d^a_3+2b(d^a_2-d^a_3)\right)\tilde{\delta}_3 + \left(e^a_4+d^a_4\right)\tilde{\delta}_4\big] = 0,\\[2ex]
  \mathbf{BBa/b}: & \sum_a N_a \big[ &\!\!\!2b(d^a_2-d^a_3)\delta_2 + 2b(d^a_3-d^a_2)\delta_3\\
     & &+ (d^a_1+2e^a_1)\tilde{\delta}_1 + \left(d^a_2+2e^a_2+2b(d^a_3+e^a_3-d^a_2-e^a_2)\right)\tilde{\delta}_2\\
     & &+ \left(d^a_3+2e^a_3+2b(d^a_2+e^a_2-d^a_3-e^a_3)\right)\tilde{\delta}_3 + (d^a_4+2e^a_4)\tilde{\delta}_4\big] = 0.
\end{array}\ee

Both bulk~\eqref{eqrrtadpoles} and exceptional~\eqref{eq:tcc_ex} tadpole conditions are used in section~\ref{Ktheory}
to discuss possible constraints from global anomalies.

\subsection{Supersymmetry}
The supersymmetry conditions on bulk cycles can be computed from
\ben
Z \equiv e^{i \phi} \left(n_1 + e^{\frac{\pi i}{3}} m_1\right)
\left(n_2 + e^{\frac{\pi i}{3}} m_2\right) \left((\frac{R_1}{R_2}+ i b) n_3 + i m_3\right)  
\een
with $\phi = 0, -\pi/6, -\pi/3$ for the {\bf AAa/b}, {\bf ABa/b} or {\bf BAa/b}, {\bf BBa/b} lattices, respectively.
Supersymmetry is preserved by the toroidal cycles for which
\ben
\Im (Z) = 0,
\quad\quad
\Re (Z) > 0.
\een
It is convenient to introduce the complex structure parameter $\varrho$ with
\ben
\frac{R_2}{R_1} = \frac{2 \varrho}{\sqrt{3}}  
\een
in order to write the necessary supersymmetry conditions $\Im(Z) =0$ for the bulk parts of the cycles as
\be{Eq:SUSY1}
\begin{aligned}
{\bf AAa/b: } & \quad \frac{3}{2\varrho} Q + (2 U +V ) +b(2 P + Q) =0 ,\\
{\bf ABa/b, \, BAa/b: } & \quad \frac{1}{2\varrho}(P - Q) -(U+ V +b(P+ Q)) =0 ,\\
{\bf BBa/b: } & \quad \frac{3}{2\varrho}  P - (U + 2 V +b(P + 2 Q))=0 .
\end{aligned}
\ee
Equation~\eqref{Eq:SUSY1} does not distinguish between supersymmetric D6--branes and their anti--D6--branes.
The anti--D6--branes are excluded by the sufficient supersymmetry conditions $\Re(Z)>0$,
\be{Eq:SUSY2}
\begin{aligned}
{\bf AAa/b: } & \quad 2P+Q -2\varrho (V+bQ) > 0,
\\
{\bf ABa/b, \,  BAa/b: } & \quad (P+Q)-\frac{2\varrho}{3}(V-U+b(Q-P)) > 0 ,
\\
{\bf BBa/b: } & \quad P+2Q +2 \varrho (U+bP) >0 .
\end{aligned}
\ee
Fractional cycles are supersymmetric if the bulk part is supersymmetric
and the exceptional cycle is composed of orbits of fixed points traversed
by the bulk part as listed in Table~\ref{T:Cycles_FixedPoints}, including
appropriate signs corresponding to the $\bZ_2$ eigenvalue
and relative Wilson lines on $T^2_1 \times T^2_3$ as in~\eqref{eq:cycleZ2expansion}.

\subsection{K--theory}\label{Ktheory}
D--branes are not fully characterised by (co)homology but rather by K--theory~\cite{wi98,mimo97} which 
imposes additional $\bZ_2$ valued constraints on model building. As in the previous analyses of intersecting D6--branes on 
$T^6/(\bZ_2\!\times\!\bZ_2)$~\cite{gbhlw05,gm05} and  $T^6/\bZ_6$~\cite{gls07} as well as of Gepner models~\cite{grs06}, 
we are not able to formulate these constraints directly, but follow the proposal of~\cite{ur00} which uses probe branes 
to check for global anomalies. This is based on the observation
that the path integral for a theory with an odd number of fermions in the fundamental representation of a $SU(2)$ gauge 
factor is ill defined~\cite{wi82}. 
The probe brane constraint is known to truly coincide with the K--theory constraint for compactifications on smooth manifolds.

In terms of intersection numbers of D6--branes, the K--theory constraint is  formulated as
\be{eq:Ktheory} 
\sum_a N_a \Pi_a \circ \Pi_{\rm probe} \stackrel{!}{=} 0 \; {\rm mod} \; 2,
\ee
where the sum over all D$6_a$--branes does not include the $\Omega{\cal R}$ images and $\Pi_{\rm probe}$ is any three--cycle wrapped
by a D6--brane carrying an $SU(2)$ or more generally $Sp(2N)$ gauge group. Although the correct identification of all $SO(2N)$ and
$Sp(2N)$ gauge groups is technically challenging and beyond the scope of this paper, one can identify all possible
D6--branes which are their own $\Omega{\cal R}$ images and therefore carry either orthogonal or symplectic gauge factors. The 
complete list for all choices of tori is given in Tables~\ref{rinvex1},~\ref{rinvex2},~\ref{rinvex3} and~\ref{rinvex4}.

Using all $\Omega{\cal R}$ invariant branes as probe branes is in general expected to be a too strong constraint since some 
of the branes may carry $SO(2N)$ gauge groups. However, as for the $T^6/\bZ_6$ orbifold, a brute computer search 
reveals that any possible constraint from~\eqref{eq:Ktheory} is automatically fulfilled.

This can also be seen analytically by reshuffling the sum in~\eqref{eq:Ktheory} using the bulk and 
exceptional tadpole cancellation conditions~\eqref{eqrrtadpoles} and~\eqref{eq:tcc_ex}, such that 
the contribution from each brane $a$ is already even. We demonstrate this in detail for the {\bf AAa/b} torus
in the following.
We start by computing the bulk and exceptional parts of the intersection numbers separately,
\ben
\Pi_a \circ \Pi_{\rm probe} = \frac{1}{4}\Pi_a^{bulk} \circ \Pi_{\rm probe}^{bulk}
+ \frac{1}{4}\Pi_a^{ex} \circ \Pi_{\rm probe}^{ex}.
\een
The bulk parts for probe branes of type 1a,b${}^{\bf b}$,c and 2a,b${}^{\bf b}$,c on {\bf AAa/b} defined in Table~\ref{rinvex1}
are listed in the third column of Table~\ref{eq:bulk_Ktheory}.
To compute these entries, the bulk tadpole conditions~\eqref{eqrrtadpoles} have been used.
For branes of type 1a,b${}^{\bf b}$ or c, the bulk contribution to the K--theory constraint is given by
\ben
\begin{aligned}
\frac{1}{4} \sum_a N_a \, \Pi_a^{bulk} \circ \Pi^{bulk}_{\rm probe} 
&= \frac{1}{2(1-b)} \sum_a N_a \,\left[2 (U_a +b P_a)  + (V_a + bQ_a)  \right] 
\\
&= \frac{1}{1-b}  \sum_a N_a \,\left(U_a + b P_a   \right) +  \frac{1}{2(1-b)} \left( -8(1-b) \right)\\
&= \sum_a N_a  \frac{U_a + b P_a}{1-b} - 4,
\end{aligned}
\een
where in the second line the tadpole condition associated to $R_2/R_1$ has been inserted. Similarly, the
entry for branes of type 2a,b${}^{\bf b}$ or c is obtained using the tadpole condition proportional to $R_1/R_2$.
The reshuffling of sums is possible since always only a finite number of D6--branes contribute to the 
tadpole conditions as discussed in detail in Section~\ref{finite}.

\renewcommand{\arraystretch}{1.3}
\begin{table}[ht]
  \begin{center}
\begin{equation*} 
\begin{array}{|c|c|c|c|}\hline
{\rm lattice} & \# & \frac{1}{4}\Pi_a^{bulk} \circ \Pi_{\rm probe}^{bulk}
& \sum_a N_a \left( \frac{1}{4}\Pi_a^{bulk} \circ \Pi_{\rm probe}^{bulk} \right)
\\\hline\hline
{\bf AAa/b} & 1 &  \frac{1}{2 (1-b)} \left(2U_a + V_a + b (2P_a+Q_a)  \right) 
& \sum_a N_a \frac{U_a + b P_a}{1-b} - 4
\\\hline
& 2 & -\frac{3}{2} Q_a 
&  3 \, \sum_a N_a P_a - 12
\\\hline
\end{array}
\end{equation*}
 \end{center}
\caption{Intersection numbers of D$6_a$--branes: bulk parts.}
\label{eq:bulk_Ktheory}
\end{table}
\renewcommand{\arraystretch}{1.3}

The exceptional contributions to the K--theory constraint~\eqref{eq:Ktheory} are computed along the same lines
as the bulk ones as displayed in Table~\ref{eq:ex_Ktheory},
where in the last column the tadpole conditions~\eqref{eq:tcc_ex} on the exceptional cycles have been used.

\renewcommand{\arraystretch}{1.3}
\begin{table}[ht]
  \begin{center}
\begin{equation*} 
\begin{array}{|c|c|c|c|}\hline
{\rm lattice} & \# & \frac{1}{4}\Pi_a^{ex} \circ \Pi_{\rm probe}^{ex}
& \sum_a N_a \left( \frac{1}{4}\Pi_a^{ex} \circ \Pi_{\rm probe}^{ex}  \right)
\\\hline\hline
{\bf AAa/b} & 1{\rm a} & \pm \frac{1}{2} (e_1^a + 2 d_1^a ) \pm \frac{1}{2} ( e_{\frac{2}{1-b}}^a + 2 d_{\frac{2}{1-b}})^a  
& \sum_a N_a (\pm d_1^a \pm  d_{\frac{2}{1-b}}^a)
\\\hline
& 1{\rm b}^{\bf b} & \pm \frac{1}{2} \left( e_3^a - e_2^a \right)
& \pm \sum_a N_a ( d_2^a - d_3^a)
\\\hline
& 1{\rm c} & \begin{array}{c}
 \pm \frac{1}{2} (e_3^a + d_3^a + d_{3-2b}^a)\\
 \pm  \frac{1}{2}(e_{4(1-b)}^a + d_{4(1-b)}^a + d_{4-2b}^a) 
\end{array}
&
\sum_a (\pm d_3^a \pm d_{4(1-b)}^a)
\\\hline\hline
 & 2{\rm a} &  \pm \frac{1}{2} (e_1^a + 2 d_1^a ) \pm \frac{1}{2} ( e_{4}^a + 2 d_{4}^a ) 
& \sum_a N_a (\pm d_1^a \pm  d_4^a)
\\\hline
 & 2{\rm b}^{\bf b} & \pm \frac{1}{2} \left( e_2^a - e_3^a \right)
& \pm \sum_a N_a (d_3^a - d_2^a)
\\\hline
 & 2{\rm c} &  \begin{array}{c}
\pm\frac{1}{2}  (e_2^a + d_2^a + d_{2+2b}^a)  \\
\pm \frac{1}{2}  (e_3^a + d_3^a + d_{3-2b}^a)  
\end{array}
&\sum_a N_a (\pm d_2^a \pm d_3^a)
\\\hline
\end{array}
\end{equation*}
 \end{center}
\caption{Intersection numbers of D$6_a$--branes and probe branes: exceptional parts. The K--theory constraint 
is evaluated for all possible choices of signs corresponding to the $\Omega{\cal R}$ invariant cycles in 
Table~\protect\ref{rinvex1}.}
\label{eq:ex_Ktheory}
\end{table}
\renewcommand{\arraystretch}{1.3}

Combining the results of Tables~\ref{eq:bulk_Ktheory} and~\ref{eq:ex_Ktheory}, for probe branes 2a, the K--theory 
constraint~\eqref{eq:Ktheory} takes the form
\be{K:example}
\sum_a N_a \left( 3 \, P_a \pm d^a_1 \pm d^a_4 \right) \stackrel{!}{=} 0 \; {\rm mod} \; 2
\ee
for any combination of signs. It turns out that every term $3 \, P_a \pm d^a_1 \pm d^a_4$ in the sum already fulfils
the constraint independently. To see this, one can analyse in which situations $P$ is even. As explained in Section~\ref{methods},
it is sufficient to assume $(n_1,m_1) = ({\rm odd},{\rm odd})$, for which 
the dependence of the first factor $n_1n_2-m_1m_2$ and $n_1 m_2 + m_1 n_2 + m_1 m_2$
in the definition of $P,U$ and $Q,V$, respectively, on the choice of wrapping numbers on $T^2_2$ is given in Table~\ref{T:nm_even-odd}.

\renewcommand{\arraystretch}{1.2}
\begin{table}[ht]
  \begin{center}
\begin{equation*} 
\begin{array}{|c|c|c|}\hline
(n_2,m_2) & n_1 m_2 + m_1 n_2 + m_1 m_1 & 
\begin{array}{c}
n_2 \pm m_2 \\
n_1 n_2 -m_1 m_2
\end{array}
\\\hline\hline
({\rm odd},{\rm odd}) & {\rm odd} & {\rm even}
\\
({\rm odd},{\rm even}) & {\rm odd} & {\rm odd}
\\
({\rm even},{\rm odd}) & {\rm even} & {\rm odd}
\\\hline
\end{array}
\end{equation*}
  \end{center}
\caption{Relation of $Q,V$ and $P,U$ and the wrapping numbers on $T^2_2$ for $(n_1,m_1)=({\rm odd},{\rm odd})$.}
\label{T:nm_even-odd}
\end{table}
\renewcommand{\arraystretch}{1.3}

For $(n_1,m_1)=({\rm odd},{\rm odd})$, the bulk part on $T^2_1$ passes either through fixed points 1 and 6 or through 
4 and 5, leading to the exceptional contributions
\ben
\begin{aligned}
d_{6j} \otimes (n_2 \pi_3 + m_2 \pi_4)  & \stackrel{\theta - {\rm orbit}}{\longrightarrow} - (n_2+m_2) \delta_j + n_2 \tilde{\delta}_j,
\\
( d_{4j} \pm d_{5j}) \otimes (n_2 \pi_3 + m_2 \pi_4) & \longrightarrow 
(n_2 \pm m_2) \delta_j + (m_2 \mp(n_2 + m_2)) \tilde{\delta}_j.
\end{aligned}
\een
From this one finds that 
\be{eq:coeffs_ex}
d_j = \pm (n_2 \pm m_2),
\quad\quad
e_j = \pm n_2 \; {\rm mod} \; 2,
\ee
if the fixed point $j$ on $T^2_3$ is traversed by the bulk cycle and zero
otherwise.

Finally, the occurrence of the fixed points $j$ for various choices of
$(n_3,m_3)$ has to be taken into account. For $n_3$ even
either $j=(1,4)$ or $(2,3)$ occur simultaneously, for
$(n_3,m_3)=({\rm odd},{\rm odd})$ it is (1,3) or (2,4) and for 
$(n_3,m_3)=({\rm odd},{\rm even})$ (1,2) or (3,4).

The terms in~\eqref{K:example} can now be shown to be always even.
For $n_3$ even both fixed points 1 and 4 on $T^2_3$ 
contribute coefficients of the form~\eqref{eq:coeffs_ex} which together are
even (or $d_1=d_4=0$), and also $P \sim n_3$ is even. If $n_3$ is odd, only
fixed points $j=1$ or 4 on $T^2_3$ add a non--vanishing contribution $d_j$,
which according to Table~\ref{T:nm_even-odd} is even or odd at the same time
when $P$ is even or odd.
This concludes the proof that probe branes of type 2a do not exclude any
solution to the tadpole cancellation condition.
The proof for type 2c branes works completely analogously. 

For branes of type 1a,c, one has to consider $M_3 = \frac{m_3 + b n_3}{1-b}$.
If $M_3$ is even, either fixed points $(1,\frac{2}{1-b})$
or $(3,4(1-b))$ on $T^2_3$ are traversed simultaneously by a bulk cycle.
If $M_3$ is odd one exceptional cycle out of each set
contributes non--trivially. The rest of the proof is identical to the
discussion above for branes of type 2a.

On the {\bf AAa} lattice, this shows that the probe brane argument does not exclude any models.
On the {\bf AAb} lattice, there are two more candidates of probe branes 1b${}^{\bf b}$ and 2b${}^{\bf b}$. In this case,
the tadpole conditions $\sum_a N_a ( d_2^a-d_3^a-e_3^a) = \sum_a N_a ( d_3^a-d_2^a-e_2^a) = 0$ serve to obtain the
constraint
\ben 
\sum_a N_a \left(P_a \pm (d_2^a - d_3^a) \right) \stackrel{!}{=} 0 \; {\rm mod} \; 2,
\een
which is always fulfilled since $P_a \pm (d_2^a - d_3^a)$ is even for any $a$.

The argumentation above can be repeated for the other lattices as well, leading in all cases to the fact that the probe brane
constraint is trivially fulfilled. The only ingredients are the bulk and exceptional 
tadpole cancellation conditions~\eqref{eqrrtadpoles} and~\eqref{eq:tcc_ex}, as well as the geometric interpretation of 
factional branes having exceptional contributions only from fixed points traversed by the bulk cycle. The argument is therefore
independent of the complex structure parameter $\varrho$ and also valid for non--supersymmetric models.

\subsection{Massless spectrum}\label{Sec:Spectrum}

\subsubsection{The closed spectrum}\label{Sec:ClosedSpectrum}

The closed string spectra for all lattices of type {\bf b} on $T^2_3$ have
been computed in~\cite{bgk00,bcs04}. Here we also
list the spectra for the {\bf a} type lattice on $T^2_3$. The complete list 
is displayed in Table~\ref{Tab:ClosedSpectrumZ6p} with $b=0,1/2$ parameterising
the {\bf a} and {\bf b} type lattice, respectively.
In terms of hodge numbers, the closed spectrum contains $h_{1,1}^+$ vector and $h_{1,1}^- + h_{2,1}$ chiral multiplets 
in addition to the axion--dilaton multiplet~\cite{tgjl05} with  Hodge numbers given in~\eqref{eqhodge} and
$h_{1,1} = h_{1,1}^+ +  h_{1,1}^-$. 

\renewcommand{\arraystretch}{1.3}
\begin{table}[ht]
  \begin{center}
    \begin{equation*}
      \begin{array}{|c||c|c||c|c|} \hline
        \multicolumn{5}{|c|}{\rule[-3mm]{0mm}{8mm}
\text{\bf Closed string spectrum $T^6/(\bZ_6' \times \Omega{\cal R})$}} \\ \hline\hline
\text{lattice}
&\multicolumn{2}{|c|}{\rule[-3mm]{0mm}{8mm}{\bf AAa/b}, {\bf BAa/b}}
&\multicolumn{2}{|c|}{\rule[-3mm]{0mm}{8mm}{\bf ABa/b}, {\bf BBa/b}}
\\\hline
\text{sector} & \text{NSNS} & \text{RR}  & \text{NSNS}  & \text{RR}  
\\\hline\hline
\text{untwisted} 
&\multicolumn{4}{|c|}{\rule[-3mm]{0mm}{8mm}
\begin{array}{c} \text{NSNS: Graviton + Dilaton + 3C + 1 scalar} 
\\
\text{RR: Axion + 1 scalar}   
\end{array}}
\\\hline
\theta+\theta^5 & (8-2b) \C & (4+2b) \V & 3(4-2b) \C & (6b) \V  
\\\hline
\theta^2+\theta^4 & 11 \C & 3\C+4\V & 15\C & 3\C 
\\\hline 
\theta^3 
& \multicolumn{4}{|c|}{\rule[-3mm]{0mm}{8mm}
\text{NSNS: (10-4b) C; RR: 2 C + (4b) V }}
\\\hline
      \end{array}
    \end{equation*}
  \end{center}
\caption{Closed string spectrum of $T^6/(\bZ_6' \times \Omega{\cal R})$. C corresponds
to the scalar degrees of freedom of a chiral multiplet while V denotes a
massless vector multiplet. The fermionic superpartners arise from the R--NS and NS--R sectors. In the
untwisted sector the two explicitly listed scalars belong to one chiral multiplet,
in the same way as the dilaton and axion.}
\label{Tab:ClosedSpectrumZ6p}
\end{table}
\renewcommand{\arraystretch}{1.3}

\subsubsection{The open spectrum}\label{Sec:OpenSpectrum}
The chiral part of the spectrum is computed from the topological intersection numbers
among the fractional cycles~\cite{bbkl02} 
of a given tadpole solution as displayed in Table~\ref{ChiralSpectrum}.

\renewcommand{\arraystretch}{1.3}
\begin{table}[ht]
  \begin{center}
    \begin{equation*}
      \begin{array}{|c|c|} \hline
        \multicolumn{2}{|c|}{\rule[-3mm]{0mm}{8mm}
\text{\bf Chiral spectrum} }\\ \hline\hline
\text{representation} & \text{net chirality} \; \chi 
\\\hline\hline
({\bf Anti}_a) & \frac{1}{2} \left(\Pi_a \circ \Pi_{a}' + \Pi_a \circ \Pi_{O6}  \right)
\\
({\bf Sym}_a) & \frac{1}{2} \left(\Pi_a \circ \Pi_{a}^{\prime} - \Pi_a \circ \Pi_{O6}  \right)
\\
({\bf N}_a,\overline{\bf N}_b) & \Pi_a \circ \Pi_b
\\
({\bf N}_a,{\bf N}_b) & \Pi_a \circ \Pi_b^{\prime}
\\ \hline
     \end{array}
    \end{equation*}
  \end{center}
\caption{Counting of net chirality $\chi \equiv \chi_L - \chi_R$ in four dimensions via
  intersection numbers of cycles $\Pi_a$. The overall orientifold cycle $\Pi_{O6}$ for
  $T^6/\bZ_6'$ is read off from Table~\protect\ref{O-planes} taking into account the multiplicity $N_{O6}/2=(1-b)$.}
\label{ChiralSpectrum}
\end{table}
\renewcommand{\arraystretch}{1.3}

In case of $T^6$~\cite{imr01} or $T^6/(\bZ_2\!\times\!\bZ_2)$~\cite{csu01} compactifications there exist, 
except from  the chiral spectrum, three chiral multiplets in 
the adjoint representation (or antisymmetric for D6--branes with gauge group
$Sp(2N)$ on $T^6/(\bZ_2\!\times\!\bZ_2)$~\cite{fhs00}) and non--chiral matter pairs if branes are parallel
on some $T^2_i$. These pairs are then counted by the intersection number
on the other tori $T^2_j \times T^2_k$ ($j,k \neq i$).

The situation is different for $T^{2n}/\bZ_M$ orbifold backgrounds
with $n=2,3$ and $M \neq 2$. Under the action of the orbifold
generator $\theta$, any $n$-cycle $a$ is mapped to its image $(\theta a)$ which
for $\bZ_6'$ has the wrapping numbers $(n_i^{(\theta a)},m_i^{(\theta a)})_{i=1 \ldots n}$ 
given in~\eqref{eqnmtransform}. For $M$ 
odd, the cycle $a$ has $M$ images. For $M=2N$, each cycle $a$ 
has only $N$ distinct images since the $\bZ_2$ subgroup maps $a$ to itself. 

Each of the orbifold images contributes to the massless spectrum, and 
open strings in the $a(\theta^k b)$ sectors can have different chiralities
for different $k$. Therefore, non--chiral pairs of massless particles can 
arise even if the two branes under consideration are not parallel on any
two--torus. 
The total number of multiplets is computed using the bulk intersection 
numbers 
$I_{a (\theta^k b)} =  \prod_{i=1}^n I_{a (\theta^k b)}^{(i)} = \prod_{i=1}^n (n_i^a m_i^{(\theta^k b)} - 
m_i^a n_i^{(\theta^k b)})$
for all $k$ and the number of intersections $I_{a (\theta^k b)}^{\bZ_2}$
which are $\bZ_2$ invariant, weighted with signs from relative Wilson lines
and $\bZ_2$ eigenvalues as defined below in~\eqref{eq:def_Ex_v2}.

The intersection number between two branes $a$ and $b$ can be split into contributions
from different orbifold images as follows. For simplicity, we start with a $T^6/\bZ_M$ orbifold with $M$ odd.
The bulk cycle is then given by
\ben
\begin{aligned}
\Pi_a =& 
\sum_{k=0}^{M-1} \theta^k \left[ \otimes_{i=1}^3 \left( n_i^a \pi_{2i-1} + m_i^a \pi_{2i} \right) \right]
\\
=&  \sum_{k=0}^{M-1} \left[ \otimes_{i=1}^3 \left( n_i^{(\theta^k a)} \pi_{2i-1} + m_i^{(\theta^k a)}  \pi_{2i} \right) \right],
\end{aligned}
\een
and the intersection number can be written as
\ben
\begin{aligned}
\Pi_a \circ \Pi_b =& - \frac{1}{M} \sum_{k,l=0}^{M-1} \,\left[\, \prod_{i=1}^3 
 \left( n_i^{(\theta^k a)} \pi_{2i-1} + m_i^{(\theta^k a)}  \pi_{2i} \right) 
\circ 
\left( n_i^{(\theta^l b)} \pi_{2i-1} + m_i^{(\theta^l b)}  \pi_{2i} \right) \right]
\\
=& - \frac{1}{M}  \sum_{k,l=0}^{M-1} I_{(\theta^k a)(\theta^l b)}.
\end{aligned}
\een
Using further that $I_{(\theta^k a)(\theta^l b)} = I_{a (\theta^{l-k} b)}$, 
the intersection number on $T^6/\bZ_M$ with $M$ odd takes the form
\be{ChiBulk}
\Pi_a \circ \Pi_b = - \sum_{m=0}^{M-1}  I_{a (\theta^m b)} .
\ee

The result is modified for  $T^6/\bZ_M$ with $M=2N$. In this case, branes wrap
fractional cycles
$\Pi^{frac} = \frac{1}{2} \Pi^{bulk} + \frac{1}{2} \Pi^{\bZ_2}$,
and the $\bZ_2$ subgroup preserves any brane position, 
$b=(\theta^N b)$. The bulk contribution to the intersection number is therefore
\ben
\left ( \frac{1}{2} \Pi^{bulk}_a \right) \circ \left(\frac{1}{2} \Pi^{bulk}_b  \right)= - \frac{1}{2}  \sum_{m=0}^{N-1}  I_{a (\theta^m b)}.
\een
The exceptional part of the factional cycle, which was given in~\eqref{eq:cycleZ2expansion} for $T^6/\bZ_6'$,
can be written for any $T^6/\bZ_{2N}$ as
\be{eq:def_Ex_v2}
\Pi^{\bZ_2}_a = 
\sum_{k=0}^{N-1} \sum_{x_ay_a} (-1)^{\tau_{x_ay_a}} \,\, d_{\theta^k (x_a) \theta^k (y_a)} 
\otimes \left(n_2^{(\theta^k a)} \pi_3 + m_2^{(\theta^k a)} \pi_4 \right).
\ee
Here $d_{x_a y_a}$ denotes the exceptional two--cycle at 
the $\bZ_2$ fixed point $(x_a,y_a)$ on $T^2_1 \times T^2_3$ 
(or some permutation of tori) 
which is traversed by the bulk part of $a$.
The coefficients $\tau_{x_ay_a}$ have to fulfill
\ben
  \sum_{x_ay_a} \tau_{x_ay_a} = 0 \mod 2,
\een
in order to account for the 
 choice of a $\bZ_2$ eigenvalue and two discrete Wilson lines.
The intersection number among exceptional 
two--cycles,
\ben
  d_{xy} \circ d_{\tilde{x}\tilde{y}} =-2 \, \delta_{x\tilde{x}} \delta_{y\tilde{y}},
\een
at $\bZ_2$ singularities leads to 
\ben
\begin{aligned}
\left( \frac{1}{2} \Pi^{\bZ_2}_a \right)  \circ \left( \frac{1}{2} \Pi^{\bZ_2}_b \right) = & 
-\frac{1}{2} \sum_{m=0}^{N-1}  \left( \sum_{x_ay_a, x_by_b} (-1)^{\tau_{x_ay_a}+\tau_{x_by_b}} \; \delta_{x_a ,\theta^m (x_b)}
\, \delta_{y_a, \theta^m (y_b)} \; I^{(2)}_{a (\theta^m b)} \right)
\\ 
\equiv &  -\frac{1}{2} \sum_{m=0}^{N-1} I^{\bZ_2}_{a (\theta^m b)}.
\end{aligned}
\een
It follwos that the intersection number for fractional cycles takes the form
\be{ChiFractional}
\chi^{ab} \equiv  \chi_L^{ab} - \chi_R^{ab} =  
\Pi_a^{frac} \circ \Pi_b^{frac}
= -  \sum_{m=0}^{N-1} \frac{ I_{a (\theta^m
    b)} 
+ I^{\bZ_2}_{a (\theta^m b)}}{2} .
\ee
Using the fact that a generic open string sector $a(\theta^k b)$ contains one
fermionic massless degree of freedom and the $(\theta^k b)a$ sector completes
this to a chiral fermion in four dimensions, the chiral plus non--chiral 
bifundamental matter is counted by
\be{VarphiDef}
\varphi^{ab} \equiv \chi_L^{ab} + \chi_R^{ab}  =\sum_{m=0}^{N-1} \left| \frac{ I_{a (\theta^m
    b)} + I^{\bZ_2}_{a (\theta^m b)}}{2}  \right| .  
\ee
The total number of symmetric and antisymmetric states is obtained in a similar manner
when $\Pi_{O6}$ is split into its $\Omega{\cal R}\theta^{-2k}$ and  $\Omega{\cal R}\theta^{-1-2k}$ 
contributions. The wrapping numbers for $T^6/\bZ_6'$ and $k=0$ are listed in Table~\ref{O-planes}, the remaining ones
are obtained by rotating the wrapping numbers by $\theta^k$ ($k=1 \ldots N-1$) within the two orbits, see~\eqref{eqnmtransform} 
for $T^6/\bZ_6'$.

\renewcommand{\arraystretch}{1.3}
\begin{table}[ht]
  \begin{center}
    \begin{equation*}
      \begin{array}{|c|c|} \hline
        \multicolumn{2}{|c|}{\rule[-3mm]{0mm}{8mm}
\text{\bf Chiral and non--chiral massless matter on } T^6/(\bZ_{2N} \times \Omega {\cal R})  }\\ \hline\hline
\text{representation} & \text{total number} =\varphi
\\\hline\hline
({\bf Adj}_a) & 1 +\frac{1}{4} \sum_{k=1}^{N-1} \left| I_{a(\theta^k a)} +I_{a(\theta^k a)}^{\bZ_2}\right|
\\
({\bf Anti}_a) &  \frac{1}{4} \sum_{k=0}^{N-1}\left|I_{a(\theta^k a')}
    +I_{a(\theta^k a')}^{\bZ_2} + I_a^{\Omega{\cal R}\theta^{-k}} + I_a^{\Omega{\cal R}\theta^{-k+N} }
 \right|
\\
({\bf Sym}_a) &  \frac{1}{4} \sum_{k=0}^{N-1}\left|I_{a(\theta^k a')}
    +I_{a(\theta^k a')}^{\bZ_2} - I_a^{\Omega{\cal R}\theta^{-k}} - I_a^{\Omega{\cal R}\theta^{-k+N}}  \right|
\\
({\bf N}_a,\overline{\bf N}_b) & \frac{1}{2} \sum_{k=0}^{N-1}\left| I_{a(\theta^k b)} +I_{a(\theta^k b)}^{\bZ_2} \right|
\\
({\bf N}_a,{\bf N}_b) & \frac{1}{2}  \sum_{k=0}^{N-1}\left| I_{a(\theta^k b')} +I_{a(\theta^k b')}^{\bZ_2} \right|
\\ \hline
     \end{array}
    \end{equation*}
  \end{center}
\caption{Counting of all chiral and non--chiral matter states $\varphi$ in $T^6/(\bZ_{2N} \times \Omega {\cal R})$ models.}
\label{NonChiralSpectrum}
\end{table}
\renewcommand{\arraystretch}{1.3}

The complete computation of open string massless spectra for $T^6/(\bZ_{2N} \times
\Omega{\cal R})$ models is displayed in Table~\ref{NonChiralSpectrum}.\footnote{The analysis of the intersection numbers of
fractional D7--branes and O7--planes and the resulting six dimensional spectrum on $T^4/\bZ_M$ is analogous to
the $T^6/\bZ_M$ case discussed here, with the exception that the adjoint representation in the $aa$ sector is
absent since fractional D7--branes in six dimensions are completely stuck at the $\bZ_2$ fixed points,
whereas the D6--brane position in four dimensional models is free on the
additional two--torus. Furthermore, D7--branes parallel on both tori with opposite $\bZ_2$ eigenvalue but no
relative Wilson line or distance contribute two hyper multiplets in the representation $({\bf N}_a,\overline{\bf N}_b)$.
The expression for D7--branes can be explicitly checked
using the topological intersection numbers of the fractional cycles since the
massless spectrum in six dimensions is chiral.}
This computation is valid for fractional branes $a$, $b$ at generic non--vanishing
angles.\footnote{For models on $T^6/\bZ_M$ with $M$ odd the simplifications are obvious, starting with the 
expansion of intersection numbers~\eqref{ChiBulk} for 
bulk cycles  instead of~\eqref{ChiFractional} for fractional cycles. The $aa$ sector contains then three multiplets in the 
adjoint representation. If branes $a,b$ are parallel along $T^2_i$, one has to replace $|I^{(i)}_{ab}|=0 \rightarrow 2$.}
On $T^6/\bZ_M$ for arbitrary $M$, there exist also supersymmetric sectors where branes are parallel on either
one or all three tori. In this case, the formulae for $M$ even are modified as follows:
\begin{itemize}
\item for D6--branes parallel along all three directions:
if some relative Wilson line or parallel displacement on $T^2_1 \times T^2_3$
exists, the massless matter is lifted. The case of identical branes is
included in Table~\ref{NonChiralSpectrum} by the universally existing chiral
multiplet in the adjoint representation.
If two branes $a$ and $b$ have opposite $\bZ_2$ eigenvalue but identical position and no
relative Wilson line, the spectrum contains $2 \times [({\bf N}_a,\overline{\bf N}_b) + c.c.]$ 
chiral multiplets. In the special case of $b=a'$, the bifundamental representation is replaced by 
the antisymmetric as in the non--chiral models in~\cite{bgk00,bcs04}. 
\item for D6--branes parallel along the two--torus on which the $\bZ_2$ subgroup acts
  trivially, i.e. $T^2_2$ for $T^6/\bZ_6'$:
as on $T^6$, there is a non--chiral matter pair at each intersection on $T^2_1 \times T^2_3$. 
The total number of states is computed by replacing
\ben  
I_{ab}^{(2)} =0 \rightarrow \left| I_{ab}^{(2)} \right| =2 \quad \text{in} \quad I_{ab} \quad \text{and}  \quad I_{ab}^{\bZ_2}.
\een
 This agrees with the result for the $({\bf 6}_i,{\bf 6}_{i+3})$ sectors of the $T^6/\bZ_6$ and $T^6/\bZ_6'$ models
in~\cite{bgk00,bcs04}. The counting of the adjoint representations in the $({\bf 6}_i,{\bf 6}_{i+2})$ 
sectors of the non--chiral $T^6/\bZ_4$ models that can be found in the papers cited above, receives a factor of 1/2 
since the $a(\theta a)$ sector has no inverse sector providing anti--particles as does $ba$ for $ab$.
\item for D6--branes parallel along one of the two two--tori with $\bZ_2$ fixed points,
  i.e. $T^2_1$ or $T^2_3$ for $T^6/\bZ_6'$:
in case of a relative Wilson line or displacement on the corresponding torus,
  the massless matter is lifted. Without Wilson line or displacement, 
the two massless states have opposite $\bZ_2$ eigenvalue, such that there
exists one chiral multiplet irrespective of the relative $\bZ_2$ eigenvalue.
The formulae for counting bifundamentals, adjoints, symmetrics and antisymmetrics simplify, e.g. for branes parallel along $T^2_3$
and no Wilson lines or displacements we obtain
\ben
\begin{aligned} 
({\bf Adj}_a):\; &  \frac{1}{4} \left|I_{a(\theta^k a)}+ I_{a(\theta^k a)}^{\bZ_2} \right| 
\longrightarrow \frac{1}{2} \left| I^{(1)}_{a(\theta^k a)} \, I^{(2)}_{a(\theta^k a)} \right| , 
\\\nonumber
({\bf Anti}_a):\; &   \frac{1}{4} \left|I_{a(\theta^k a')}
    +I_{a(\theta^k a')}^{\bZ_2} + I_a^{\Omega{\cal R}\theta^{-k}} + I_a^{\Omega{\cal R}\theta^{-k+N} } \right| \\
&\qquad\longrightarrow \frac{1}{2} \left| I^{(1)}_{a(\theta^k a')} \, I^{(2)}_{a(\theta^k a')} + I^{\Omega{\cal R}\theta^{-k}}_{a; T_1 \times T_2}
 \right| , 
\\\nonumber
({\bf Sym}_a):\; &   \frac{1}{4} \left|I_{a(\theta^k a')}
    +I_{a(\theta^k a')}^{\bZ_2} - I_a^{\Omega{\cal R}\theta^{-k}} - I_a^{\Omega{\cal R}\theta^{-k+N}}  \right| \\
&\qquad\longrightarrow \frac{1}{2} \left|I^{(1)}_{a(\theta^k a')} \, I^{(2)}_{a(\theta^k a')} - I^{\Omega {\cal R}\theta^{-k}}_{a; T_1 \times T_2}
 \right| , 
\\\nonumber
({\bf N}_a,\overline{\bf N}_b):\; & \frac{1}{2} \left|I_{a(\theta^k b)}+ I_{a(\theta^k b)}^{\bZ_2} \right|
\longrightarrow \left| I^{(1)}_{a(\theta^k b)} \, I^{(2)}_{a(\theta^k b)} \right| , 
\\\nonumber
({\bf N}_a,{\bf N}_b):\; & \frac{1}{2} \left|I_{a(\theta^k b')}+ I_{a(\theta^k b')}^{\bZ_2} \right|
\longrightarrow \left| I^{(1)}_{a(\theta^k b')} \, I^{(2)}_{a(\theta^k b')} \right| .
\end{aligned}
\een
The modifications for parallel branes along $T^2_1$ are obvious.
These formulae apply  to the $({\bf 6}_i,{\bf 6}_{i+2})$ sectors of the non--chiral $T^6/\bZ_6'$ models 
in~\cite{bgk00,bcs04}.
\end{itemize}

%
%
\section{Methods of analysis}\label{methods}
To study the solution space of the constraining equations at a statistical level, there are two basic possibilities.
Either we can find a suitable approximation to the distribution of solutions in the parameter space,
or we can study a set of explicitly calculated solutions.
We follow the second approach and use a computer generated ensemble of solutions. In fact we have been able to
construct \emph{all} possible solutions for this geometry and do therefore not have to rely on a choice of random subsets of
parameters as in the case of $T^6/\bZ_6$ in~\cite{gls07}.
The computer algorithm we used combines methods developed for the similar analyses in~\cite{gbhlw05} and~\cite{gls07}.

%
%
\subsection{Finiteness of solutions}\label{finite}
Before starting to analyse the space of four--dimensional solutions, one would like to know whether there are only finitely
many. If this were not the case, a systematic study could give no results and one should better use a method based on random
subsets instead.
In the following we give a proof that the number of solutions to the constraining equations from
supersymmetry, K--theory and RR--tadpoles is finite.
The proof is similar in structure to the one given for the $\bZ_6$--case
in~\cite{gls07}, but some subtle differences arise due to the different
structure of the tadpole equations as well as the complex structure parameter $\varrho$. 
Nevertheless it is also possible to show in
the case at hand that the contributions from individual D6--brane stacks to
the left hand side of the bulk tadpole conditions~\eqref{eqrrtadpoles} 
are always positive and the total number of possible
solutions is therefore bounded from above.

For the question of a finite number of solutions we care only about the
tadpole conditions for the bulk cycles~\eqref{eqrrtadpoles} and the
supersymmetry constraints~\eqref{Eq:SUSY1} and~\eqref{Eq:SUSY2}. This is
justified by two results from Section~\ref{secrrtadpoles}. Firstly the fact
that the tadpole conditions for the exceptional part of the fractional cycles
are decoupled and receive no contribution from the orientifold planes.
Secondly by the notion that the number of possible combinations of exceptional
cycles that can be used to ``dress'' one bulk cycle is always finite.
The K--theory conditions could only reduce the total number of solutions and
we do not need them for the proof. As discussed in section~\ref{Ktheory} they
are trivially fulfilled in the present case anyway.

Showing the finiteness of solutions therefore consists of two remaining
steps.
In a first step we show that the contribution of a single supersymmetric 
brane stack to
the tadpole condition for the bulk cycles~\eqref{eqrrtadpoles} in terms of the
variables $P, Q, U, V$, defined by~\eqref{eqpquvdef}, is always
positive.
In a second step it is demonstrated that the variables we use are ``good variables''
in the sense that no infinite series of values for the wrapping numbers
$\{n_i,m_i\}$ can occur for constant values of $P,Q,U$ or $V$.
To simplify the discussion, we deal only with the
$\mathbf{AAa/b}$--geometry in the following, the proof for the other
possibilities can be obtained by analogy.

Using the first supersymmetry condition~\eqref{Eq:SUSY1} and
$U+V=m_3/n_3(P+Q)$ for $n_3 \neq 0$, we obtain the following expression for the 
 contribution to the first bulk 
tadpole of one brane stack,
\be{eqfin1}
  2P+Q=\frac{-3Q}{2\varrho(b+\frac{m_3}{n_3})}.
\ee
We have to impose the constraints $n_3\ge 0$ and \mbox{$m_3+bn_3 \ge 0$} on the
wrapping numbers on the third torus in order to avoid overcounting of solutions
obtained by trivial geometric symmetries, as explained in detail in
Section~\ref{algorithm}.
This restriction constrains \mbox{$b+m_3/n_3$} to be always positive, which is
also trivially true for the complex modulus $\varrho$.
Negative contributions to the tadpole equation can therefore only occur
for branes with $Q>0$.

Combining~\eqref{eqfin1} with the second supersymmetry condition~\eqref{Eq:SUSY2} leads
to the inequality
\ben
  -3Q > 4\varrho^2\left(b+\frac{m_3}{n_3}\right)Q,
\een
which can only be fulfilled for $Q<0$. Therefore all terms in the first
sum of the bulk tadpole constraints~\eqref{eqrrtadpoles} are positive. The values
of $(2P+Q)$ are bounded from above by the right hand side of the equation,
which is given by the constant orientifold charge.\footnote{This constant might
be modified in the presence of fluxes (cf. also the discussion
in~\cite{bghlw04}), but it is always positive and bounded from above.}

Having established that the value of $Q$ is bounded, the only possibility to
obtain an infinite series of solutions arises if the
values of \mbox{$(n_1,m_1,n_2,m_2,n_3,m_3)\in\bZ$} are unbounded for a constant value of $Q$.

An infinite series can only arise in the part $n_1m_2+m_1n_2+m_1m_2$ of the
definition~\eqref{eqpquvdef}. There are two possible cases: Either all
$\{n_1,m_1,n_2,m_2\}$ are unbounded, or only two of them. In the first case it
would be impossible to satisfy~\eqref{Eq:SUSY2}, therefore we are left with the
second case. An infinite series of solutions can arise, if we have that
$n_1=n_2\in\bN$ and $m_1=-m_2=\mathrm{const}$. But this is no valid solution
since $P=(n_1n_2-m_1m_2)n_3$ is unbounded from above and therefore the
tadpole conditions cannot be fulfilled. 

The second bulk tadpole condition is treated similarly: $n_3 >0$ and $Q <0$ 
imply $n_1m_2+m_1n_2+m_1m_2 <0$, which together with $m_3+bn_3 \ge 0$ produces
$V+bQ \le 0$.

The case $n_3=0$ leads to a vanishing contribution to the first bulk tadpole
and a positive contribution to the second one, as well as to the simplified supersymmetry constraint
$-V > 0$.
This completes the proof.

\subsection{Algorithm}\label{algorithm}
The computer algorithm we used to generate the ensemble of solutions is divided into three parts.
Thereby we make use of the fact that the model building constraints can be treated separately for bulk and exceptional
cycles.

In a first step we determine all supersymmetric bulk branes, i.e. those satisfying the constraints~\eqref{Eq:SUSY1}
and~\eqref{Eq:SUSY2}, for all possible values of the complex structure modulus~$\varrho$ and all choices of tori
such that the bulk tadpole conditions~\eqref{eqrrtadpoles} are not exceeded.\footnote{To determine the
supersymmetric cycles for all possible values of $\varrho$ is non--trivial and has shown to be impossible in the case
of $T^6/(\bZ_2\!\times\!\bZ_2)$~\cite{gbhlw05}.}
In order to count models only once which lie in the same orbit of the orbifold and orientifold projections, we use the
additional constraints\footnote{There are several possibilities of constraints one could choose, singling out
different representatives of the orbits of the orbifold and orientifold projections. However, this choice does not
affect the properties of the models.}
\ben
  (n_1,m_1) \equiv (1,1) \mod 2,\qquad n_1,n_3,m_3+bn_3 \ge 0.
\een
In more detail, the first constraint restricts the choice of representatives for equivalent bulk--cycles under the orbifold
projection to just one. The second and third constraint ($n_1, n_3 \ge 0$) take care of the fact that one might flip the signs
of bulk cycles on two of the three two--tori simultaneously, without changing the model. The last constraint finally makes
sure that we count models only once under the exchange of branes with their orientifold images.

The bulk branes obtained in this way are used in the second step to construct all combinations of stacks of branes
that fulfil the bulk tadpole conditions~\eqref{eqrrtadpoles}.
At this point we allow for combinations of different bulk cycles only, which leads to a rather restricted set of
bulk configurations. The properties of these models without exceptional branes are presented in Section~\ref{resbulk}.

Finally, all consistent combinations of exceptional cycles, i.e. those for which the tadpoles of the exceptional cycles
cancel, are computed for each bulk configuration.
In this way we obtain all possible fractional brane solutions to the combined tadpole and supersymmetry constraints.
More concretely, the exceptional cycles are obtained by considering all 128 combinations of shifts
$(\sigma_1,\sigma_2,\sigma_5,\sigma_6)\in\{0,1/2\}$ and signs $(\tau_0,\tau_1,\tau_3)\in\{0,1\}$, which 
together with the wrapping numbers on the second torus $(n_2,m_2)$
determine the cycle unambiguously.

%
%
\section{Results}\label{results}
In this section we present the results of a complete survey of supersymmetric models on $T^6/\bZ_6'$.
In analogy to the way the models are constructed, we begin with combinations of bulk cycles that fulfil the
bulk tadpole constraints~\footnote{Note that we are always referring to the bulk part of fractional cycles
when we talk about bulk cycles in this section.}.
In a second part the full models, including exceptional cycles, are analysed.
Finally we look in detail at solutions that resemble the gauge sector and chiral matter content of the
standard model and Pati--Salam or $SU(5)$ models, respectively.

%
%
\subsection{Solutions for the bulk branes}\label{resbulk}
In a first step, we construct all possible supersymmetric bulk cycles and combine them to models fulfilling the bulk
tadpole constraints~\eqref{eqrrtadpoles}.
In total there are 4416 bulk cycles, out of which one can construct 13416 different models.
The value of the complex structure modulus $\varrho$ varies between $1/96$ and $135$.

\renewcommand{\arraystretch}{1.3}
\tabfig{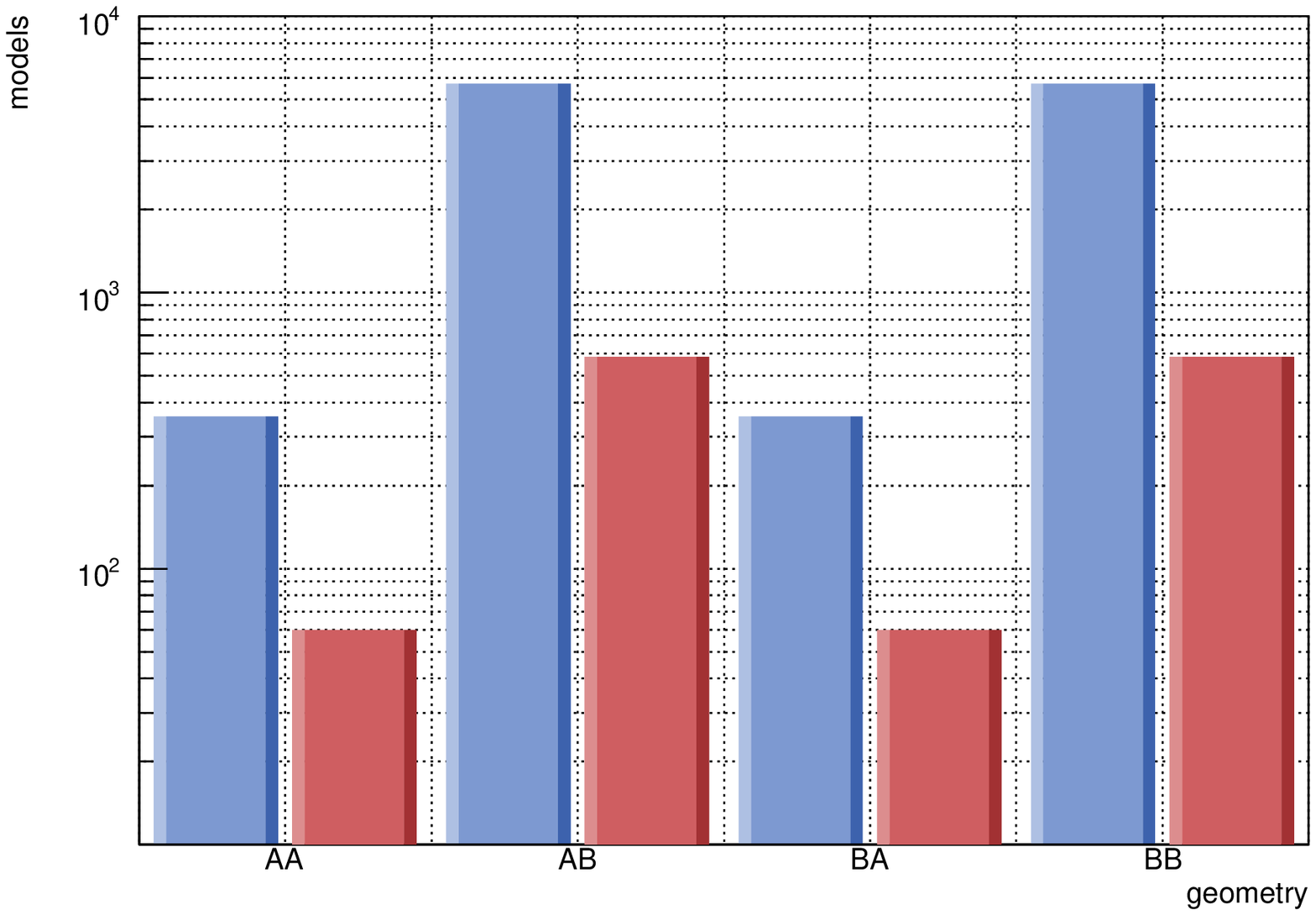}{fig_bulk_num}{%
Number of pure bulk solutions for the different geometries.
The four groups of bars represent the geometry on the first two tori, while the geometry of the third torus is represented by the blue
bars on the left ($b=0$) and the red bars on the right ($b=1/2$) in each group.}{%
\begin{tabular}{|l||r|r|}\hline
\textbf{b}&0&1/2\\\hline\hline
\textbf{AA}&  $356$&   $60$\\\hline
\textbf{AB}& $5706$&  $586$\\\hline
\textbf{BA}&  $356$&   $60$\\\hline
\textbf{BB}& $5706$&  $586$\\\hline
\end{tabular}%
}{0.5}{0.5}
\renewcommand{\arraystretch}{1.3}

In Figure~\ref{fig_bulk_num} the total number of solutions to the bulk tadpole equations for the different
geometries on the three two--tori is given.
Obviously the geometry of the first torus has no influence on the results, while the geometries of the second and
third torus change the number of solutions significantly.

\twofig{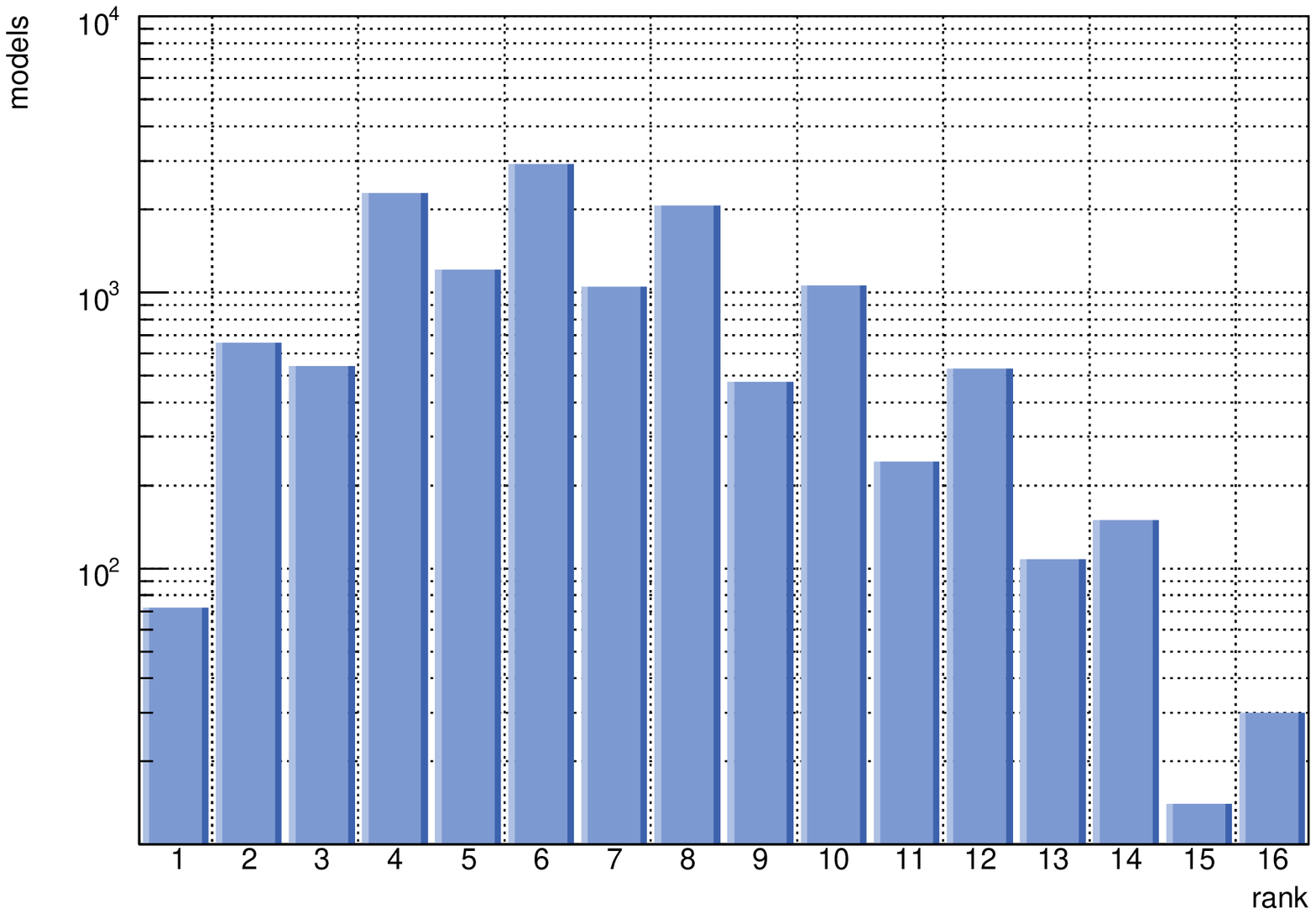}{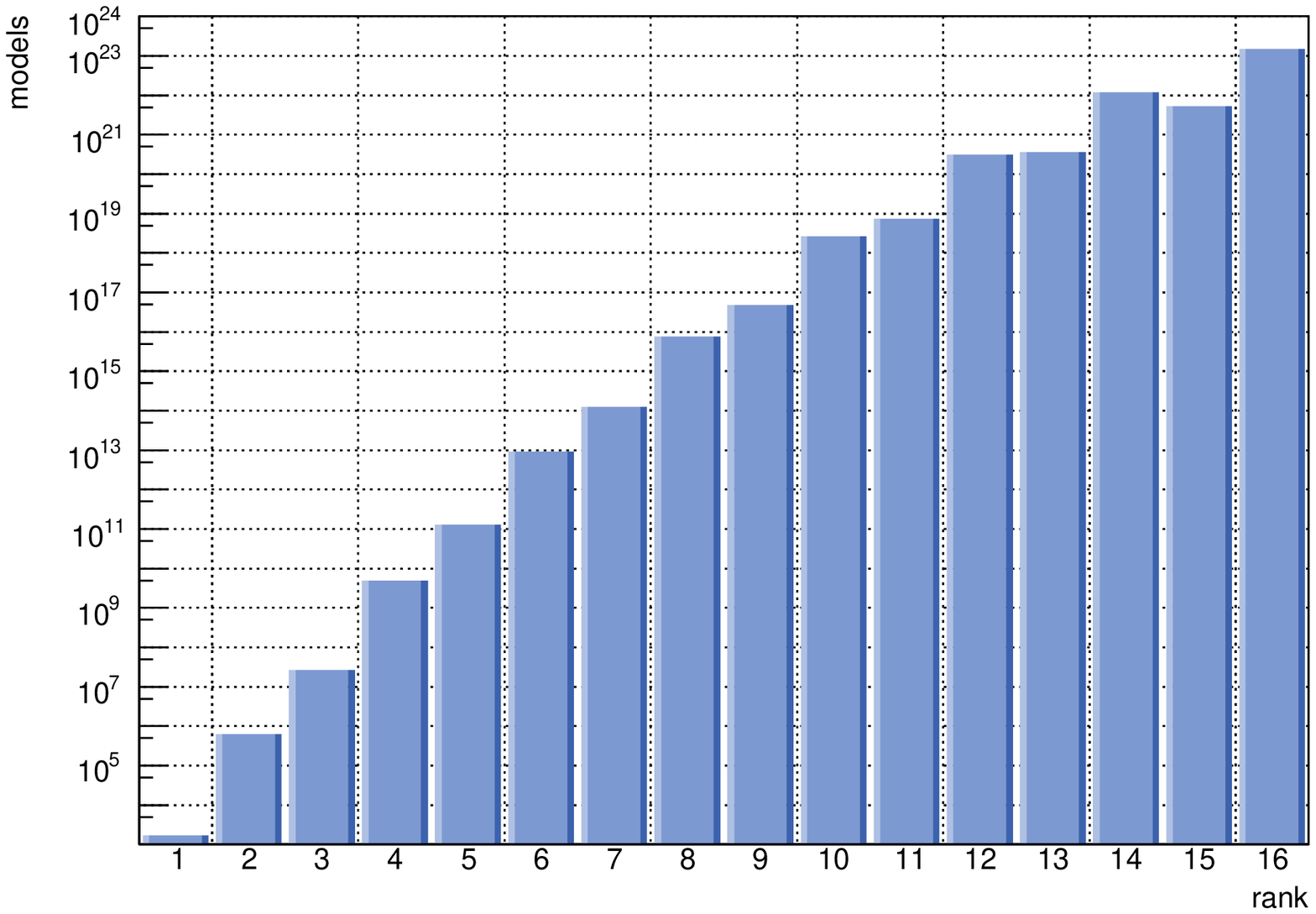}{fig_rank}{Frequency distributions of the total rank of the gauge group for (a) pure bulk solutions and (b) the full set of solutions.}

Another interesting feature of the solutions that shows up in the bulk part and that parallels the observations we
made for~$\bZ_6$ in~\cite{gls07}, concerns the rank of the total gauge group.
The distribution shown in Figure~\ref{fig_rank_a} is quite similar to the results on the $T^6/(\bZ_2\!\times\!\bZ_2)$--orbifold.
We expect this behaviour to change drastically after the inclusion of exceptional cycles, in a similar manner as
in the~$\bZ_6$ case. The number of possible combinations of exceptional cycles goes roughly as
$128^k$, where $k$ is the number of brane--stacks, and therefore models with a large number of stacks
make up for the main contribution to the statistics. We show in the next section that this is indeed the observed behaviour.

\twofig{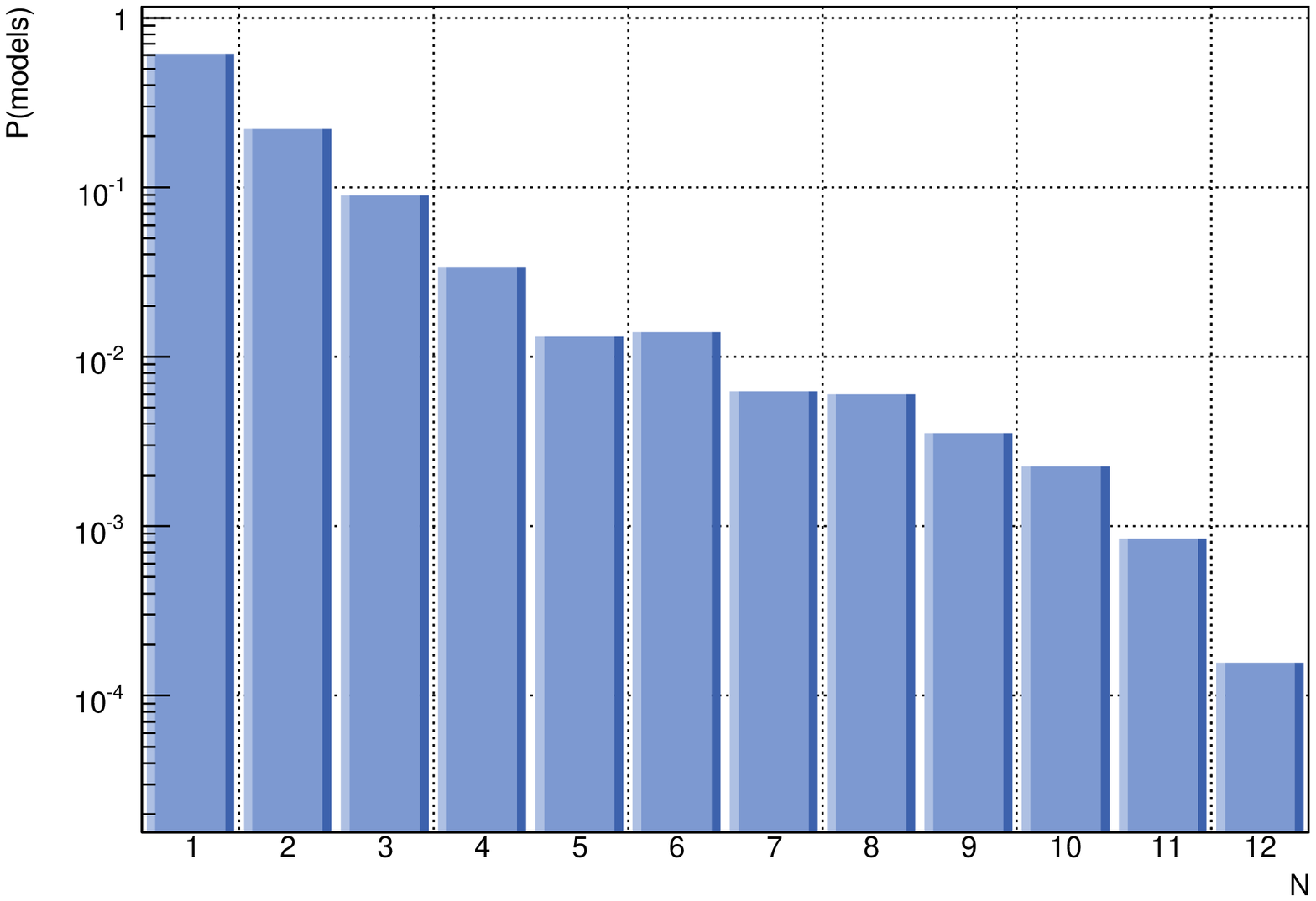}{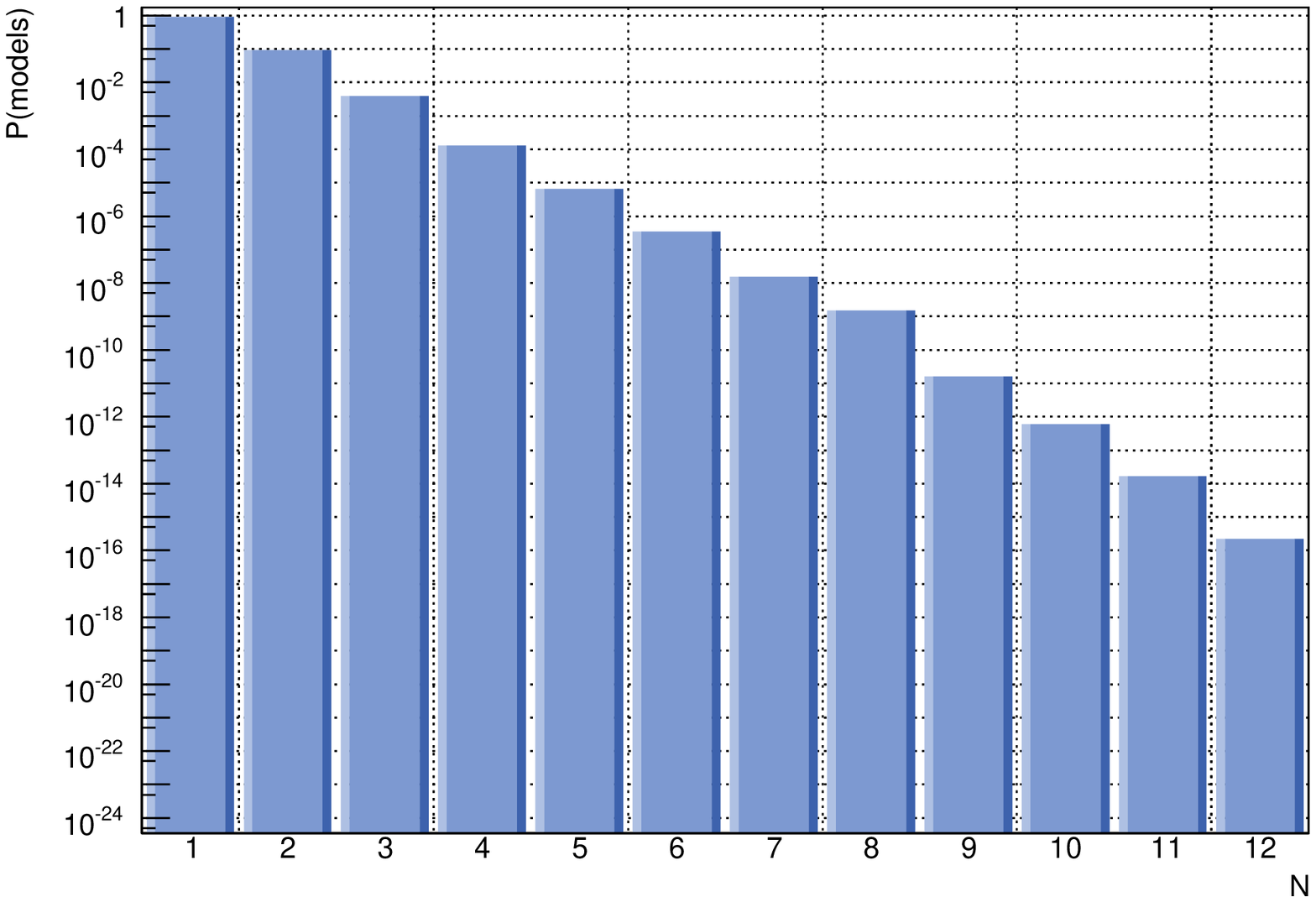}{fig_un}{Frequency distributions of the probability to find a gauge factor of rank $N$
for (a) pure bulk solutions and (b) the full ensemble of solutions.}

Finally we consider the probability distribution to find a single gauge group factor $G_i$ of rank $N$ within the total gauge
group~$G=\bigotimes G_i$.
The plot shown in Figure~\ref{fig_un_a} describes the probability distribution for pure bulk models.
The number of solutions $\cN(N)$ can be approximated very well by
\be{eq:bulk_un_approx}
  \cN(N) \approx \sum_{k=1}^{T+1-N}\frac{T^4}{N^2} = \frac{T^4}{N^2}(T+1-N),
\ee
where $T$ is the orientifold charge appearing on the right hand side of the tadpole constraint and can be read off for
the different geometries from~\eqref{eqrrtadpoles}.
The factor of $\frac{T^4}{N^2}$ has been derived in~\cite{dota06} for the $T^6/(\bZ_2\!\times\!\bZ_2)$ orbifold and holds
in the present case as well. The additional factor of $(T+1-N)$ has to be included, because the result
of~\cite{dota06} has been derived for a fixed number of stacks. Since we are showing the contributions from all
stacks, we have to sum over all possibilities for the gauge factor in question to appear. This is given approximately
by the maximum number of stacks, which is fixed by $T$, minus the contribution of the brane factor, which is
proportional to $N$.

%
%
\subsection{Complete solutions}\label{resfull}
After the inclusion of all possible combinations of exceptional cycles, we obtain a much larger number of solutions,
in total $1.67\times 10^{23}$. Note that this number is five orders of magnitude smaller then the total number of
solutions we estimated\footnote{Since it was not possible to compute all models explicitly in the $\bZ_6$--case, a
random subset method has been used to estimate the full frequency distributions. The results have been checked for
convergence and the estimated error is smaller then 1\%.} in the case of~$\bZ_6$ to be~$3.4\times 10^{28}$~\cite{gls07}.
In Figure~\ref{fig_frac_num} the total number of full solutions to the tadpole and supersymmetry constraints is
shown for the different geometries.

\renewcommand{\arraystretch}{1.3}
\tabfig{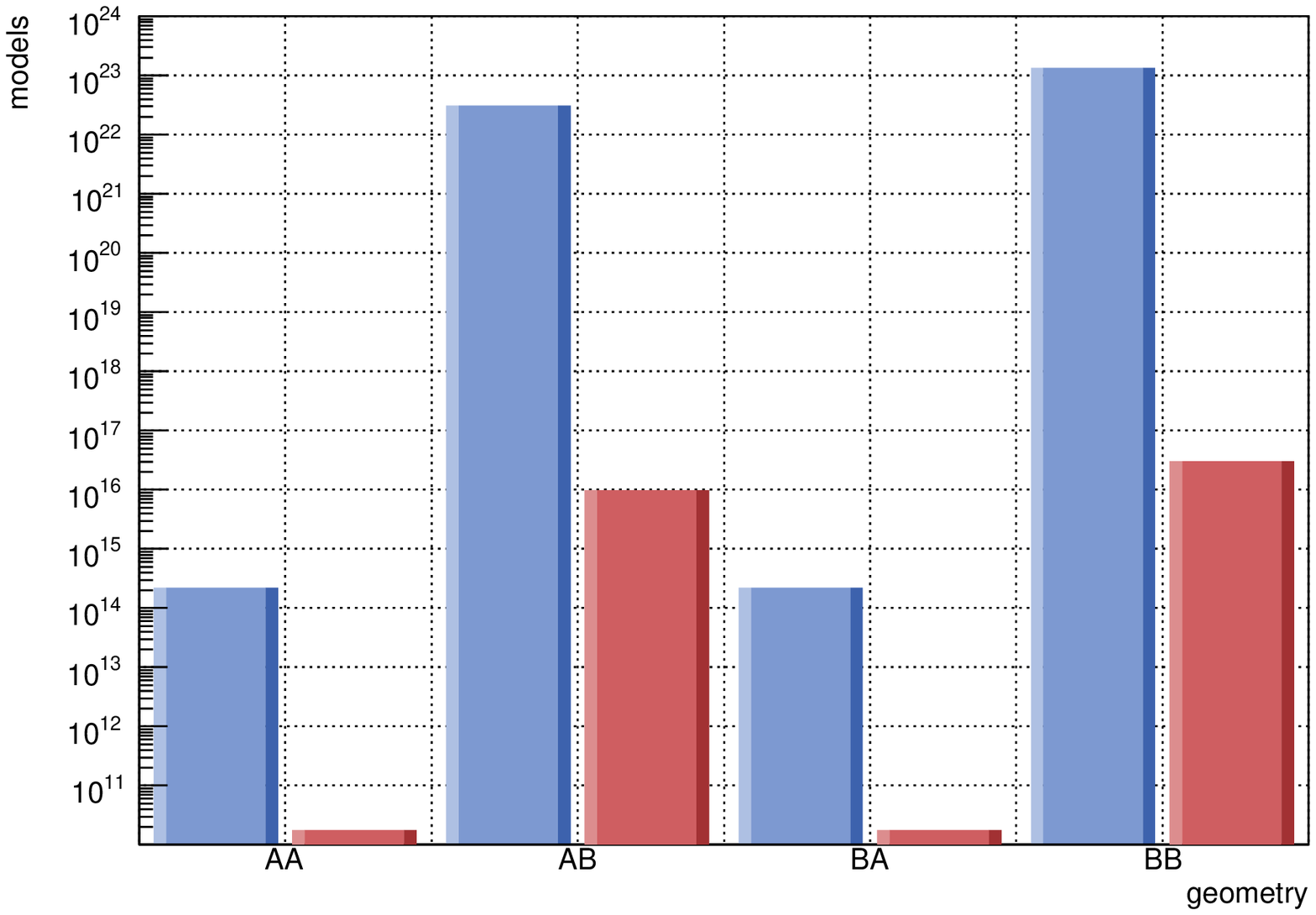}{fig_frac_num}{%
Frequency distribution of the total number of solutions for the different geometries.
As in Figure~\protect\ref{fig_bulk_num} the four groups of bars represent the geometry on the first two tori,
while the geometry of the third torus is represented by the blue bars on the left ($b=0$) and the red bars on the
right ($b=1/2$) in each group.}{%
\begin{tabular}{|l||r|r|}\hline
\textbf{b} &                   0&                 1/2\\\hline\hline
\textbf{AA}&$2.21\times 10^{14}$&$1.79\times 10^{10}$\\\hline
\textbf{AB}&$3.09\times 10^{22}$&$9.93\times 10^{15}$\\\hline
\textbf{BA}&$2.21\times 10^{14}$&$1.79\times 10^{10}$\\\hline
\textbf{BB}&$1.37\times 10^{23}$&$3.08\times 10^{16}$\\\hline
\end{tabular}%
}{0.5}{0.5}
\renewcommand{\arraystretch}{1.3}

In contrast to the pure bulk results, we obtain different numbers of solutions for the \textbf{AB} and \textbf{BB} geometries,
although the numbers for the \textbf{AA} and \textbf{BA} variants are again identical.
The symmetry between different geometries is therefore lifted if the second torus is tilted.

In Figure~\ref{fig_rank_b} the frequency distribution of the total rank of the gauge group is shown. As already
advertised in the last section, we find a very different behaviour compared to the bulk result in Figure~\ref{fig_rank_a}.
Solutions with a large total
rank of the gauge group are greatly enhanced. This is due to the fact that many more possibilities for different
combinations of exceptional cycles exist for configurations with a large number of stacks, each consisting of a small
number of branes. In fact those configurations with the maximum number of stacks, each consisting of a single brane,
dominate the statistics, since the number of combinations of exceptional cycles scales roughly as $n_e^k$, where $k$ is
the number of stacks and $n_e$ the number of exceptional cycles that can be chosen for one bulk cycle. In principle there
are $2^7$ possibilities for each bulk cycle, considering all combinations of $\sigma$ and $\tau$, but not all of them
lead to solutions of the tadpole and supersymmetry constraints.
This behaviour is completely analogous to the case of $T^6/\bZ_6$, where very similar results have been found~\cite{gls07}.

Another statistical distribution that gets enhanced by the contribution of the exceptional part of the fractional cycles
is given by the probability distribution of single gauge group factors of rank $N$.
In Figure~\ref{fig_un_b} we show the distribution for the full set of models. Comparing with the distribution for the
pure bulk solutions in Figure~\ref{fig_un_a}, we see an enhancement of smaller factors.
Including the factor of $n_e^k$ in~\eqref{eq:bulk_un_approx}, we find
\be{eq:frac_un_approx}
  \cN(N) \approx \sum_{k=1}^{T+1-N}\frac{T^4}{N^2}n_e^k = \frac{T^4}{N^2}\left(\frac{n_e^{T+2-N}-1}{n_e-1}\right)
    \approx\frac{T^4}{N^2}n_e^{T+1-N}.
\ee
Using this formula to fit the results, we obtain $n_e\approx 32$, which means that actually only $1/4$ of all possible
combinations of exceptional cycles lead to a consistent model.

%
%
\subsection{Standard models}\label{ressm}
After the general analysis in the last two sections, we are now going to restrict our attention to a specific subset
of solutions. In the context of a survey of parts of the landscape we are of course interested in the probability to find
vacua which possess as many properties of the standard model, or more precisely the MSSM since we are dealing with supersymmetric
solutions only, as possible.
To single out solutions which come close to this goal, we will firstly consider only those which contain factors of
$SU(3)\times SU(2)\times U(1)$ in their gauge group $G$.
Secondly we require the right amount of chiral matter, appropriately charged under the standard model gauge group. Since in
the type II orientifold setup only $U(N)$, $SO(2N)$ or $Sp(2N)$ gauge groups can occur, we will always have at least one
additional $U(1)$ factor\footnote{This is only true if the $SU(2)$ factor is realised by a $Sp(2)$ brane, otherwise there are
at least two additional $U(1)$s.}.

\renewcommand{\arraystretch}{1.3}
\begin{table}[ht]
\begin{center}
\begin{tabular}{|l||c|c|c|c|}\hline
      & \textbf{(i)}                  & \textbf{(ii)}                 & \textbf{(iii)}             & \textbf{(iv)}            \\\hline\hline
$G$   & $\quad U(3)_a \times Sp(2)_b$ & \muc{3}{c|}{$\quad U(3)_a \times U(2)_b$}                                             \\
      & $\times U(1)_c \times U(1)_d$ & \muc{3}{c|}{$\times U(1)_c \times U(1)_d$}                                            \\\hline\hline
$Q_L$ & $\chi^{ab}$                   & $\chi^{ab}+\chi^{ab'}$        & $\chi^{ab}$                & $\chi^{ab}$              \\\hline
$u_R$ & $\chi^{a'c}+\chi^{a'd}$       & $\chi^{a'c}+\chi^{a'd}$       & $\chi^{\Anti_a}$           & $\chi^{\Anti_a}$         \\\hline
$d_R$ & $\chi^{a'c'}+\chi^{a'd'}$     & $\chi^{a'c'}+\chi^{a'd'}$     & $\chi^{a'c}+\chi^{a'd}$    & $\chi^{a'c}+\chi^{a'c'}$ \\
      & $+\chi^{\Anti_a}$             & $+\chi^{\Anti_a}$             & $+\chi^{a'c'}+\chi^{a'd'}$ &                          \\\hline
  $L$ & $\chi^{bc}+\chi^{bd}$         & $\chi^{bc}+\chi^{bd}$         & $\chi^{bc}+\chi^{bd}$      & $\chi^{b'd}$             \\
      &                               & $+\chi^{b'c}+\chi^{b'd}$      & $+\chi^{b'c}+\chi^{b'd}$   &                          \\\hline
$e_R$ & $\chi^{\Sym_c}+\chi^{\Sym_d}$ & $\chi^{\Sym_c}+\chi^{\Sym_d}$ & $-\chi^{\Sym_b}$           & $\chi^{cd'}+\chi^{c'd'}$ \\
      & $+\chi^{cd'}$                 & $+\chi^{cd'}$                 &                            & $-\chi^{\Sym_b}$         \\\hline\hline
$q_Y$ & $\frac{1}{6}q_a+\half q_c+\half q_d$ & $\frac{1}{6}q_a+\half q_c+\half q_d$
         & $-\frac{1}{3}q_a-\half q_b$ & $-\frac{1}{3}q_a-\half q_b+q_d$ \\\hline
\end{tabular}
\caption{Chiral matter spectrum and definition of the hypercharge $q_Y$ in terms of the four $U(1)$ charges for the four different embeddings of the standard model. The amount of chiral matter is given in terms of the intersection numbers as defined in Table~\protect\ref{ChiralSpectrum}. The constraints on possible models can be read off by requiring the number of generations for all sectors to be equal.}
\label{tab_smmatter}
\end{center}
\end{table}
\renewcommand{\arraystretch}{1.3}

Note that we use the same constructions as in previous works on intersecting brane statistics~\cite{gbhlw05,gls07}.
The standard model sector is implemented with four stacks of branes ($a$, $b$, $c$, $d$)~\cite{imr01}.
To realise the matter spectrum of the standard model, we allow for four different possibilities.
In terms of the intersection numbers between the four brane stacks, the chiral matter sector is summarised in Table~\ref{tab_smmatter}.
We do not restrict our search to models with three generations of quarks and leptons, but we require the spectrum to be self--consistent
in the sense that the number of generations for all matter species is identical and we do not obtain additional chiral matter from the
visible sector.

We do allow for a hidden sector, i.e. additional gauge group factors and chiral matter, which ideally should be completely
decoupled from the visible sector of the standard model in the sense that no chiral matter charged under both, visible and
hidden sectors occurs (so called chiral exotics).
Furthermore we have to find a combination of the several $U(1)$s in the
game which has the right properties to serve as the standard model hypercharge. Since $U(1)$ factors in these models might
receive a mass through the Green--Schwarz mechanism, we will have to make sure that at least the hypercharge stays massless.
This is exactly the case if the hypercharge is non--anomalous under mixed gauge anomalies.
It is interesting to see how large the suppression of solutions by this condition will be,
since it has been found in~\cite{gbhlw05} that this additional constraint is quite weak compared to the much
stronger requirements of obtaining the right gauge group and matter spectrum.
To check if this is true in the present case as well, we left the property of having a non--anomalous hypercharge as an
open parameter.

Explicitly, the mixed gauge anomaly for $U(1)_a$ and $SU(N)_b$ is given by
\ben
  \cA \sim N_a\left(\chi^{ab}+\chi^{ab'}\right).
\een
To obtain a massless hypercharge $U(1)_Y$, this anomaly has to vanish for all possible non--abelian gauge factors.
This results in the condition that the corresponding cycle,
\ben
  \Pi_Y = \sum_{i\in\{a,b,c,d\}} N_i x_i \Pi_i,
\een
with coefficients $x_i$ as given in Table~\ref{tab_smmatter}, has to be invariant under the orientifold projection.

\renewcommand{\arraystretch}{1.3}
\tabfig{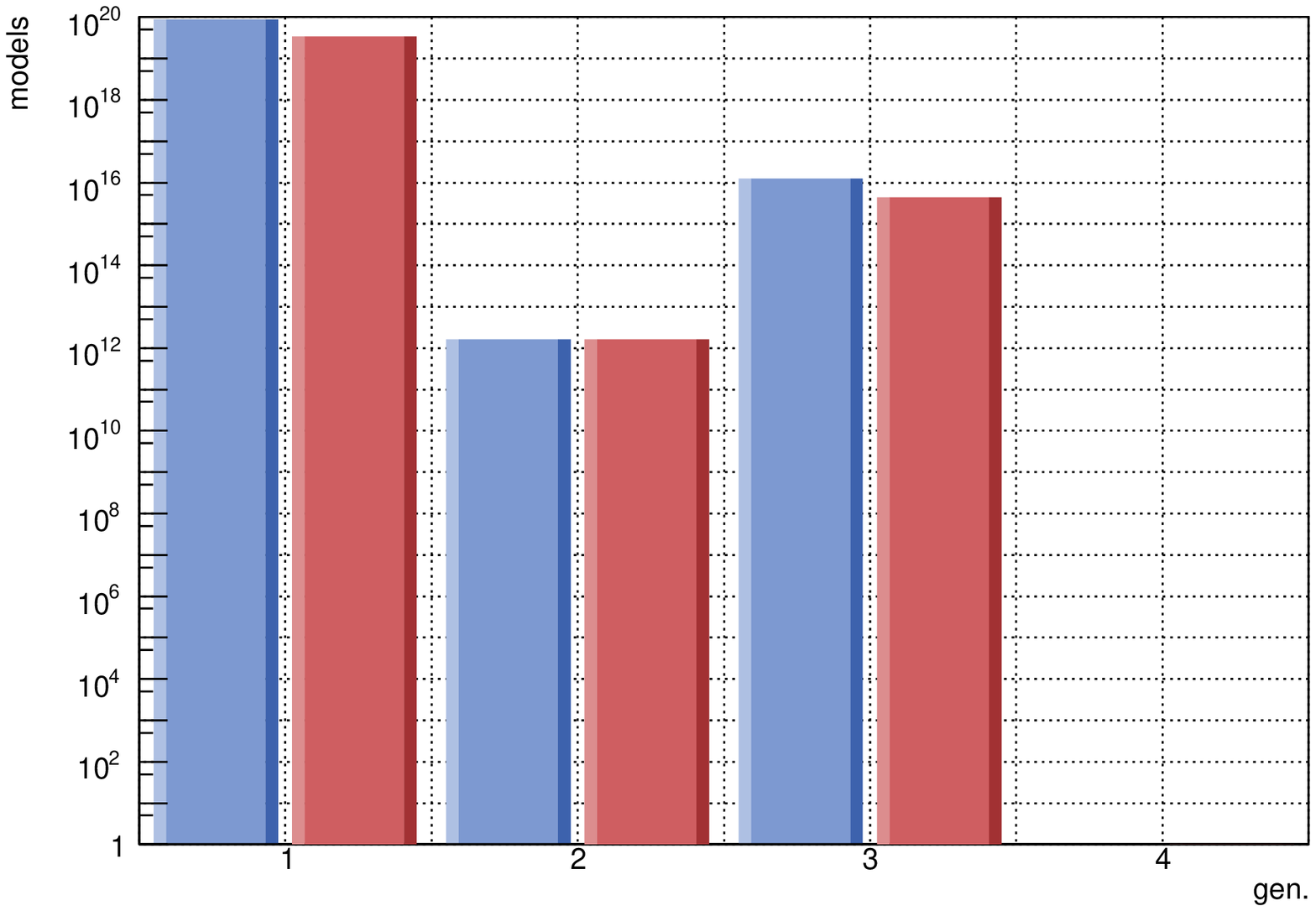}{fig_sm_a}{Frequency distribution of models with the gauge group and the chiral matter
spectrum of the MSSM for different numbers of generations in the visible sector.
For each generation the left blue bars represent models with a massive hypercharge, while the red bars on
the right count models with a massless hypercharge only.}{%
\begin{tabular}{|c|c|c|}\hline
\textbf{\# gen.}&\textbf{\# models}   &\textbf{\# models w/}\\[-1ex]
                &                     &\textbf{massless U(1)}\\\hline\hline
$1$ & $8.79\times 10^{19}$ & $3.42\times 10^{19}$ \\\hline
$2$ & $1.63\times 10^{12}$ & $1.63\times 10^{12}$ \\\hline
$3$ & $1.28\times 10^{16}$ & $4.43\times 10^{15}$ \\\hline
\end{tabular}%
}{0.5}{0.5}
\renewcommand{\arraystretch}{1.3}

The results of a search for possible standard models are shown in Figure~\ref{fig_sm_a}.
All models that have been found belong to the embedding of type (ii) from Table~\ref{tab_smmatter}.
Demanding a massless hypercharge does change the
number of solutions by a factor of~$0.4$, which is quite insignificant compared to the suppression factor
of~$7.3\times 10^{-4}$ for models that contain a gauge factor of $U(3) \times U(2)/Sp(2) \times U(1)$ and
$n$ generations of matter. It is also interesting to observe that in models with two generations of chiral matter
the hypercharge is always massless.

Taking the number of generations into account, we find a suppression factor of standard model configurations
of~$2.6\times 10^{-8}$, which  comes quite close to the value of $10^{-9}$ found for models
on the $T^6/(\bZ_2\!\times\!\bZ_2)$ orbifold~\cite{gbhlw05}. This is somewhat surprising, because the
geometry we are considering and the results for the frequency distributions of gauge sector properties
are much more similar to $T^6/\bZ_6$. In that case however the suppression factor has been found to be
$10^{-22}$~\cite{gls07}.

\subsubsection{Chiral exotics}\label{sec:sm_exotics}
What has not yet been taken into account in our analysis is the amount of chiral exotics in these models.
With this term we refer to chiral matter that is charged under both, the visible and the hidden sector gauge group.
When it comes to phenomenology these chiral multiplets are highly undesirable, so it is an interesting question how
many of them are present in the models we found. Ideally one would like to find solutions without any chiral
exotics.

In the following we restrict our attention to models with three generations of standard model particles.
To quantify the amount of chiral exotics, we define a total amount
\be{eq:totex}
  \xi := \sum_{\substack{v\in V\\h\in H}}\left|\chi^{vh}-\chi^{v'h}\right|,
\ee
where $V=\{a,b,c,d\}$ defines the set of branes in the visible sector and $H$ contains all stacks of the hidden sector.

\widefig{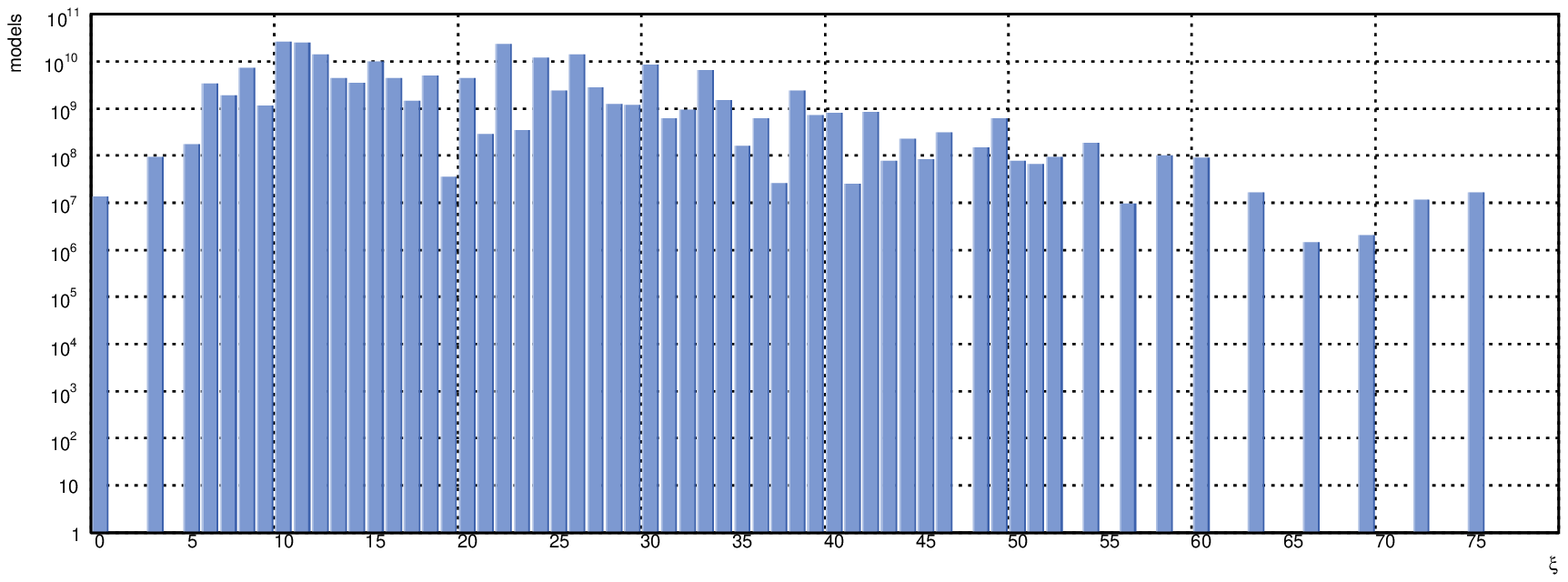}{fig_sm_b}{Frequency distribution of the total amount of chiral exotic matter for a subset of solutions
with geometry $\mathbf{ABa}$, three generations of MSSM matter and a massless hypercharge.
For the definition of $\xi$ see~\protect\eqref{eq:totex}.}

The possible solutions with geometry $\mathbf{ABa}$ for different values of $\xi$ are displayed in Figure~\ref{fig_sm_b}.
We find a broad spectrum of different amounts of chiral exotics, with values for $\xi$ between $10$ and $30$ being the most common.
Compared to the large number of solutions however, the variety is rather restricted, which follows from the fact that
many of the models are quite similar in structure.
This can be understood by noticing that most of them have only slightly different configurations of exceptional
cycles that very often lead to the same chiral matter content.
Note that there is a particularly interesting set of $1.3\times10^7$ models that have no chiral exotics at all.

\subsubsection{Example}\label{sec:sm_example}
To give an explicit example of this class of constructions,
we choose one with complex structure $\varrho=\frac{1}{2}$ and five stacks of branes.
The gauge group is $U(3)\times U(2)\times U(1)_Y\times U(1)^2$ and has total rank $8$.
The three--cycles wrapped by the five stacks of branes are given in Table~\ref{tab:sm_ex_branes}, where $a$ refers to the
$U(3)$ stack, $b$ to the first $U(2)$, $c$ and $d$ to the two $U(1)$s which together with the $U(1)$ of stack
$a$ combine to the massless hypercharge. The remaining stack $e$ is the additional $U(1)$ hidden sector gauge
group.
The full matter spectrum can be computed as outlined in section~\ref{Sec:OpenSpectrum} from the intersection numbers listed
in Tables~\ref{tab:sm_ex_fullspec} and~\ref{tab:sm_ex_chiralspec} for the full and chiral spectrum, respectively.

In terms of representations of $\left(SU(3)\times SU(2)\right)_{U(1)_Y}$
the matter spectrum contains the following multiplets,
\bea{eq:smexpectrum}\nonumber
&& 2 \times \left(\rep{8},\rep{1}\right)_0
  + 10 \times \left(\rep{1},\rep{3}\right)_0
  + 36 \times \left(\rep{1},\rep{1}\right)_0
\\\nonumber
&+& 3\times\left[
    \left(\rep{3},\rep{2}\right)_{1/6}
  + \left(\rep{\bar{3}},\rep{1}\right)_{1/3}
  + \left(\rep{\bar{3}},\rep{1}\right)_{-2/3}
  + 5 \times \left(\rep{1},\rep{2}\right)_{-1/2}
  + 4 \times  \left(\rep{1},\rep{2}\right)_{1/2}
  + \left(\rep{1},\rep{1}\right)_{1}
  +  \left(\rep{1},\rep{1}\right)_{0}
\right]
\\\nonumber
&+& \left[
    \left(\rep{3},\rep{2}\right)_{1/6}
  + 6 \times \left(\rep{3},\rep{1}\right)_{-1/3}
  + 3 \times \left(\rep{3},\rep{1}\right)_{2/3}
  + 4 \times \left(\rep{1},\rep{2}\right)_{-1/2}
  + 8 \times \left(\rep{1},\rep{2}\right)_{0} 
  + 4 \times \left(\rep{1},\rep{1}_2 \right)_{0}
\right.
\\\nonumber
&& \left.
   + 6 \times \left(\rep{1},\rep{3}_2\right)_{0}
  + 4 \times \left(\rep{1},\rep{1}\right)_{0}
  + 6 \times \left(\rep{1},\rep{1}\right)_{1/2} 
  + 4 \times \left(\rep{1},\rep{1}\right)_{1}
  \; + \cc \; \right].
\eea

\renewcommand{\arraystretch}{1.3}
\setlength{\tabcolsep}{0.5\tabcolsep}
\begin{table}[ht]
\begin{center}%
\begin{tabular*}{\linewidth}{@{\extracolsep{\fill}}|c||r|rrrr|rrrr|rrrr|rrrr|rrr|}\hline
\muc{21}{|c|}{\textbf{Standard model example: brane configuration}}\\\hline
\textbf{brane}&$N$&$P,$&$Q,$&$U,$&$V$&$d_1,$&$d_2,$&$d_3,$&$d_4$&$e_1,$&$e_2,$&$e_3,$&$e_4$&$\sigma_1,$&$\sigma_2,$&$\sigma_5,$&$\sigma_6$&$\tau_0,$&$\tau_1,$&$\tau_3$\\\hline\hline
a & 3 & 0,& 0,& 1,&-1 & 0,& 1,& -1,&0 & 0,& 1,& -1,& 0 & $\half$,& 0,& $\half$,& 0 & 0,&1,&1 \\\hline
b & 2 & 3,& 0,& 3,& 0 & -3,& 0,& -3,& 0 &  0,& 0,& 0,& 0 & $\half$,& 0,&   0,&0 & 1,&1,&0 \\\hline
c & 1 & 1,& 1,& 0,& 0 &  0,& 0,& 3,& -3 & 0,&0,& -3,& 3 & $\half$,& 0,&   0,& $\half$ & 0,&1,& 1 \\\hline
d & 1 & 0,& 0,& 3,&-3 &  0,& -1,& 1,& 0 &  0,& -1,& 1,& 0 & 0,&0,&   $\half$,&0 & 0,&1,&1 \\\hline\hline
e & 1 & 0,& 0,& 3,&-3 & 3,& 0,& 0,& 3 & -3,& 0,& 0,& -3 & $\half$,& 0,&   0,&0 & 0,&1,&0 \\\hline
\end{tabular*}%
\caption{Brane configuration for one particular standard model with one brane in the hidden sector. The cycles are given in terms
of the basis of bulk--cycles~\protect\eqref{eqbulkexp} and exceptional cycles~\protect\eqref{eq:delta_expansion}. In addition we list
the displacement and Wilson line coefficients, $\sigma$ and $\tau$, as defined in Section~\protect\ref{fraccycsec}.
For an explicit list of the torus wrapping numbers $\{n_i,m_i\}$, see Appendix~\protect\ref{app:exwrap_sm}.}
\label{tab:sm_ex_branes}
\end{center}
\end{table}
\setlength{\tabcolsep}{2\tabcolsep}
\renewcommand{\arraystretch}{1.3}

\renewcommand{\arraystretch}{1.3}
\begin{table}[ht]
\begin{center}%
\begin{tabular*}{\linewidth}{@{\extracolsep{\fill}}|c||r||r|r||r|r|r|r|r|r|r|r|}\hline
\muc{12}{|c|}{\textbf{Standard model example: complete matter spectrum}}\\\hline
\textbf{brane} &
$\mathbf{\varphi^{\Adj}}$ &
$\mathbf{\varphi^{\Anti}}$ & $\mathbf{\varphi^{\Sym}}$ &
$\mathbf{\varphi^{\cdot b}}$ & $\mathbf{\varphi^{\cdot c}}$ & $\mathbf{\varphi^{\cdot d}}$ & $\mathbf{\varphi^{\cdot e}}$ &
$\mathbf{\varphi^{\cdot b'}}$ & $\mathbf{\varphi^{\cdot c'}}$ & $\mathbf{\varphi^{\cdot d'}}$ & $\mathbf{\varphi^{\cdot e'}}$ \\\hline\hline
$a$ &  2 & 6 &  0 & 0 &  3        & 6 & 0 & 5 &  3         & 6 & 0 \\\hline
$b$ & 10 & 8 & 12  &   & 11        & 8 & 8 &   & 11         & 5 & 8 \\\hline
$c$ & 4 & 10 &  0 & \muc{2}{r|}{} & 5 & 6 & \muc{2}{r|}{}  & 5 & 6 \\\hline
$d$ & 10 & 16 &  6 & \muc{3}{r|}{}     & 0 & \muc{3}{r|}{}      & 0 \\\hline
$e$ & 10 & 16 &  6 & \muc{8}{r|}{}                                 \\\hline
\end{tabular*}
\caption{Number of chiral and non--chiral multiplets for the standard model example. The branes are defined in
Table~\protect\ref{tab:sm_ex_branes}, for the definition of $\varphi$ see Equation~\protect\eqref{VarphiDef}.}
\label{tab:sm_ex_fullspec}
\end{center}%
\end{table}
\renewcommand{\arraystretch}{1.3}

\renewcommand{\arraystretch}{1.3}
\begin{table}[ht]
\begin{center}%
\begin{tabular*}{\linewidth}{@{\extracolsep{\fill}}|c||r|r||r|r|r|r|r|r|r|r|}\hline
\muc{11}{|c|}{\textbf{Standard model example: chiral matter spectrum}}\\\hline
\textbf{brane} &
$\mathbf{\chi^{\Anti}}$ & $\mathbf{\chi^{\Sym}}$ &
$\mathbf{\chi^{\cdot b}}$ & $\mathbf{\chi^{\cdot c}}$ & $\mathbf{\chi^{\cdot d}}$ & $\mathbf{\chi^{\cdot e}}$ &
$\mathbf{\chi^{\cdot b'}}$ & $\mathbf{\chi^{\cdot c'}}$ & $\mathbf{\chi^{\cdot d'}}$ & $\mathbf{\chi^{\cdot e'}}$ \\\hline\hline
$a$ &  0 &  0 &  0 &  -3 &       0 &  0 &  3 &  -3      &  0 &  0 \\\hline
$b$ & 0 &  0 &    & -9 &       6 &  0 &    & -9     &  3 &  0  \\\hline
$c$ &  0 &  0 & \muc{2}{r|}{} & -3 &  0 & \muc{2}{r|}{} &  3 & 0 \\\hline
$d$ &  0 &  0 & \muc{3}{r|}{}      &  0 & \muc{3}{r|}{}      &  0 \\\hline
$e$ &  0 &  0 & \muc{8}{r|}{}                                     \\\hline
\end{tabular*}
\caption{Number of chiral multiplets for the standard model example. The branes are defined in
Table~\protect\ref{tab:sm_ex_branes}, for the definition of $\chi$ see Equation~\protect\eqref{ChiFractional}.}
\label{tab:sm_ex_chiralspec}
\end{center}%
\end{table}
\renewcommand{\arraystretch}{1.3}

\subsection{$SU(5)$ models}
Another interesting class of models are those which do not contain the MSSM directly at the string scale, but the gauge group
and chiral matter content of a grand unified theory.
In the following we present a statistical analysis of solutions with the characteristics of $SU(5)$ and Pati--Salam
models within the full ensemble of solutions.

To obtain $SU(5)$ models we checked for configurations containing at least two stacks of branes, one (referred to as stack $a$)
containing five branes that form a $U(5)$ gauge group, and one stack ($b$) with only one brane to give a $U(1)$ gauge group.
The chiral matter of $SU(5)$ models is contained in the $\mathbf{10}$ and $\mathbf{\bar{5}}$ representations of $SU(5)$.
The number of these representations is given in our construction by the number of antisymmetric matter of the first stack
$\chi^{\Anti_a}$, and the bifundamental matter in $\chi^{a'b}$. These two numbers should be the same and give the number of
generations of chiral matter, which we use as a free parameter.
In addition there might be matter in the symmetric representation of $SU(5)$, which is not very desirable from a phenomenological
point of view, but as it turns out there do not exist any solutions where this matter is absent\footnote{Moreover, there might
exist more exotic matter between the visible sector and the hidden sector group, which is also undesirable phenomenologically.
Since there do not exist models without symmetric representations, we did not perform the (computationally more complex) check for
other exotic matter multiplets.}.


\renewcommand{\arraystretch}{1.3}
\tabfig{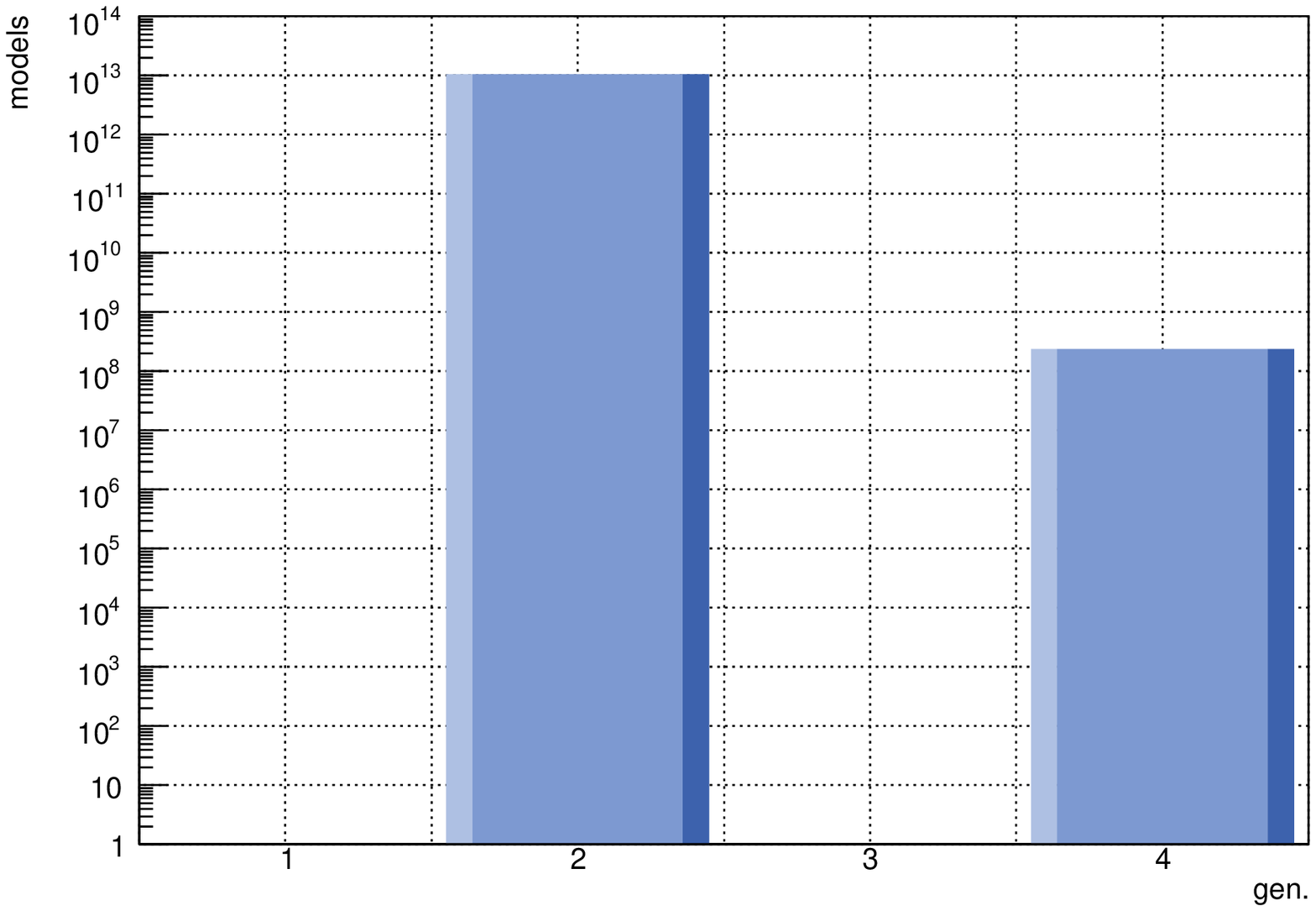}{fig_su5_a}{
Frequency distributions of $SU(5)$ models according to the number of generations ($=\chi^{\Anti_a}=\chi^{a'b}$).
The table shows in addition the number of multiplets in the symmetric representations of $SU(5)$.}{%
\begin{tabular}{|c|c|c|}\hline
\textbf{\# gen.}&$\mathbf{\left|\chi^{\Sym_a}\right|}$&\textbf{\# models}\\\hline\hline
$2$ & $1$ & $7.05\times 10^{12}$ \\\hline
$2$ & $2$ & $3.46\times 10^{12}$ \\\hline
$4$ & $2$ & $2.39\times 10^{8}$  \\\hline
\end{tabular}%
}{0.5}{0.5}
\renewcommand{\arraystretch}{1.3}

The result of our survey is shown in Figure~\ref{fig_su5_a}.
There exist only solutions with 2 or 4 generations of chiral matter and all of them contain at least one symmetric representation
of $SU(5)$.
This result shows much less possibilities for $SU(5)$ models than in the $T^6/(\bZ_2\!\times\!\bZ_2)$ case, studied
in~\cite{cps02,gmst06}.
In the systematic analysis of~\cite{gmst06} more variety in the number of generations has been found
(solutions with 1,2,4 and 8 generations) and, more importantly,
models without symmetric representations are possible.
Although the total number of solutions is quite large, this is mainly an effect due to the exceptional cycles, as explained earlier.
None of the models is very attractive from a phenomenological point of view, in particular because of the presence of symmetric matter.

\subsection{Pati--Salam models}
To obtain Pati--Salam models with gauge group $SU(4)\times SU(2)_L\times SU(2)_R$, we constrained our search to models with at least
three stacks. One with four branes forming a $U(4)$ gauge group (referred to as stack $a$), and two with two branes that form $U(2)$
gauge groups (stacks $b$ and $c$). In principle it is also possible to allow for $Sp(2)$ groups here (which can be obtained from branes
sitting on top of the orientifold planes), but we focus on those with gauge group $U(4)\times U(2)\times U(2)\times H$ in the
following. $H$ denotes the hidden sector gauge group, which also might be absent.

The chiral matter in the visible sector that we would like to obtain is
\ben
  Q_L=(\rep{4},\rep{2},\rep{1}),\qquad Q_R=(\rep{\bar{4}},\rep{1},\rep{2}),
\een
written in terms of representations of the visible sector gauge group. This amounts to the following constraints on the intersection
numbers of the three stacks $a$, $b$ and $c$ introduced above,
\be{eq_psconst}
  \chi^{ab}=\chi^{a'c'}=g,\qquad\chi^{\Anti_a}=\chi^{\Sym_a}=0,
\ee
where $g$ denotes the number of generations.
The second set of constraints excludes unwanted symmetric and antisymmetric representations of $SU(4)$.
An additional constraint can be imposed on the amount of matter in the bifundamental representation of $SU(2)_L$ and $SU(2)_R$.
These multiplets might however also be candidates for Higgses, which is why we leave their number as an open parameter,
concretely defined as
\be{eq:ps_bifundamental}
  \chi^{(\rep{2},\rep{2})} := \left|\chi^{bc}\right|+\left|\chi^{bc'}\right|.
\ee

\renewcommand{\arraystretch}{1.3}
\tabfig{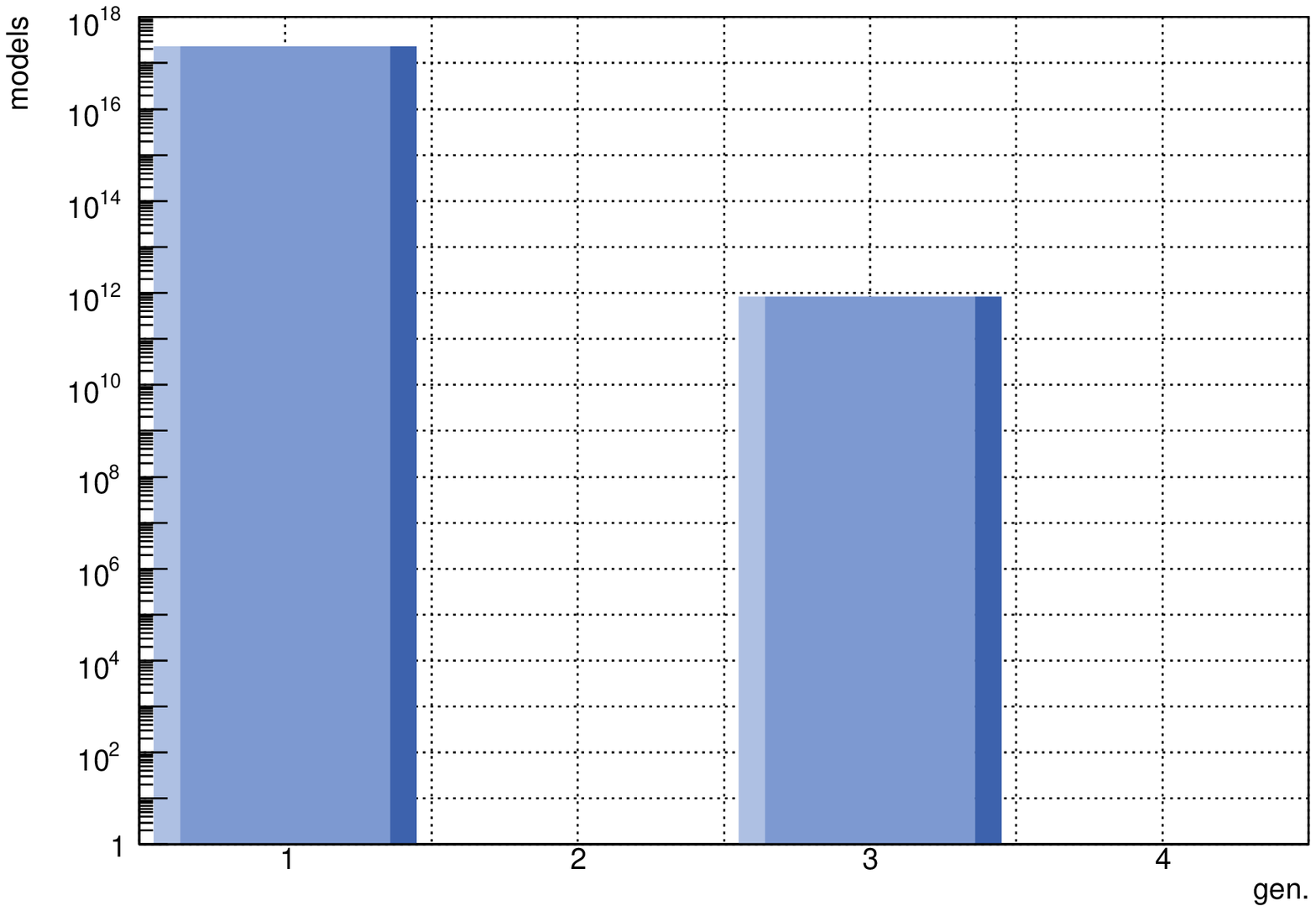}{fig_ps_a}{%
Frequency distribution of Pati--Salam models according to the number of generations in the visible sector ($=\chi^{ab}=\chi^{a'c'}$).
In the table we list additionally the amount of matter in the bifundamental representation $(\mathbf{2},\mathbf{2})$ given by
$\chi^{(\rep{2},\rep{2})}$, as defined in~\protect\eqref{eq:ps_bifundamental}.}{%
\begin{tabular}{|c|c|c|}\hline
\textbf{\# gen.}&$\mathbf{\chi^{(\rep{2},\rep{2})}}$&\textbf{\# models}\\\hline\hline
$1$ & $0$ & $3.23\times 10^{9}$  \\\hline
$1$ & $1$ & $1.38\times 10^{17}$ \\\hline
$1$ & $2$ & $9.05\times 10^{16}$ \\\hline
$3$ & $0$ & $8.19\times 10^{11}$ \\\hline
$3$ & $3$ & $3.55\times 10^{9}$  \\\hline
\end{tabular}%
}{0.5}{0.5}

The results, shown in Figure~\ref{fig_ps_a}, are interesting for at least two reasons. Firstly,
the number of possible generations is one or three, exactly those numbers which did not appear in the $SU(5)$ case.
Secondly there exist models that contain three generations of chiral matter, with or without additional chiral Higgs
candidates\footnote{There might or might not exist additional non--chiral multiplets, which can also be used as possible
Higgs candidates.}.
These are given by the intersection number between the two $SU(2)$ branes.

\subsubsection{Chiral exotics}\label{sec:ps_exotic}
As in the case of the standard model constructions, we would like to know how much chiral exotics arise in these constructions.
Unfortunately it turns out that this number is generically very large.
For a more detailed analysis we chose a subset of Pati--Salam models, consisting of
all constructions with three generations of chiral matter in the visible
sector and a maximum number of two stacks in the hidden sector.

\renewcommand{\arraystretch}{1.3}
\widefig{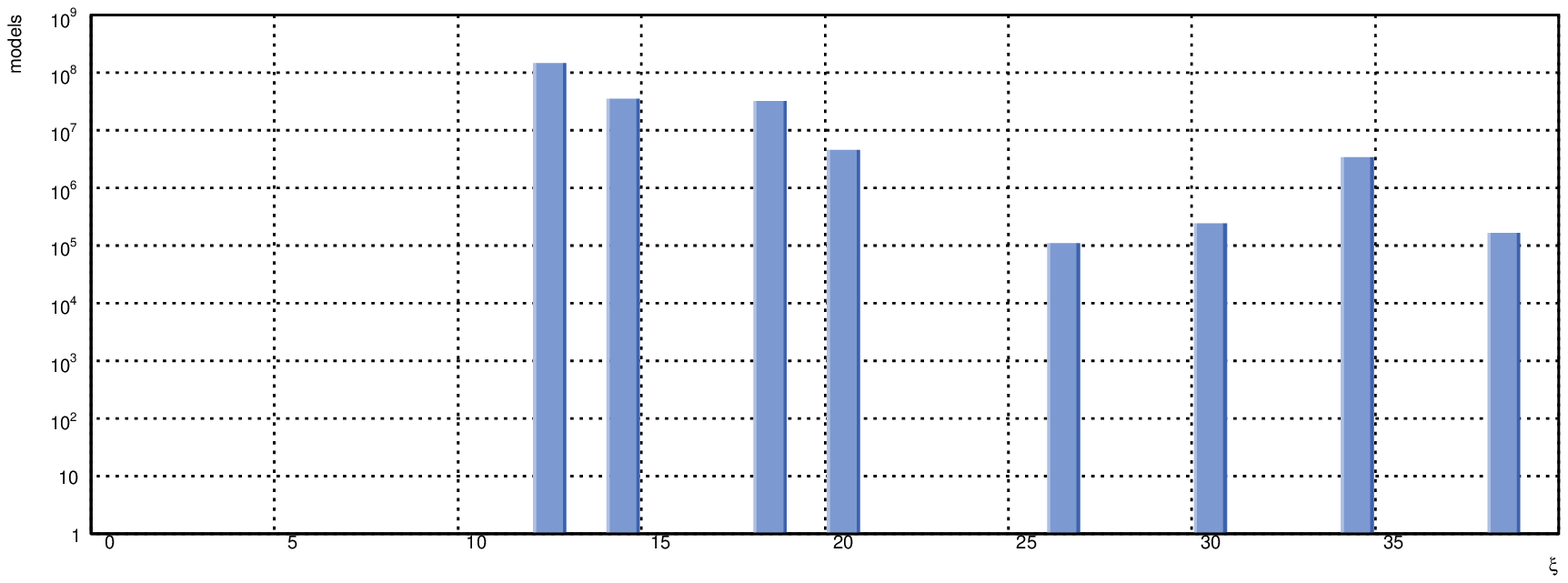}{fig_ps_b}{Frequency distribution of Pati--Salam models with three generations of matter in the
visible sector and not more then two stacks in the hidden sector,
divided according to the total amount of chiral exotic matter $\xi$, as defined by~\protect\eqref{eq:totex}.}

To quantify the amount of chiral exotics, we use their absolute number $\xi$, as defined in~\eqref{eq:totex}.
The results are shown in Figure~\ref{fig_ps_b}.
They reveal that in all possible constructions not only a non--vanishing number of chiral exotics is present, but there are always
a lot of them. The small number of different values for $\xi$ shows the similar structure of constructions of this type. The
number of different values is much smaller than in the standard model case, due to the effect that $U(4)$ factors are suppressed
compared to $U(3)$ factors. Moreover, we are only considering models with a maximum number of two branes in the hidden sector, a
condition that severely reduces the number of possible configurations.

\subsubsection{Example}\label{sec:ps_example}
Let us look at one example of a special subset of these solutions, consisting only of models with three generations of chiral
matter without chiral Higgs candidates. The geometry of the model is $\mathbf{ABa}$, the complex structure modulus is given
by $\rho=\frac{3}{2}$ and we have a total number of five stacks, which are given in detail in Table~\ref{tab_psexample}.
The gauge group is $U(4)\times U(2)_L \times U(2)_R \times U(1)^2$ with total rank $10$.
The intersection numbers between stacks $a$, $b$ and $c$ are fixed by the constraints~\eqref{eq_psconst}, but as already mentioned
above, there is quite an amount of chiral exotic matter in bifundamental representations of the first three stacks and stacks $d$ and $e$.
The relevant intersections to compute the
full chiral and non--chiral spectrum are listed in Tables~\ref{tab_psspectot} and~\ref{tab_psspec}.
In terms of representations of $U(4)\times U(2)_L\times U(2)_R$ the spectrum contains
the following matter multiplets,
\bea{eq_ps_fullspec}\nonumber
  &&           2\times \left(\rep{16},          \rep{1},      \rep{1}      \right)
             + 4\times \left(\rep{1},           \rep{4},      \rep{1}      \right)
             + 4\times \left(\rep{1},           \rep{1},      \rep{4}      \right)
             + 6\times \left(\rep{1},           \rep{1},      \rep{1}      \right)
\\\nonumber
  &+&  3\times\left[   \left(\rep{4},           \rep{2},      \rep{1}      \right)
                     + \left(\rep{\bar{4}},     \rep{1},      \rep{2}      \right)
                     + \left(\rep{4},           \rep{1},      \rep{1}      \right)_{1,0}
                     + \left(\rep{\bar{4}},     \rep{1},      \rep{1}      \right)_{0,1}  \right]
\\\nonumber
  &+&  6\times\left[   \left(\rep{1},           \rep{1},      \rep{2}      \right)_{0,1}
                     + \left(\rep{1},           \rep{1},      \rep{\bar{2}}\right)_{1,0}
                     + \left(\rep{1},           \rep{1}_{2}, \rep{1}      \right)
                     + \left(\rep{1},           \rep{1},      \rep{1}_{2} \right)
                     + \rep{1}_{1,\mbox{-}1}
                     + \rep{1}_{\mbox{-}2,0} \right]
\\\nonumber
  &+&        \left[ 3 \times  \left(\rep{4},           \rep{1},      \rep{1}      \right)_{\mbox{-}1,0}
                     +  \left(\rep{4},   \rep{\bar{2}},   \rep{1}      \right)
                     +  \left(\rep{4},   \rep{2},   \rep{1}      \right)
                     + 2 \times \left(\rep{4},           \rep{1},      \rep{1}      \right)_{1,0}
                     + 5 \times \left(\rep{1},           \rep{2},      \rep{\bar{2}}\right)
                     + 4 \times \left(\rep{1},           \rep{2},      \rep{2}\right) \right.
\\\nonumber
  &&        \left.
                     + 3 \times \left(\rep{1},           \rep{2},      \rep{1}      \right)_{0,\mbox{-}1}
                     + 7 \times \left(\rep{1},           \rep{2},      \rep{1}      \right)_{1,0}
                     + 2 \times \left(\rep{1},           \rep{2},      \rep{1}      \right)_{0,1}
                     + 3 \times \left(\rep{1},           \rep{1},      \rep{2}      \right)_{0,\mbox{-}1}
                     + 5 \times \left(\rep{1},           \rep{1},      \rep{2}      \right)_{1,0} 
 \right.
\\\nonumber
  &&        \left.
                     + 2 \times \left(\rep{1},           \rep{1},      \rep{2}      \right)_{\mbox{-}1,0} 
                     + 3 \times \rep{1}_{1,1}
                     + 3 \times \left(\rep{6},           \rep{1},      \rep{1}      \right)
                     + 2 \times \left(\rep{1},           \rep{3},      \rep{1}      \right)
                     + 2 \times \left(\rep{1},           \rep{1},      \rep{3}      \right)
\; + \cc \;\right],
\eea
where the subscripts denote the charge under the $U(1)$ factors of branes $d$ and $e$.
If no charges are given, they are zero and if the non-abelian part is written just as $\rep{1}$,
the corresponding multiplet is understood to be a singlet under the full visible sector gauge group.

\renewcommand{\arraystretch}{1.3}
\setlength{\tabcolsep}{0.6\tabcolsep}
\begin{table}[ht]
\begin{center}%
\begin{tabular*}{\linewidth}{@{\extracolsep{\fill}}|c||r|rrrr|rrrr|rrrr|rrrr|rrr|}\hline
\muc{21}{|c|}{\textbf{Pati--Salam example: brane configuration}}\\\hline
\textbf{brane}&$N$&$P,$&$Q,$&$U,$&$V$&$d_1,$&$d_2,$&$d_3,$&$d_4$&$e_1,$&$e_2,$&$e_3,$&$e_4$&$\sigma_1,$&$\sigma_2,$&$\sigma_5,$&$\sigma_6$&$\tau_0,$&$\tau_1,$&$\tau_3$\\\hline\hline
a & 4 & 0,& 0,&1,&-1 &  0,&1,& 1,&0 &  0,&1,& 1,&0 &$\half$,& 0,& $\half$,&0 & 0,&1,&0 \\\hline
b & 2 & 2,&-1,&2,&-1 & -2,&0,&-2,&0 &  1,&0,& 1,&0 &$\half$,& 0,&      0,&0 & 1,&1,&0 \\\hline
c & 2 & 2,&-1,&2,&-1 &  2,&0,& 2,&0 & -1,&0,&-1,&0 &$\half$,&  0,&     0,&0 & 0,&1,&0 \\\hline\hline
d & 1 & 3,& 0,&1,& 0 &  1,&0,& 1,&0 & -2,&0,&-2,&0 &$\half$,& 0,&      0,&0 & 1,&1,&0 \\\hline
e & 1 & 2,&-1,&2,&-1 & -2,&0,&-2,&0 &  1,&0,& 1,&0 & 0,&      0,&      0,&0 & 0,&0,&0 \\\hline
\end{tabular*}%
\caption{Brane configuration for one Pati--Salam model with two branes in the hidden sector. The cycles are given in terms
of the basis of bulk--cycles~\protect\eqref{eqbulkexp} and exceptional cycles~\protect\eqref{eq:delta_expansion}. In addition we list
the displacement and Wilson line coefficients, $\sigma$ and $\tau$, as defined in Section~\protect\ref{fraccycsec}.
For an explicit list of the torus wrapping numbers $\{n_i,m_i\}$, see Appendix~\protect\ref{app:exwrap_ps}.}
\label{tab_psexample}
\end{center}
\end{table}
\setlength{\tabcolsep}{2\tabcolsep}
\renewcommand{\arraystretch}{1.3}

\renewcommand{\arraystretch}{1.3}
\begin{table}[ht]
\begin{center}%
\begin{tabular*}{\linewidth}{@{\extracolsep{\fill}}|c||r||r|r||r|r|r|r|r|r|r|r|}\hline
\muc{12}{|c|}{\textbf{Pati--Salam example: complete matter spectrum}}\\\hline
\textbf{brane} &
$\mathbf{\varphi^{\Adj}}$ &
$\mathbf{\varphi^{\Anti}}$ & $\mathbf{\varphi^{\Sym}}$ &
$\mathbf{\varphi^{\cdot b}}$ & $\mathbf{\varphi^{\cdot c}}$ & $\mathbf{\varphi^{\cdot d}}$ & $\mathbf{\varphi^{\cdot e}}$ &
$\mathbf{\varphi^{\cdot b'}}$ & $\mathbf{\varphi^{\cdot c'}}$ & $\mathbf{\varphi^{\cdot d'}}$ & $\mathbf{\varphi^{\cdot e'}}$ \\\hline\hline
$a$ &  2  & 6 &  0 & 3 &  2        & 6 & 3 & 0 & 5         & 7 & 0 \\\hline
$b$ &  4 & 6 &  4 &   &  10       &10 & 6 &   & 8         & 10& 6 \\\hline
$c$ &  4 & 6 &  4 & \muc{2}{r|}{} & 0 & 6 & \muc{2}{r|}{} & 14& 4 \\\hline
$d$ &  2 & 8 &  6 & \muc{3}{r|}{}     & 6 & \muc{3}{r|}{}     & 6 \\\hline
$e$ &  4 &10 &  0& \muc{8}{r|}{}                                 \\\hline
\end{tabular*}
\caption{Number of chiral and non--chiral multiplets for the Pati--Salam example. The branes are defined in
Table~\protect\ref{tab_psexample}, for the definition of $\varphi$ see Equation~\protect\eqref{VarphiDef}.}
\label{tab_psspectot}
\end{center}%
\end{table}
\renewcommand{\arraystretch}{1.3}

\renewcommand{\arraystretch}{1.3}
\begin{table}[ht]
\begin{center}%
\begin{tabular*}{\linewidth}{@{\extracolsep{\fill}}|c||r|r||r|r|r|r|r|r|r|r|}\hline
\muc{11}{|c|}{\textbf{Pati--Salam example: chiral matter spectrum}}\\\hline
\textbf{brane} &
$\mathbf{\chi^{\Anti}}$ & $\mathbf{\chi^{\Sym}}$ &
$\mathbf{\chi^{\cdot b}}$ & $\mathbf{\chi^{\cdot c}}$ & $\mathbf{\chi^{\cdot d}}$ & $\mathbf{\chi^{\cdot e}}$ &
$\mathbf{\chi^{\cdot b'}}$ & $\mathbf{\chi^{\cdot c'}}$ & $\mathbf{\chi^{\cdot d'}}$ & $\mathbf{\chi^{\cdot e'}}$ \\\hline\hline
$a$ &  0 &  0 & -3 & 0 &         0 & -3 &  0 &  3       &  3 &  0 \\\hline
$b$ &  6 &  0 &    & 0 &        -6 &  0 &    &  0       &  0 &  6 \\\hline
$c$ &  6 &  0 & \muc{2}{r|}{} &  0 &  0 & \muc{2}{r|}{} &  0 &  0 \\\hline
$d$ &  0 & -6 & \muc{3}{r|}{}      &  6 & \muc{3}{r|}{}      &  0 \\\hline
$e$ &  6 &  0 & \muc{8}{r|}{}                                     \\\hline
\end{tabular*}
\caption{Number of chiral multiplets for the Pati--Salam example. The branes are defined in
Table~\protect\ref{tab_psexample}, for the definition of $\chi$ see Equation~\protect\eqref{ChiFractional}.}
\label{tab_psspec}
\end{center}%
\end{table}
\renewcommand{\arraystretch}{1.3}

%
%
\section{Conclusions}\label{conclusions}

In this article we have performed a complete analysis of all possible $N=1$ supersymmetric solutions for
the $T^6/\bZ'_6$ orbifold background in the context of intersecting D6--brane models on type IIA orientifolds.
To analyse all $\cO(10^{23})$ solutions, we reformulated the constraining equations as well as the formulae to calculate the amount
of chiral and non--chiral matter in terms of algebraic equations with rational coefficients. The algebraic formalism to compute the
non--chiral spectrum is new and can be used for all toroidal orbifold backgrounds.

The full ensemble of solutions, as well as special subsets of interesting configurations, have been analysed from a statistical point
of view. The main results are the following. Concerning the full set of solutions we found that the frequency distributions of the
total rank of the gauge group behaves at a qualitative level exactly as it has been found in an earlier study of
$T^6/\bZ_6$~\cite{gls07}, showing an enhancement of models with bigger total rank,
compared to constructions without exceptional cycles, such as~$T^6/(\bZ_2\!\times\!\bZ_2)$.
Concerning the probability distribution of individual factors of the gauge group with rank $N$,
we confirm a result from~\cite{dota06}, showing that the distribution scales as $N^{-2}$ if one ignores contributions
from exceptional cycles. Taking them into account amounts to an additional exponential factor in~\eqref{eq:frac_un_approx}.

Things are different from $T^6/\bZ_6$ if one starts to look for specific configurations.
We have done an analysis of possible configurations that contain the gauge group of the standard model,
as well as $SU(5)$ and Pati--Salam GUT models.
In general, the number of solutions for such configurations is larger than in the $\bZ_6$ case, although the total number of
solutions is smaller by five orders of magnitude.
In particular, there does exist a considerable amount of interesting standard model and Pati--Salam configurations
with three generations of chiral matter. We provided an example for both cases.
We analysed these subsets in some detail, focussing on the existence of a massless hypercharge and
the presence of chiral exotics. As in other orbifold constructions, requiring a massless hypercharge is not a
very strong restriction, but chiral exotics are omnipresent in Pati--Salam models and in the vast majority
of Standard Model constructions.
In our analysis, which has been restricted to a subset of all possible constructions, we find
a class of $\cO(10^7)$ models without chiral exotics. This is certainly a very interesting set of
solutions that deserves further study. We hope to come back to this in future work.

\acknowledgments
We would like to thank Pascal Anastasopoulos, Ralph Blumenhagen, Keith Dienes, Bert Schellekens and Washington Taylor
for interesting discussions and David Bailin for useful correspondence.
It is a pleasure to thank the Benasque Center for Science for hospitality.
The work of F.~G. is supported by the Foundation for Fundamental Research of Matter (FOM)
and the National Organisation for Scientific Research (NWO).

\appendix

\section{The non--chiral models in cycle language}

As a cross--check of the formulae given in section~\ref{setup}, the non--chiral spectra with D6--branes on top of the O6--planes
passing through the origin computed in~\cite{bgk00,bcs04} are reproduced 
by two types of fractional branes for each choice of lattice, 
\be{Cycles_nonchiral}
\begin{aligned}
{\bf AAa/b:} & \left\{\begin{array}{l}
\Pi_1 =\frac{1}{2}\frac{\rho_1-b \rho_3}{1-b} +\frac{1}{2}\left(\delta_1 + (1-2b) \delta_2 + (2b) \delta_4 \right)
\\
\Pi_2 =\frac{1}{2} (\rho_3 - 2 \rho_4) +\frac{1}{2}\left(-\delta_1 -\delta_4 \right)
\end{array}\right.  ,
\\
{\bf ABa/b:} & \left\{\begin{array}{l}
\Pi_1 =\frac{1}{2}\frac{\rho_1+\rho_2 -b (\rho_3+\rho_4)}{1-b} 
+\frac{1}{2}\left(\delta_1 +\tilde{\delta}_1 + (1-2b) [\delta_2 +\tilde{\delta}_2] + (2b)[\delta_4 +\tilde{\delta}_4] \right)
\\
\Pi_2 =\frac{1}{2} 3(\rho_3 - \rho_4)+\frac{1}{2}\left(-\delta_1 -\tilde{\delta}_1 -\delta_4 -\tilde{\delta}_4 \right)
\end{array}\right.  ,
\\
{\bf BAa/b:} &  \left\{\begin{array}{l}
\Pi_1 =\frac{1}{2}\frac{\rho_1+\rho_2 -b (\rho_3+\rho_4)}{1-b}+
\frac{1}{2}\left(\tilde{\delta}_1 - \delta_1 + (1-2b)[\tilde{\delta}_2 - \delta_2 ] + (2b) [\tilde{\delta}_4 - \delta_4] \right)
\\
\Pi_2 =\frac{1}{2}(\rho_3 - \rho_4)+\frac{1}{2}\left(\delta_1 - \tilde{\delta}_1 +\delta_4 -\tilde{\delta}_4 \right)
\end{array}\right.  ,
\\
{\bf BBa/b:} & \left\{\begin{array}{l}
\Pi_1 =\frac{1}{2}\frac{3(\rho_2-b \rho_4)}{1-b}+
\frac{1}{2}\left(\tilde{\delta}_1 - 2\delta_1 + (1-2b) [\tilde{\delta}_2 - 2\delta_2] + (2b) [\tilde{\delta}_4 - 2\delta_4]\right)
\\
\Pi_2 =\frac{1}{2}( 2\rho_3 -\rho_4)+\frac{1}{2}\left(2\delta_1 -\tilde{\delta}_1 + 2 \delta_4 -\tilde{\delta}_4 \right)
\end{array}\right.  .
\end{aligned}
\ee
The choices {\bf A} and {\bf B}  for the lattice $T^2_3$ in~\cite{bgk00,bcs04}
correspond to setting the complex structure parameter $\varrho_{\bf A} = 3$, $\varrho_{\bf B}=1$ in the tilted torus {\bf b} with $b=1/2$. 
Each model has gauge group $U(2) \times U(2)$ and a completely non--chiral spectrum as required to fit with the previously computed 
spectra.

\section{Search for $SO(2N)$ or $Sp(2N)$ gauge groups}

If some fractional cycle is $\Omega{\cal R}$ invariant, besides from $U(N)$ gauge groups,  $SO(2N)$ or $Sp(2N)$ gauge factors can occur.
In Tables~\ref{rinvex1}, \ref{rinvex2}, \ref{rinvex3} and~\ref{rinvex4}, we examine the candidate cycles which have bulk parts parallel to the O6--planes.

As listed in the tables, fractional cycles which sit on top of the O6--planes, i.e.
\ben
  (\sigma_1,\sigma_2;\sigma_5,\sigma_6) =(0,0;0,0),
\een 
for the $T^2_3$ {\bf a}-type lattice also
\ben
  (\sigma_1,\sigma_2;\sigma_5,\sigma_6) =(0,0;\underline{\frac{1}{2},0}),
\een
are never  $\Omega{\cal R}$ invariant. All other
$\Omega{\cal R}$ invariant bulk cycle positions can serve as probe brane candidates for the K--theory constraint discussed in 
section~\ref{Ktheory} for appropriate choices of Wilson lines.

For the K--theory constraints, only $Sp(2)$--branes are relevant,
but since we do not find a fast method to distinguish $SO$ and $Sp$ groups,
we take all $\Omega{\cal R}$ invariant cycles as probe branes.
Even with this larger set, it can be shown that the constraint is always
trivially fulfilled (cf. Section~\ref{Ktheory}).

\section{The lattice of three--cycles}\label{App:Basis}

The twelve dimensional lattice of fractional cycles consisting of bulk and $\bZ_2$ exceptional cycles only is spanned by 
\begin{equation}
\begin{array}{ll}
\alpha_1 = \frac{1}{2} \left( \rho_1 + \delta_1 + \delta_2\right) ,
& \alpha_7 = \frac{1}{2} \left( \delta_1 + \delta_2 + \delta_3 - \delta_4 \right)  ,
\\
\alpha_2 = \frac{1}{2} \left(\rho_4 - \tilde{\delta}_1 - \tilde{\delta}_4 \right) ,
& \alpha_8=- \frac{1}{2} \left(\tilde{\delta}_1- \tilde{\delta}_2 +\tilde{\delta}_3 -\tilde{\delta}_4\right) ,
\\
\alpha_3 =\frac{1}{2} \left( \rho_1 - \delta_1 - \delta_2\right) ,
& \alpha_9=\frac{1}{2} \left( \delta_1 + 3 \,\delta_2 + \delta_3 - \delta_4 \right) ,
\\
\alpha_4 =\frac{1}{2} \left(\rho_4 + \tilde{\delta}_1 + \tilde{\delta}_4 \right) ,
& \alpha_{10}=\frac{1}{2} \left(\tilde{\delta}_1- \tilde{\delta}_2 +\tilde{\delta}_3 +\tilde{\delta}_4\right) ,
\\
\alpha_5 =  \frac{1}{2} \left(-2\, \rho_1 + \rho_2 +\tilde{\delta}_3 +\tilde{\delta}_4 \right) ,
& \alpha_{11}= \frac{1}{2} \left(2\, \rho_1 - \rho_2 -3\, \tilde{\delta}_3 -3\,\tilde{\delta}_4 \right) ,
\\
\alpha_6 = \frac{1}{2} \left( -\rho_3 + 2 \, \rho_4 - \delta_2 - \delta_3\right) ,
& \alpha_{12}= \frac{1}{2} \left( - \rho_3 + 2 \, \rho_4 - 3\, \delta_2 - 3\, \delta_3\right) .
\end{array}
\end{equation}
Defining 
\begin{equation}
\varepsilon = \left(\begin{array}{cc} 0 & 1 \\ -1 & 0
\end{array}\right),
\end{equation}
the intersection matrix is
\begin{equation}
I^{bulk + \bZ_2} = \text{Diag} ( \varepsilon, \varepsilon,  \varepsilon, \varepsilon,  \varepsilon, 3 \,\varepsilon).
\end{equation}
An unimodular basis of three--cycles is obtained if also the exceptional cycles at $\bZ_3$ fixed points are taken into account 
as follows,
\begin{equation}
\begin{aligned}
& \alpha_{13} = \frac{1}{3} \, \rho_1 - \frac{1}{6} \, \rho_2 -\frac{1}{2} \left(\tilde{\delta}_3 + \tilde{\delta}_4 \right)
+\frac{1}{3} \left((\gamma^{(1)}_1-\gamma^{(2)}_1 ) - (\gamma^{(1)}_2 -\gamma^{2}_2) \right),
\\
& \alpha_{14} = \frac{1}{3}  \left((\gamma^{(1)}_1-\gamma^{(2)}_1 ) + (\gamma^{(1)}_2 -\gamma^{2}_2) 
+  (\gamma^{(1)}_3 -\gamma^{2}_3)\right),
\\
& \alpha_{15} = \frac{1}{6} \, \rho_3 - \frac{1}{3} \, \rho_4 +\frac{1}{2} \left(\delta_2 +\delta_3 \right)
-\frac{1}{3} \left( (\tilde{\gamma}^{(1)}_1 - \tilde{\gamma}^{(2)}_1) - (\tilde{\gamma}^{(1)}_2 - \tilde{\gamma}^{(2)}_2)\right),
\\
& \alpha_{16} = \frac{1}{3} \left( (\tilde{\gamma}^{(1)}_1 - \tilde{\gamma}^{(2)}_1) 
+ (\tilde{\gamma}^{(1)}_2 - \tilde{\gamma}^{(2)}_2)+ (\tilde{\gamma}^{(1)}_3 - \tilde{\gamma}^{(2)}_3)\right),
\\
& \begin{array}{llll}
\alpha_{17} = \gamma_1^{(1)} - \gamma_1^{(2)}, \quad\quad
& \alpha_{18} = \gamma^{(1)}_1, \quad\quad 
& \alpha_{19} = \gamma^{(1)}_2, \quad\quad
& \alpha_{20} = \gamma^{(1)}_3, 
\\ \alpha_{21} =  \tilde{\gamma}^{(1)}_1 - \tilde{\gamma}^{(2)}_1, \quad\quad
& \alpha_{22} =  \tilde{\gamma}^{(2)}_1, \quad\quad
& \alpha_{23} =  \tilde{\gamma}^{(2)}_2, \quad\quad
& \alpha_{24} =  \tilde{\gamma}^{(2)}_3.
\end{array}
\end{aligned}
\end{equation}

\section{Bulk wrapping numbers}

In the following we list the torus wrapping numbers of the branes in the two examples
we gave in Sections~\ref{sec:sm_example} and~\ref{sec:ps_example}.

\subsection{Standard model example}
\label{app:exwrap_sm}
An explicit realisation of the bulk branes in the standard model example is given by
\bea{eq:sm_ex_tw}\nonumber
  &a: & (n_1,m_1;n_2,m_2;n_3,m_3)=(1,-1; 1,0; 0,1) \to (P,Q,U,V)=(0,0,1,-1),\\\nonumber
  &b: & (n_1,m_1;n_2,m_2;n_3,m_3)=(1,1; 2,-1; 1,1) \to (P,Q,U,V)=(3,0,3,0),\\\nonumber
  &c: & (n_1,m_1;n_2,m_2;n_3,m_3)=(1,-1; -1,2; 1,0) \to (P,Q,U,V)=(1,1,0,0),\\\nonumber
  &d: & (n_1,m_1;n_2,m_2;n_3,m_3)=(1,-1; 1,-2; 0,1) \to (P,Q,U,V)=(0,0,3,-3),\\\nonumber
  &e: & (n_1,m_1;n_2,m_2;n_3,m_3)=(1,-1; 1,-2; 0,1) \to (P,Q,U,V)=(0,0,3,-3).
\eea

\subsection{Pati--Salam example}
\label{app:exwrap_ps}
An explicit realisation of the bulk branes in the Pati--Salam example is given by
\bea{eq:ps_ex_tw}\nonumber
  &a: & (n_1,m_1;n_2,m_2;n_3,m_3)=(1,-1; 1,0; 0,1) \to (P,Q,U,V)=(0,0,1,-1),\\\nonumber
  &b: & (n_1,m_1;n_2,m_2;n_3,m_3)=(1,1; 1,-1; 1,1) \to (P,Q,U,V)=(2,-1,2,-1),\\\nonumber
  &c: & (n_1,m_1;n_2,m_2;n_3,m_3)=(1,1; 1,-1; 1,1) \to (P,Q,U,V)=(2,-1,2,-1),\\\nonumber
  &d: & (n_1,m_1;n_2,m_2;n_3,m_3)=(1,-1; 0,1; 3,1) \to (P,Q,U,V)=(3,0,1,0),\\\nonumber
  &e: & (n_1,m_1;n_2,m_2;n_3,m_3)=(1,-1; 1,1; 1,1) \to (P,Q,U,V)=(2,-1,2,-1).
\eea

\newpage

\renewcommand{\arraystretch}{1.3}
\begin{sidewaystable}[H]
    \begin{equation*}
      \begin{array}{|c||c||c||c|c||c|} \hline
        \multicolumn{6}{|c|}{\rule[-3mm]{0mm}{8mm}
\text{\bf Fractional cycles parallel to O6--planes for $T^6/\bZ_6'$}}\\ \hline\hline
\begin{array}{c} \text{lattice} \\ \# \end{array}
& \Pi^{bulk} &(\sigma_1,\sigma_2;\sigma_5,\sigma_6)& \Pi^{\bZ_2} & \Omega{\cal R} (\Pi^{\bZ_2})
&\begin{array}{c} \Omega{\cal R} \\ \text{inv.}\end{array} \\\hline\hline
{\bf AAa/b} & \frac{\rho_1 -b\rho_3}{1-b} &(0,0;0,0) & \tilde{\tau}_1 \delta_1 + \tilde{\tau}_2 \delta_{\frac{2}{1-b}}
& -\tilde{\tau}_1 \delta_1 - \tilde{\tau}_2 \delta_{\frac{2}{1-b}}
& \text{no}
\\\hline
(1{\rm a}) & & (0,\frac{1}{2};0,0) & 
\begin{array}{c} -\tilde{\tau}_2 \delta_1 + (\tilde{\tau}_2 - \tilde{\tau}_1) \tilde{\delta}_1\\
 -\tilde{\tau}_4 \delta_{\frac{2}{1-b}} + (\tilde{\tau}_4 -\tilde{\tau}_3) \tilde{\delta}_{\frac{2}{1-b}}\end{array} 
 & 
\begin{array}{c}\tilde{\tau}_1 \delta_1  + (\tilde{\tau}_2 - \tilde{\tau}_1) \tilde{\delta}_1\\
+ \tilde{\tau}_3  \delta_{\frac{2}{1-b}} + (\tilde{\tau}_4 -\tilde{\tau}_3) \tilde{\delta}_{\frac{2}{1-b}}
\end{array}
& 
\begin{array}{c} \tilde{\tau}_2 = -\tilde{\tau}_1 \\ \tilde{\tau}_4 = -\tilde{\tau}_3
\end{array}
\\\hline
(1{\rm b}^{\bf b}) & & (0,0;b,\frac{1-2b}{2}) & \tilde{\tau}_1 \delta_3 + \tilde{\tau}_2 \delta_{4(1-b)}
& -\tilde{\tau}_1 \delta_{3-2b} - \tilde{\tau}_2 \delta_{4-2b}
& 
\begin{array}{c} {\bf a:} \text{no} \\ {\bf b:} \tilde{\tau}_2 =-\tilde{\tau}_1
\end{array}
\\\hline
(1{\rm c}) & & (0,\frac{1}{2};b,\frac{1-2b}{2}) &   
\begin{array}{c} -\tilde{\tau}_2 \delta_3 + (\tilde{\tau}_2 - \tilde{\tau}_1) \tilde{\delta}_3 \\
-\tilde{\tau}_4 \delta_{4(1-b)} + (\tilde{\tau}_4 -\tilde{\tau}_3) \tilde{\delta}_{4(1-b)}\end{array}
 & 
\begin{array}{c} \tilde{\tau}_1 \delta_{3-2b}+ (\tilde{\tau}_2 - \tilde{\tau}_1) \tilde{\delta}_{3-2b} \\
+\tilde{\tau}_3 \delta_{4-2b}   + (\tilde{\tau}_4 -\tilde{\tau}_3) \tilde{\delta}_{4-2b}
\end{array}
& 
\begin{array}{c} \tilde{\tau}_1 = -\tilde{\tau}_{\frac{2}{1-b}} \\ \tilde{\tau}_3 = -\tilde{\tau}_{4(1-b)}
\end{array}
\\\hline\hline
{\bf AAa/b} & \rho_3 - 2\rho_4 & (0,0;0,0) & -\tilde{\tau}_1 \delta_1 - \tilde{\tau}_2 \delta_4 
& \tilde{\tau}_1 \delta_1 + \tilde{\tau}_2 \delta_4 
& \text{no}
\\\hline
(2{\rm a}) & & (0,\frac{1}{2};0,0) &  
\begin{array}{c} \tilde{\tau}_1  \delta_1+(\tilde{\tau}_2 -\tilde{\tau}_1) \tilde{\delta}_1 \\
+\tilde{\tau}_3 \delta_4 +(\tilde{\tau}_4-\tilde{\tau}_3) \tilde{\delta}_4\end{array} 
 & 
\begin{array}{c}-\tilde{\tau}_2 \delta_{1}+(\tilde{\tau}_2 -\tilde{\tau}_1) \tilde{\delta}_{1} 
\\ -\tilde{\tau}_4 \delta_{4} +(\tilde{\tau}_4-\tilde{\tau}_3) \tilde{\delta}_{4}
\end{array}
& \begin{array}{c} \tilde{\tau}_1 = -\tilde{\tau}_2 \\ \tilde{\tau}_3 = -\tilde{\tau}_4
\end{array}
\\\hline
(2{\rm b}^{\bf b}) & & (0,0;\frac{1}{2},0) & -\tilde{\tau}_1 \delta_2 - \tilde{\tau}_2 \delta_3
&\tilde{\tau}_1 \delta_{2+2b} + \tilde{\tau}_2 \delta_{3-2b}
& 
\begin{array}{c} {\bf a:} \text{no} \\ {\bf b:} \tilde{\tau}_2 =-\tilde{\tau}_1
\end{array}
\\ \hline
(2{\rm c}) & & (0,\frac{1}{2};\frac{1}{2},0) &   
\begin{array}{c} \tilde{\tau}_1 \delta_2+(\tilde{\tau}_2- \tilde{\tau}_1) \tilde{\delta}_2 \\
+ \tilde{\tau}_3 \delta_3 +( \tilde{\tau}_4-\tilde{\tau}_3) \tilde{\delta}_3\end{array}
 & 
\begin{array}{c} -\tilde{\tau}_2 \delta_{2+2b} +(\tilde{\tau}_2-\tilde{\tau}_1) \tilde{\delta}_{2+2b} 
\\ -\tilde{\tau}_4 \delta_{3-2b} +(\tilde{\tau}_4-\tilde{\tau}_3) \tilde{\delta}_{3-2b}
\end{array}
&  \begin{array}{c} \tilde{\tau}_1 = -\tilde{\tau}_{\frac{2}{1-b}}
\\ \tilde{\tau}_3 = -\tilde{\tau}_{4(1-b)}
\end{array}
\\\hline
     \end{array}
    \end{equation*}
\caption{Search for $\Omega{\cal R}$ invariant cycles $\frac{1}{2} \Pi^{bulk} + \frac{1}{2} \Pi^{\bZ_2}$. 
The prefactors $\tilde{\tau}_i=\pm 1$ corresponding to the choice of a $\bZ_2$ eigenvalue and two Wilson lines 
are subject to the constraint $\prod_{i=1}^4\tilde{\tau}_i = 1$. 
The numbering in the first column corresponds to the probe brane candidates discussed in section~\protect\ref{Ktheory}.
In the last column, the combinations of $\tilde{\tau}_i$ for which $\Omega{\cal R}$ invariant cycles occur are listed. Part 1. }
\label{rinvex1}
\end{sidewaystable}

\begin{sidewaystable}[H]
    \begin{equation*}
      \begin{array}{|c||c||c||c|c||c|} \hline
        \multicolumn{6}{|c|}{\rule[-3mm]{0mm}{8mm}
\text{\bf Fractional cycles parallel to O6--planes for $T^6/\bZ_6'$}}\\ \hline\hline
\begin{array}{c} \text{lattice} \\ \# \end{array}
& \Pi^{bulk} &(\sigma_1,\sigma_2;\sigma_5,\sigma_6)& \Pi^{\bZ_2} & \Omega{\cal R} (\Pi^{\bZ_2})
&\begin{array}{c} \Omega{\cal R} \\ \text{inv.}\end{array} \\\hline\hline
{\bf ABa/b} &\begin{array}{c}\frac{\rho_1+\rho_2}{1-b}\\ -\frac{b(\rho_3+\rho_4)}{1-b} \end{array}& (0,0;0,0) 
&  \tilde{\tau}_1 (\delta_1+\tilde{\delta}_1) + \tilde{\tau}_2 (\delta_{\frac{2}{1-b}} + \tilde{\delta}_{\frac{2}{1-b}})
& -\tilde{\tau}_1 (\delta_1+\tilde{\delta}_1) - \tilde{\tau}_2 (\delta_{\frac{2}{1-b}} + \tilde{\delta}_{\frac{2}{1-b}})
& \text{no}
\\\hline
(1{\rm a}) & &  (0,\frac{1}{2};0,0) &  
\begin{array}{c} (\tilde{\tau}_1-2\tilde{\tau}_2) \delta_1 + (\tilde{\tau}_2 - 2\tilde{\tau}_1) \tilde{\delta}_1\\
+(\tilde{\tau}_3 - 2\tilde{\tau}_4) \delta_{\frac{2}{1-b}} + (\tilde{\tau}_4 -2\tilde{\tau}_3) \tilde{\delta}_{\frac{2}{1-b}}\end{array}
& 
\begin{array}{c} (2\tilde{\tau}_1-\tilde{\tau}_2) \delta_1 + (2\tilde{\tau}_2 - \tilde{\tau}_1) \tilde{\delta}_1\\
+(2\tilde{\tau}_3 - \tilde{\tau}_4) \delta_{\frac{2}{1-b}} + (2\tilde{\tau}_4 -\tilde{\tau}_3) \tilde{\delta}_{\frac{2}{1-b}}\end{array}
&  \begin{array}{c}\tilde{\tau}_1 = -\tilde{\tau}_2
\\\tilde{\tau}_3 = -\tilde{\tau}_4
\end{array}
\\\hline
(1{\rm b}^{\bf b}) & & (0,0;b,\frac{1-2b}{2}) &\tilde{\tau}_1 (\delta_3+\tilde{\delta}_3) + \tilde{\tau}_2 (\delta_{4(1-b)} + \tilde{\delta}_{4(1-b)})
& 
\begin{array}{c}  -\tilde{\tau}_1 (\delta_{3-2b}+\tilde{\delta}_{3-2b}) \\- \tilde{\tau}_2 (\delta_{4-2b} + \tilde{\delta}_{4-2b})\end{array}
&  \begin{array}{c} {\bf a:} \text{no}
\\ {\bf b:} \tilde{\tau}_2 = -\tilde{\tau}_1
\end{array}
\\\hline
(1{\rm c}) & & (0,\frac{1}{2};b,\frac{1-2b}{2}) 
&
\begin{array}{c} (\tilde{\tau}_1-2\tilde{\tau}_2) \delta_3 + (\tilde{\tau}_2 - 2\tilde{\tau}_1) \tilde{\delta}_3\\
+(\tilde{\tau}_3 - 2\tilde{\tau}_4) \delta_{4(1-b)} + (\tilde{\tau}_4 -2\tilde{\tau}_3) \tilde{\delta}_{4(1-b)}\end{array}
&
\begin{array}{c} (2\tilde{\tau}_1-\tilde{\tau}_2) \delta_{3-2b} + (2\tilde{\tau}_2 - \tilde{\tau}_1) \tilde{\delta}_{3-2b}\\
+(2\tilde{\tau}_3 - \tilde{\tau}_4) \delta_{4-2b} + (2\tilde{\tau}_4 -\tilde{\tau}_3) \tilde{\delta}_{4-2b}\end{array}
&  \begin{array}{c} \tilde{\tau}_1 = -\tilde{\tau}_{\frac{2}{1-b}}
\\ \tilde{\tau}_3 = -\tilde{\tau}_{4(1-b)}
\end{array}
\\\hline\hline
{\bf ABa/b} & 3( \rho_3 - \rho_4) &   (0,0;0,0) & -\tilde{\tau}_1 (\delta_1+\tilde{\delta}_1) - \tilde{\tau}_2 (\delta_4 + \tilde{\delta}_4) 
& \tilde{\tau}_1 (\delta_1+\tilde{\delta}_1) + \tilde{\tau}_2 (\delta_4 + \tilde{\delta}_4) 
& \text{no}
\\\hline
(2{\rm a}) & & (0,\frac{1}{2};0,0) &  
\begin{array}{c}(2\tilde{\tau}_2 - \tilde{\tau}_1) \delta_1 + (2\tilde{\tau}_1 - \tilde{\tau}_2) \tilde{\delta}_1 
\\+(2\tilde{\tau}_4 - \tilde{\tau}_3) \delta_4 + (2\tilde{\tau}_3 - \tilde{\tau}_4) \tilde{\delta}_4 
\end{array}
&\begin{array}{c}(\tilde{\tau}_2 - 2\tilde{\tau}_1) \delta_1 + (\tilde{\tau}_1 - 2\tilde{\tau}_2) \tilde{\delta}_1 
\\+(\tilde{\tau}_4 - 2\tilde{\tau}_3) \delta_4 + (\tilde{\tau}_3 - 2\tilde{\tau}_4) \tilde{\delta}_4 
\end{array}
&  \begin{array}{c}\tilde{\tau}_1 = -\tilde{\tau}_2
\\\tilde{\tau}_3 = -\tilde{\tau}_4
\end{array}
\\\hline
(2{\rm b}^{\bf b}) & & (0,0;\frac{1}{2},0) &  -\tilde{\tau}_1 (\delta_2+\tilde{\delta}_2) - \tilde{\tau}_2 (\delta_3 + \tilde{\delta}_3) 
& 
\begin{array}{c} \tilde{\tau}_1 (\delta_{2+2b}+\tilde{\delta}_{2+2b})\\ + \tilde{\tau}_2 (\delta_{3-2b} + \tilde{\delta}_{3-2b}) \end{array}
&  \begin{array}{c} {\bf a:} \text{no}
\\ {\bf b:} \tilde{\tau}_2 = -\tilde{\tau}_1
\end{array}
\\\hline
(2{\rm c}) & & (0,\frac{1}{2};\frac{1}{2},0) &  
\begin{array}{c}(2\tilde{\tau}_2 - \tilde{\tau}_1) \delta_2 + (2\tilde{\tau}_1 - \tilde{\tau}_2) \tilde{\delta}_2
\\+(2\tilde{\tau}_4 - \tilde{\tau}_3) \delta_3 + (2\tilde{\tau}_3 - \tilde{\tau}_4) \tilde{\delta}_3 
\end{array}
&\begin{array}{c}(\tilde{\tau}_2 - 2\tilde{\tau}_1) \delta_{2+2b} + (\tilde{\tau}_1 - 2\tilde{\tau}_2) \tilde{\delta}_{2+2b}
\\+(\tilde{\tau}_4 - 2\tilde{\tau}_3) \delta_{3-2b} + (\tilde{\tau}_3 - 2\tilde{\tau}_4) \tilde{\delta}_{3-2b} 
\end{array}
&  \begin{array}{c} \tilde{\tau}_1 = -\tilde{\tau}_{\frac{2}{1-b}}
\\ \tilde{\tau}_3 = -\tilde{\tau}_{4(1-b)}
\end{array}
\\\hline
     \end{array}
    \end{equation*}
\caption{Search for $\Omega{\cal R}$ invariant cycles. Part 2.}
\label{rinvex2}
\end{sidewaystable}

\begin{sidewaystable}[H]
    \begin{equation*}
      \begin{array}{|c||c||c||c|c||c|} \hline
        \multicolumn{6}{|c|}{\rule[-3mm]{0mm}{8mm}
\text{\bf Fractional cycles parallel to O6--planes for $T^6/\bZ_6'$}}\\ \hline\hline
 \begin{array}{c}
\text{lattice} \\ \#
\end{array}
& \Pi^{bulk} &(\sigma_1,\sigma_2;\sigma_5,\sigma_6)& \Pi^{\bZ_2} & \Omega{\cal R} (\Pi^{\bZ_2})
& \begin{array}{c} \Omega{\cal R} \\
\text{inv.}
\end{array}
\\\hline\hline
{\bf BAa/b} &\begin{array}{c}\frac{\rho_1+\rho_2}{1-b}\\ -\frac{b(\rho_3+\rho_4)}{1-b} \end{array}&(0,0;0,0) & 
\tilde{\tau}_1 (\tilde{\delta}_1 - \delta_1) + \tilde{\tau}_2 (\tilde{\delta}_{\frac{2}{1-b}} - \delta_{\frac{2}{1-b}})
& \tilde{\tau}_1( \delta_1 - \tilde{\delta}_1) +  \tilde{\tau}_2 (\delta_{\frac{2}{1-b}} - \tilde{\delta}_{\frac{2}{1-b}})
& \text{no}
\\\hline
(1{\rm a}) & & (0,\frac{1}{2};0,0) &\tilde{\tau}_1 \delta_1- \tilde{\tau}_2 \tilde{\delta}_1  + \tilde{\tau}_3 \delta_{\frac{2}{1-b}} - \tilde{\tau}_4 \tilde{\delta}_{\frac{2}{1-b}}
& - \tilde{\tau}_2  \delta_1 + \tilde{\tau}_1\tilde{\delta}_1- \tilde{\tau}_4 \delta_{\frac{2}{1-b}}  + \tilde{\tau}_3 \tilde{\delta}_{\frac{2}{1-b}}
& \begin{array}{c}\tilde{\tau}_1 = -\tilde{\tau}_2
\\\tilde{\tau}_3 = -\tilde{\tau}_4
\end{array}
\\\hline
(1{\rm b}^{\bf b}) & & (0,0;b,\frac{1-2b}{2}) & \tilde{\tau}_1 (\tilde{\delta}_3 - \delta_3) + \tilde{\tau}_2 (\tilde{\delta}_{4(1-b)} - \delta_{4(1-b)})
& \tilde{\tau}_1( \delta_{3-2b} - \tilde{\delta}_{3-2b}) +  \tilde{\tau}_2 (\delta_{4-2b} - \tilde{\delta}_{4-2b})
&\begin{array}{c} {\bf a:} \text{no} \\ {\bf b:}  \tilde{\tau}_2 = -\tilde{\tau}_1
\end{array}
\\\hline
(1{\rm c}) & & (0,\frac{1}{2};b,\frac{1-2b}{2}) &\tilde{\tau}_1 \delta_3- \tilde{\tau}_2 \tilde{\delta}_3  + \tilde{\tau}_3 \delta_{4(1-b)} - \tilde{\tau}_4 \tilde{\delta}_{4(1-b)}
&- \tilde{\tau}_2  \delta_{3-2b} + \tilde{\tau}_1\tilde{\delta}_{3-2b}- \tilde{\tau}_4 \delta_{4-2b}  + \tilde{\tau}_3 \tilde{\delta}_{4-2b}
& \begin{array}{c} \tilde{\tau}_1 = -\tilde{\tau}_{\frac{2}{1-b}}
\\ \tilde{\tau}_3 = -\tilde{\tau}_{4(1-b)}
\end{array}
\\\hline\hline
{\bf BAa/b} & \rho_3 - \rho_4 & (0,0;0,0) & \tilde{\tau}_1 (\delta_1 - \tilde{\delta}_1) +  \tilde{\tau}_2 (\delta_4 - \tilde{\delta}_4)
 & -\tilde{\tau}_1 (\delta_1 - \tilde{\delta}_1) -  \tilde{\tau}_2 (\delta_4 - \tilde{\delta}_4)
& \text{no}
\\\hline
(2{\rm a}) & & (\frac{1}{2},0;0,0) & -\tilde{\tau}_1\delta_1 + \tilde{\tau}_2 \tilde{\delta}_1 - \tilde{\tau}_3  \delta_4 +\tilde{\tau}_4  \tilde{\delta}_4
& \tilde{\tau}_2\delta_1 -\tilde{\tau}_1\tilde{\delta}_1 +\tilde{\tau}_4 \delta_4 - \tilde{\tau}_3  \tilde{\delta}_4
& \begin{array}{c}\tilde{\tau}_1 = -\tilde{\tau}_2
\\\tilde{\tau}_3 = -\tilde{\tau}_4
\end{array}
\\\hline
(2{\rm b}^{\bf b})  & & (0,0;\frac{1}{2},0) &\tilde{\tau}_1 (\delta_2 - \tilde{\delta}_2) +  \tilde{\tau}_2 (\delta_3 - \tilde{\delta}_3)
& -\tilde{\tau}_1 (\delta_{2+2b} - \tilde{\delta}_{2+2b}) -  \tilde{\tau}_2 (\delta_{3-2b} - \tilde{\delta}_{3-2b})
& \begin{array}{c} {\bf a:} \text{no} \\ {\bf b:} \tilde{\tau}_1 = -\tilde{\tau}_2
\end{array}
\\\hline
(2{\rm c}) & & (\frac{1}{2},0;\frac{1}{2},0) &-\tilde{\tau}_1\delta_2 + \tilde{\tau}_2 \tilde{\delta}_2 - \tilde{\tau}_3  \delta_3 +\tilde{\tau}_4  \tilde{\delta}_3
&\tilde{\tau}_2\delta_{2+2b} -\tilde{\tau}_1\tilde{\delta}_{2+2b} +\tilde{\tau}_4 \delta_{3-2b} - \tilde{\tau}_3  \tilde{\delta}_{3-2b}
& \begin{array}{c} \tilde{\tau}_1 = -\tilde{\tau}_{\frac{2}{1-b}}
\\ \tilde{\tau}_3 = -\tilde{\tau}_{4(1-b)}
\end{array}
\\\hline
     \end{array}
    \end{equation*}
\caption{Search for $\Omega{\cal R}$ invariant cycles. Part 3.}
\label{rinvex3}
\end{sidewaystable}

\begin{sidewaystable}[H]
    \begin{equation*}
      \begin{array}{|c||c||c||c|c||c|} \hline
        \multicolumn{6}{|c|}{\rule[-3mm]{0mm}{8mm}
\text{\bf Fractional cycles parallel to O6--planes for $T^6/\bZ_6'$}}\\ \hline\hline
 \begin{array}{c}
\text{lattice} \\ \#
\end{array}
& \Pi^{bulk} &(\sigma_1,\sigma_2;\sigma_5,\sigma_6)& \Pi^{\bZ_2} & \Omega{\cal R} (\Pi^{\bZ_2})
& \begin{array}{c} \Omega{\cal R} \\
\text{inv.}
\end{array}
\\\hline\hline
{\bf BBa/b} & \frac{3(\rho_2 - b \rho_4)}{1-b} &(0,0;0,0) & 
\tilde{\tau}_1 (\tilde{\delta}_1 -2 \delta_1) + \tilde{\tau}_2 (\tilde{\delta}_{\frac{2}{1-b}} -2 \delta_{\frac{2}{1-b}}) 
& \tilde{\tau}_1( 2\delta_1 - \tilde{\delta}_1) +  \tilde{\tau}_2 (2\delta_{\frac{2}{1-b}} - \tilde{\delta}_{\frac{2}{1-b}})
& \text{no}
\\\hline
(1{\rm a}) & & (0,\frac{1}{2};0,0) & \begin{array}{c}  (\tilde{\tau}_1 + \tilde{\tau}_2 ) \delta_1 + (\tilde{\tau}_1 - 2 \tilde{\tau}_2) \tilde{\delta}_1
\\ + (\tilde{\tau}_3 + \tilde{\tau}_4) \delta_{\frac{2}{1-b}} + (\tilde{\tau}_3 - 2 \tilde{\tau}_4) \tilde{\delta}_{\frac{2}{1-b}}
\end{array} 
& \begin{array}{c}-(\tilde{\tau}_1 + \tilde{\tau}_2 ) \delta_1+ (2\tilde{\tau}_1 - \tilde{\tau}_2) \tilde{\delta}_1
\\- (\tilde{\tau}_3 + \tilde{\tau}_4) \delta_{\frac{2}{1-b}} + (2\tilde{\tau}_3 -  \tilde{\tau}_4) \tilde{\delta}_{\frac{2}{1-b}}
\end{array} 
&\begin{array}{c}\tilde{\tau}_1 = -\tilde{\tau}_2
\\\tilde{\tau}_3 = -\tilde{\tau}_4
\end{array}
\\\hline
(1{\rm b}^{\bf b}) & & (0,0;b,\frac{1-2b}{2}) &\tilde{\tau}_1 (\tilde{\delta}_3 -2 \delta_3) + \tilde{\tau}_2 (\tilde{\delta}_{4(1-b)} -2 \delta_{4(1-b)}) 
& \tilde{\tau}_1( 2\delta_{3-2b} - \tilde{\delta}_{3-2b}) +  \tilde{\tau}_2 (2\delta_{4-2b} - \tilde{\delta}_{4-2b})
&\begin{array}{c} {\bf a:} \text{no} \\ {\bf b:} \tilde{\tau}_1 = -\tilde{\tau}_2
\end{array}
\\\hline
(1{\rm c}) & & (0,\frac{1}{2};b,\frac{1-2b}{2}) & \begin{array}{c}  (\tilde{\tau}_1 + \tilde{\tau}_2 ) \delta_3 + (\tilde{\tau}_1 - 2 \tilde{\tau}_2) \tilde{\delta}_3
\\ + (\tilde{\tau}_3 + \tilde{\tau}_4) \delta_{4(1-b)} + (\tilde{\tau}_3 - 2 \tilde{\tau}_4) \tilde{\delta}_{4(1-b)}
\end{array} 
& \begin{array}{c}-(\tilde{\tau}_1 + \tilde{\tau}_2 ) \delta_{3-2b}+ (2\tilde{\tau}_1 - \tilde{\tau}_2) \tilde{\delta}_{3-2b}
\\- (\tilde{\tau}_3 + \tilde{\tau}_4) \delta_{4-2b} + (2\tilde{\tau}_3 -  \tilde{\tau}_4) \tilde{\delta}_{4-2b}
\end{array} 
& \begin{array}{c} \tilde{\tau}_1 = -\tilde{\tau}_{\frac{2}{1-b}}
\\ \tilde{\tau}_3 = -\tilde{\tau}_{4(1-b)}
\end{array}
\\\hline\hline
{\bf BBa/b} & 2\rho_3 - \rho_4 & (0,0;0,0) & \tilde{\tau}_1 (2\delta_1 - \tilde{\delta}_1) +  \tilde{\tau}_2 (2\delta_4 - \tilde{\delta}_4)
& -\tilde{\tau}_1 (2\delta_1 - \tilde{\delta}_1) -  \tilde{\tau}_2 (2\delta_4 - \tilde{\delta}_4)
& \text{no}
\\\hline
(2{\rm a}) & & (\frac{1}{2},0;0,0) & \begin{array}{c} -(\tilde{\tau}_1 + \tilde{\tau}_2)\delta_1 + (2\tilde{\tau}_1 - \tilde{\tau}_2) \tilde{\delta}_1
\\-(\tilde{\tau}_3 + \tilde{\tau}_4)\delta_4 + (2\tilde{\tau}_3 - \tilde{\tau}_4) \tilde{\delta}_4
\end{array} 
& \begin{array}{c} (\tilde{\tau}_1 + \tilde{\tau}_2)\delta_1 + (\tilde{\tau}_1 - 2\tilde{\tau}_2) \tilde{\delta}_1
\\+(\tilde{\tau}_3 + \tilde{\tau}_4)\delta_4 + (\tilde{\tau}_3 - 2\tilde{\tau}_4) \tilde{\delta}_4
\end{array} 
&\begin{array}{c}\tilde{\tau}_1 = -\tilde{\tau}_2
\\\tilde{\tau}_3 = -\tilde{\tau}_4
\end{array}
\\\hline
(2{\rm b}^{\bf b}) & & (0,0;\frac{1}{2},0) &\tilde{\tau}_1 (2\delta_2 - \tilde{\delta}_2) +  \tilde{\tau}_2 (2\delta_3 - \tilde{\delta}_3)
&-\tilde{\tau}_1 (2\delta_{2+2b} - \tilde{\delta}_{2+2b}) -  \tilde{\tau}_2 (2\delta_{3-2b} - \tilde{\delta}_{3-2b})
&\begin{array}{c} {\bf a:} \text{no} \\ {\bf b:} \tilde{\tau}_1 = -\tilde{\tau}_2
\end{array}
\\\hline
(2{\rm c}) & & (\frac{1}{2},0;\frac{1}{2},0) &\begin{array}{c} -(\tilde{\tau}_1 + \tilde{\tau}_2)\delta_2 + (2\tilde{\tau}_1 - \tilde{\tau}_2) \tilde{\delta}_2
\\-(\tilde{\tau}_3 + \tilde{\tau}_4)\delta_3 + (2\tilde{\tau}_3 - \tilde{\tau}_4) \tilde{\delta}_3
\end{array} 
&\begin{array}{c} (\tilde{\tau}_1 + \tilde{\tau}_2)\delta_{2+2b} + (\tilde{\tau}_1 - 2\tilde{\tau}_2) \tilde{\delta}_{2+2b}
\\+(\tilde{\tau}_3 + \tilde{\tau}_4)\delta_{3-2b} + (\tilde{\tau}_3 - 2\tilde{\tau}_4) \tilde{\delta}_{3-2b}
\end{array} 
& \begin{array}{c} \tilde{\tau}_1 = -\tilde{\tau}_{\frac{2}{1-b}}
\\ \tilde{\tau}_3 = -\tilde{\tau}_{4(1-b)}
\end{array}
\\\hline
     \end{array}
    \end{equation*}
\caption{Search for $\Omega{\cal R}$ invariant cycles. Part 4.}
\label{rinvex4}
\end{sidewaystable}
\renewcommand{\arraystretch}{1.3}
\clearpage

%
%
\addcontentsline{toc}{section}{References}
\bibliographystyle{JHEP}
\bibliography{refs_z6prime}

\providecommand{\href}[2]{#2}\begingroup\raggedright\begin{thebibliography}{10}

\bibitem{su03}
L.~Susskind, {\it The anthropic landscape of string theory},
  \href{http://xxx.lanl.gov/abs/hep-th/0302219}{{\tt hep-th/0302219}}.

\bibitem{sc06}
A.~N. Schellekens, {\it The landscape ``avant la lettre''},
  \href{http://xxx.lanl.gov/abs/physics/0604134}{{\tt physics/0604134}}.

\bibitem{lu07}
D.~L{\"u}st, {\it String {L}andscape and the {S}tandard {M}odel of {P}article
  {P}hysics},  \href{http://xxx.lanl.gov/abs/0707.2305}{{\tt 0707.2305}}.

\bibitem{do03}
M.~R. Douglas, {\it The statistics of string / {M} theory vacua},  {\em JHEP}
  {\bf 05} (2003) 046, [\href{http://xxx.lanl.gov/abs/hep-th/0303194}{{\tt
  hep-th/0303194}}].

\bibitem{dedo04}
F.~Denef and M.~R. Douglas, {\it Distributions of flux vacua},  {\em JHEP} {\bf
  05} (2004) 072, [\href{http://xxx.lanl.gov/abs/hep-th/0404116}{{\tt
  hep-th/0404116}}].

\bibitem{bghlw04}
R.~Blumenhagen, F.~Gmeiner, G.~Honecker, D.~L{\"u}st, and T.~Weigand, {\it The
  statistics of supersymmetric {D}-brane models},  {\em Nucl. Phys.} {\bf B713}
  (2005) 83--135, [\href{http://xxx.lanl.gov/abs/hep-th/0411173}{{\tt
  hep-th/0411173}}].

\bibitem{dedo05}
F.~Denef and M.~R. Douglas, {\it Distributions of nonsupersymmetric flux
  vacua},  {\em JHEP} {\bf 03} (2005) 061,
  [\href{http://xxx.lanl.gov/abs/hep-th/0411183}{{\tt hep-th/0411183}}].

\bibitem{adv05}
B.~S. Acharya, F.~Denef, and R.~Valandro, {\it Statistics of {M} theory vacua},
   {\em JHEP} {\bf 06} (2005) 056,
  [\href{http://xxx.lanl.gov/abs/hep-th/0502060}{{\tt hep-th/0502060}}].

\bibitem{ku06}
J.~Kumar, {\it A review of distributions on the string landscape},  {\em Int.
  J. Mod. Phys.} {\bf A21} (2006) 3441--3472,
  [\href{http://xxx.lanl.gov/abs/hep-th/0601053}{{\tt hep-th/0601053}}].

\bibitem{kuwe05}
J.~Kumar and J.~D. Wells, {\it Surveying standard model flux vacua on
  ${T}^6/\mathbb{Z}_2\times\mathbb{Z}_2$},  {\em JHEP} {\bf 09} (2005) 067,
  [\href{http://xxx.lanl.gov/abs/hep-th/0506252}{{\tt hep-th/0506252}}].

\bibitem{gbhlw05}
F.~Gmeiner, R.~Blumenhagen, G.~Honecker, D.~L{\"u}st, and T.~Weigand, {\it One
  in a billion: {MSSM}-like {D}-brane statistics},  {\em JHEP} {\bf 01} (2006)
  004, [\href{http://xxx.lanl.gov/abs/hep-th/0510170}{{\tt hep-th/0510170}}].

\bibitem{gm05}
F.~Gmeiner, {\it Standard model statistics of a type {II} orientifold},  {\em
  Fortsch. Phys.} {\bf 54} (2006) 391--398,
  [\href{http://xxx.lanl.gov/abs/hep-th/0512190}{{\tt hep-th/0512190}}].

\bibitem{gmst06}
F.~Gmeiner and M.~Stein, {\it Statistics of {SU}(5) {D}-brane models on a type
  {II} orientifold},  {\em Phys. Rev.} {\bf D73} (2006) 126008,
  [\href{http://xxx.lanl.gov/abs/hep-th/0603019}{{\tt hep-th/0603019}}].

\bibitem{dota06}
M.~R. Douglas and W.~Taylor, {\it The landscape of intersecting brane models},
  {\em JHEP} {\bf 01} (2007) 031,
  [\href{http://xxx.lanl.gov/abs/hep-th/0606109}{{\tt hep-th/0606109}}].

\bibitem{gls07}
F.~Gmeiner, D.~L{\"u}st, and M.~Stein, {\it Statistics of orientifold models on
  ${T}^6/\mathbb{Z}_6$},  {\em JHEP} {\bf 05} (2007) 018,
  [\href{http://xxx.lanl.gov/abs/hep-th/0703011}{{\tt hep-th/0703011}}].

\bibitem{ftz07}
S.~F{\"o}rste, C.~Timirgaziu, and I.~Zavala, {\it Orientifold's {L}andscape:
  {N}on-{F}actorisable {S}ix-{T}ori},
  \href{http://xxx.lanl.gov/abs/0707.0747}{{\tt 0707.0747}}.

\bibitem{dhs04a}
T.~P.~T. Dijkstra, L.~R. Huiszoon, and A.~N. Schellekens, {\it Chiral
  supersymmetric standard model spectra from orientifolds of {G}epner models},
  {\em Phys. Lett.} {\bf B609} (2005) 408--417,
  [\href{http://xxx.lanl.gov/abs/hep-th/0403196}{{\tt hep-th/0403196}}].

\bibitem{dhs04b}
T.~P.~T. Dijkstra, L.~R. Huiszoon, and A.~N. Schellekens, {\it Supersymmetric
  standard model spectra from {RCFT} orientifolds},  {\em Nucl. Phys.} {\bf
  B710} (2005) 3--57, [\href{http://xxx.lanl.gov/abs/hep-th/0411129}{{\tt
  hep-th/0411129}}].

\bibitem{adks06}
P.~Anastasopoulos, T.~P.~T. Dijkstra, E.~Kiritsis, and A.~N. Schellekens, {\it
  Orientifolds, hypercharge embeddings and the standard model},  {\em Nucl.
  Phys.} {\bf B759} (2006) 83--146,
  [\href{http://xxx.lanl.gov/abs/hep-th/0605226}{{\tt hep-th/0605226}}].

\bibitem{di06}
K.~R. Dienes, {\it Statistics on the heterotic landscape: {G}auge groups and
  cosmological constants of four-dimensional heterotic strings},  {\em Phys.
  Rev.} {\bf D73} (2006) 106010,
  [\href{http://xxx.lanl.gov/abs/hep-th/0602286}{{\tt hep-th/0602286}}].

\bibitem{le06}
O.~Lebedev {\em et.~al.}, {\it A mini-landscape of exact {MSSM} spectra in
  heterotic orbifolds},  {\em Phys. Lett.} {\bf B645} (2007) 88--94,
  [\href{http://xxx.lanl.gov/abs/hep-th/0611095}{{\tt hep-th/0611095}}].

\bibitem{dlsw07}
K.~R. Dienes, M.~Lennek, D.~Senechal, and V.~Wasnik, {\it Supersymmetry versus
  {G}auge {S}ymmetry on the {H}eterotic {L}andscape},  {\em Phys. Rev.} {\bf
  D75} (2007) 126005, [\href{http://xxx.lanl.gov/abs/0704.1320}{{\tt
  0704.1320}}].

\bibitem{bkls06}
R.~Blumenhagen, B.~K{\"o}rs, D.~L{\"u}st, and S.~Stieberger, {\it
  Four-dimensional {S}tring {C}ompactifications with {D}-{B}ranes,
  {O}rientifolds and {F}luxes},  {\em Phys. Rept.} {\bf 445} (2007) 1--193,
  [\href{http://xxx.lanl.gov/abs/hep-th/0610327}{{\tt hep-th/0610327}}].

\bibitem{gm06}
F.~Gmeiner, {\it Gauge sector statistics of intersecting {D}-brane models},
  {\em Fortsch. Phys.} {\bf 55} (2007) 111--160,
  [\href{http://xxx.lanl.gov/abs/hep-th/0608227}{{\tt hep-th/0608227}}].

\bibitem{acdo06}
B.~S. Acharya and M.~R. Douglas, {\it A finite landscape?},
  \href{http://xxx.lanl.gov/abs/hep-th/0606212}{{\tt hep-th/0606212}}.

\bibitem{dile06}
K.~R. Dienes and M.~Lennek, {\it Fighting the floating correlations:
  {E}xpectations and complications in extracting statistical correlations from
  the string theory landscape},  {\em Phys. Rev.} {\bf D75} (2007) 026008,
  [\href{http://xxx.lanl.gov/abs/hep-th/0610319}{{\tt hep-th/0610319}}].

\bibitem{balo06}
D.~Bailin and A.~Love, {\it Towards the supersymmetric standard model from
  intersecting {D}6-branes on the {Z}'(6) orientifold},  {\em Nucl. Phys.} {\bf
  B755} (2006) 79--111, [\href{http://xxx.lanl.gov/abs/hep-th/0603172}{{\tt
  hep-th/0603172}}].

\bibitem{balo07}
D.~Bailin and A.~Love, {\it Almost the supersymmetric standard model from
  intersecting {D}6-branes on the {Z}'(6) orientifold},  {\em Phys. Lett.} {\bf
  B651} (2007) 324--328, [\href{http://xxx.lanl.gov/abs/0705.0646}{{\tt
  0705.0646}}].

\bibitem{hoot04}
G.~Honecker and T.~Ott, {\it Getting just the supersymmetric standard model at
  intersecting branes on the $\mathbb{Z}_6$-orientifold},  {\em Phys. Rev.}
  {\bf D70} (2004) 126010, [\href{http://xxx.lanl.gov/abs/hep-th/0404055}{{\tt
  hep-th/0404055}}].

\bibitem{klra00}
M.~Klein and R.~Rabad{\'a}n, {\it D = 4, {N} = 1 orientifolds with vector
  structure},  {\em Nucl. Phys.} {\bf B596} (2001) 197,
  [\href{http://xxx.lanl.gov/abs/hep-th/0007087}{{\tt hep-th/0007087}}].

\bibitem{bgk00}
R.~Blumenhagen, L.~G{\"o}rlich, and B.~K{\"o}rs, {\it Supersymmetric 4{D}
  orientifolds of type {IIA} with {D}6-branes at angles},  {\em JHEP} {\bf 01}
  (2000) 040, [\href{http://xxx.lanl.gov/abs/hep-th/9912204}{{\tt
  hep-th/9912204}}].

\bibitem{bcs04}
R.~Blumenhagen, J.~P. Conlon, and K.~Suruliz, {\it Type {IIA} orientifolds on
  general supersymmetric {Z}({N}) orbifolds},  {\em JHEP} {\bf 07} (2004) 022,
  [\href{http://xxx.lanl.gov/abs/hep-th/0404254}{{\tt hep-th/0404254}}].

\bibitem{bbkl02}
R.~Blumenhagen, V.~Braun, B.~K{\"o}rs, and D.~L{\"u}st, {\it Orientifolds of
  {K}3 and {C}alabi-{Y}au manifolds with intersecting {D}-branes},  {\em JHEP}
  {\bf 07} (2002) 026, [\href{http://xxx.lanl.gov/abs/hep-th/0206038}{{\tt
  hep-th/0206038}}].

\bibitem{blkFB02}
R.~Blumenhagen, B.~K{\"o}rs, and D.~L{\"u}st, {\it Type {I} strings with {F}-
  and {B}-flux},  {\em JHEP} {\bf 02} (2001) 030,
  [\href{http://xxx.lanl.gov/abs/hep-th/0012156}{{\tt hep-th/0012156}}].

\bibitem{wi98}
E.~Witten, {\it D-branes and {K}-theory},  {\em JHEP} {\bf 12} (1998) 019,
  [\href{http://xxx.lanl.gov/abs/hep-th/9810188}{{\tt hep-th/9810188}}].

\bibitem{mimo97}
R.~Minasian and G.~W. Moore, {\it K-theory and {R}amond-{R}amond charge},  {\em
  JHEP} {\bf 11} (1997) 002,
  [\href{http://xxx.lanl.gov/abs/hep-th/9710230}{{\tt hep-th/9710230}}].

\bibitem{grs06}
B.~Gato-Rivera and A.~N. Schellekens, {\it Remarks on global anomalies in
  {RCFT} orientifolds},  {\em Phys. Lett.} {\bf B632} (2006) 728--732,
  [\href{http://xxx.lanl.gov/abs/hep-th/0510074}{{\tt hep-th/0510074}}].

\bibitem{ur00}
A.~M. Uranga, {\it D-brane probes, {R}{R} tadpole cancellation and {K}-theory
  charge},  {\em Nucl. Phys.} {\bf B598} (2001) 225--246,
  [\href{http://xxx.lanl.gov/abs/hep-th/0011048}{{\tt hep-th/0011048}}].

\bibitem{wi82}
E.~Witten, {\it An {SU}(2) anomaly},  {\em Phys. Lett.} {\bf B117} (1982)
  324--328.

\bibitem{tgjl05}
T.~W. Grimm and J.~Louis, {\it The effective action of type {IIA}
  {C}alabi-{Y}au orientifolds},  {\em Nucl. Phys.} {\bf B718} (2005) 153--202,
  [\href{http://xxx.lanl.gov/abs/hep-th/0412277}{{\tt hep-th/0412277}}].

\bibitem{imr01}
L.~E. Ib{\'a}{\~n}ez, F.~Marchesano, and R.~Rabad{\'a}n, {\it Getting just the
  standard model at intersecting branes},  {\em JHEP} {\bf 11} (2001) 002,
  [\href{http://xxx.lanl.gov/abs/hep-th/0105155}{{\tt hep-th/0105155}}].

\bibitem{csu01}
M.~Cveti{\v{c}}, G.~Shiu, and A.~M. Uranga, {\it Three-family supersymmetric
  standard like models from intersecting brane worlds},  {\em Phys. Rev. Lett.}
  {\bf 87} (2001) 201801, [\href{http://xxx.lanl.gov/abs/hep-th/0107143}{{\tt
  hep-th/0107143}}].

\bibitem{fhs00}
S.~F{\"o}rste, G.~Honecker, and R.~Schreyer, {\it Supersymmetric {Z(N) x Z(M)}
  orientifolds in 4{D} with {D}-branes at angles},  {\em Nucl. Phys.} {\bf
  B593} (2001) 127--154, [\href{http://xxx.lanl.gov/abs/hep-th/0008250}{{\tt
  hep-th/0008250}}].

\bibitem{cps02}
M.~Cveti{\v{c}}, I.~Papadimitriou, and G.~Shiu, {\it Supersymmetric three
  family {SU(5)} grand unified models from type {IIA} orientifolds with
  intersecting {D6}-branes},  {\em Nucl. Phys.} {\bf B659} (2003) 193--223,
  [\href{http://xxx.lanl.gov/abs/hep-th/0212177}{{\tt hep-th/0212177}}].

\end{thebibliography}\endgroup

\end{document}